\documentclass[12pt,a4paper]{article}
\usepackage[backend=biber,natbib=true,style=authoryear,uniquename=false,maxcitenames=2,maxbibnames=10,hyperref=auto,uniquelist=minyear,labeldateparts=true, nohashothers=true, useprefix=true, autocite=footnote, mergedate=true]{biblatex}
\usepackage[utf8]{inputenc}
\usepackage{amsmath}
\usepackage{amsfonts}
\usepackage{biblatex}
\usepackage{amssymb}
\usepackage{graphicx}
\usepackage{pdflscape}
\usepackage{rotating}
\usepackage{booktabs,caption}
\usepackage{booktabs}
\usepackage[left=2.3cm,top=2.4cm,right=2.3cm,bottom=2.4cm]{geometry}
\usepackage{array}
\usepackage{tabularx}
\usepackage{hyperref}
\usepackage{caption}
\usepackage{float}
\usepackage{threeparttable}
\usepackage{setspace}
\usepackage{etoolbox}
\hypersetup{colorlinks, breaklinks=true, setpagesize=true, linkcolor=[rgb]{0,0.3,1}, citecolor=[rgb]{0,0.3,1}, urlcolor=[rgb]{0,0.3,1} }

\DeclareNameAlias{sortname}{family-given}
\providetoggle{blx@skipbiblist} 
 
\begin{document}
\author{DUYGU BUYUKYAZICI}
\title{Religiosity and Innovation Attitudes: An Instrumental Variables Analysis}
\author{Duygu Buyukyazici\footnote{\href{mailto:duygu.buyukyazici@imtlucca.it.}{duygu.buyukyazici@imtlucca.it.} Laboratory for the Analysis of Complex Economic Systems, IMT School for Advanced Studies, piazza San Francesco 19 - 55100 Lucca, Italy.}  \\  Francesco Serti\footnote{IMT School for Advanced Studies Lucca, Italy}}
\date{}

\onehalfspacing

\maketitle
{\normalsize }
\begin{abstract}
\noindent Estimating the influence of religion on innovation is challenging because of both complexness and endogeneity. In order to untangle these issues, we use several measures of religiosity, adopt an individual-level approach to innovation and employ the instrumental variables method. We analyse the effect of religiosity on individual attitudes that are either favourable or unfavourable to innovation, presenting an individual's propensity to innovate. We instrument one's religiosity with the average religiosity of people of the same sex, age range, and religious affiliation who live in countries with the same dominant religious denomination. The results strongly suggest that each measure of religiosity has a somewhat negative effect on innovation attitudes. The diagnostic test results and sensitivity analyses support the main findings. We propose three causality channels from religion to innovation: time allocation, the fear of uncertainty, and conventional roles reinforced by religion.\\
\medskip

\noindent JEL-Codes: Z12, O31.     
\end{abstract}

\newpage


\section{Introduction} 

There have been numerous studies on the economic consequences of religion since Max Weber's \textit{The Protestant Ethic and the Spirit of Capitalism} (1905/\cite*{Weber}), where he discusses that the emergence of modern capitalism in Europe was a product of Protestant Reformation that fostered certain traits of people including work ethic and thrift. Despite Weber's thesis being more of a verbal observation rather than an empirical argument, and many empirical studies prove Weber to be wrong (see \cite{Samuelsson}; \cite{BeckerWoessmann}), the idea that religion affects economic outcomes through shaping and transforming individual preferences and behaviours is far from being a myth. The most important connections between economics and religion happen through the effects of religion on economically important individual behaviours —including consumption patterns, saving patterns, time allocation decisions, marriage, fertility, gender roles in family and society― and traits ―such as trust, honesty, thriftiness, tolerance to dissimilarity, willingness to work hard, openness to strangers, being prone to crime. The present paper explores one dimension of this connection. It examines the effect of religion on innovation at the individual level by focusing on some economically important individual beliefs, behaviours, and traits that we call innovation attitudes. 


To our knowledge, \textcite{Benabouetal2013} is the first empirical study at the intersection of religion and innovation. Following their contribution, the research in this field has begun to flourish (\cite{Perret2014}; \cite{Chenetal2014}; \cite{Benabouetal2015}; \cite{Huang16}; \cite{CinnirellaStreb2017}; \cite{Assouad18}; \cite{Recio19}). Previous studies have consistently demonstrated a correlation between religion and innovation yet recognise that observed associations probably do not identify whether religion has a causal impact on innovation due to the potential endogeneity of religion. Given the difficulty of defining and measuring all potentially non-ignorable factors, some unobservable components of the error term are likely to be correlated both with religion and innovation, leading to an omitted variable bias. Moreover, the process of causation is generally considered from religion to innovation. Nevertheless, also higher levels of innovation can affect religion through economic growth and development as conceptualised in the secularisation hypothesis (\cite{Iyer2016}), raising the question of reverse causality.

Motivated by these premises, the present study introduces the first attempt to untangle the endogeneity of religion with respect to innovation, aiming to provide a causal interpretation of the effect of religion by focusing on individual religiosity and innovation attitudes. 
We use eight waves ―between 2002 and 2016― of the European Social Survey (ESS), an academically driven, cross-sectional, and an individual-level data set containing observations for 36 European countries. To our knowledge, the ESS has not been used by prior studies on innovation and religion. We define four different measures of religiosity —the degree of being religious, the frequency of attendance religious activities, the frequency of praying, and religiosity index, which is the principal component of the first three measures— since prior work has shown that different religiosity measures are likely to have different effects on economic outcomes (\cite{BarroMcCleary2006}). Innovation is measured at the individual level with self-reported personal traits and beliefs related to the propensity to innovate of an individual. These innovation attitudes are broken down into positive and negative innovation attitudes. Positive innovation attitudes (PIA) are creativity, being free, being different, and being adventurous; negative innovation attitudes (NIA) are following traditions and following rules. Using attitudes, instead of an aggregate measure of innovation output such as patents per capita, allows the influence of institutional level confounding factors to be mitigated. Moreover, a large set of fixed effects is used to minimise the effect of (1) country-level economic and institutional confounding factors, (2) individual-level unobservables such as human capital and ability, and (3) global shocks. 
Finally, we instrument each measure of religiosity with the average religiosity of people of the same sex, age range, and religious affiliation who live in countries with the same dominant religious denomination. In other words, the religiosity of individual $i$ from country $c$ who belongs to a given religion (does not belong to any religion) is instrumented by the average religiosity of people who belong to the same religion (do not belong to any religion), share the same age range and gender with $i$, and live in countries that have the same dominant religious denomination with country $c$. By doing so, we aim to eliminate the effect of individual-level unobservables. Furthermore, we use religious affiliation along with religious intensity to mitigate reverse causality since religious affiliation is inherited and thus is not responsive to changes in other factors.

We start by estimating a linear model with OLS; then we apply our IV strategy. The first-stage results show that all the excluded instruments are strongly and significantly correlated with the religiosity measures. The OLS results show that each measure of religiosity is significantly and positively correlated with NIA, suggesting that religiosity has a negative effect on innovativeness. Regarding PIA, religiosity measures are positively correlated with creativity and being different, while negatively correlated with being free. In the IV results, the significant and positive association remains robust between the measures of religiosity and NIA. But the overall picture changes with PIA. The religiosity measures are no longer significantly related to creativity and being different. In contrast, the negative relation between religiosity and being free persists, meaning the OLS results suggesting that religiosity fosters some favourable attitudes to innovation may be driven by endogeneity. 


The negative effect of religiosity is robust to a series of sensitivity analyses. First, we introduce a different dependent variable, an overall summary measure of innovation attitudes: \textit{innosum}. OLS and IV results of \textit{innosum} support the negative effect of religiosity on innovativeness. Second, we reconsider the religiosity index (\textit{religiosity}) and drop the degree of being religious (\textit{degree}) from its components, thinking that \textit{degree} might be prone to self-report bias since it is a self-evaluation of a belief, a value. In contrast, the other components of \textit{religiosity} (frequency of pray and attendance to religious services) are measures of an activity, an action that is difficult to over or underestimate. The estimated coefficients are pretty close to the initial one, meaning that the initial instrument of \textit{religiosity} is not significantly affected by the potential self-report bias of \textit{degree}. Third, we re-estimate the baseline model by reducing the main sample in four ways to analyse if minorities and/or outliers affect the main findings. Fourth, we consider two possible violations of the exogeneity condition. Sensitivity analyses, overall, demonstrate that coefficient values might change depending on the specification choices, but the main results stay robust.

Lastly, we discuss three possible causality channels from religion to innovation: time allocation, the fear of uncertainty, and roles reinforced by religion (individualism- collectivism and conventional gender roles). First, we consider the opportunity cost of time spent on religious activities, arguing that religious participation might negatively affect the human capital formation, which is essential to innovation. Since time is scarce, if an individual allocates a certain amount of time to religious activities, then there will be a decrease in the maximum amount of potential time devoted to human capital formation. Our empirical results support this argument. \textit{Attendance in religious services} is the most robust religiosity measure throughout the analysis. Second, we argue that for religious people, personal uncertainty is mitigated by their faith and by the support of religious institutions such as churches, mosques, or religious social organisations. Therefore, they have less experience dealing with risks and uncertainty inherent to the innovation process. Third, we assert that high religiosity may foster a collectivist culture that highly values established rules and might leave limited space for reformist and creative endeavours, negatively affecting innovation. We also consider the long-standing argument that religions foster gender discrimination by imposing stricter rules for women, influencing the access of women to education, liberty, labour force, and social and legal rights. We empirically test these hypotheses and find supporting evidence. 

The remaining of the paper is structured as follows. Section \ref{section: relecon} briefly reviews the existing literature on religion and innovation. Section \ref{section: inno} looks at innovation from a behavioural point of view and describes innovation as a mindset, establishing the theoretical foundation for our innovation measures. Section \ref{section: strategy} extensively describes the estimation strategy and discusses possible endogeneity channels together with our strategies to overcome them. Section \ref{section: result} reports OLS and IV results and discusses the main findings. Section \ref{section: robust} introduces the sensitivity analyses. Section \ref{section: chan} discusses three possible channels from religion to innovation. Section \ref{section: conc} provides concluding remarks. In the Appendix, we present data summary tables and a detailed variable index (Section A), the first stage results for IV estimates (Section B), the OLS results for each innovation attitude (Section C), and the IV results for each innovation attitude (Section D).

\section{Literature}
The present study mainly relates to the expanding literature on the effects of religion on economic outcomes. The first subsection briefly overviews the literature on the relation between religion and economic outcomes, emphasizing innovation. The second subsection summarizes the literature on the individual and behavioural aspects of innovation, presenting a theoretical framework for our innovation measures.

\subsection{Religion, Economic Outcomes, and Innovation}
\label{section: relecon}

The interaction of economics and religion has long been a topic of sociological analyses. Until recently, economists have not focused on the topic, possibly due to a lack of reliable data on religion and potential methodological issues such as endogeneity. Nevertheless, recent decades have witnessed expanding literature on the economics of religion\footnote{For an in-depth literature review, please refer to \textcite{Iannaccone1998} and \textcite{Iyer2016}.} in which religion has been integrated into economic research in three different ways: (i) applying the methodology of economics to religion (\textit{e.g.,} microeconomic theory to analyse religious behaviour of individuals, groups, or institutions); (ii) analyzing economic outcomes of religion (\textit{e.g.,} the effect of religiosity on economic growth); (iii) making use of holy books and theological norms to praise or criticize economic behaviours and economic policies (\cite{Iannaccone1998}), (\textit{e.g.,} Islamic economics is critical about interest rates since Islamic law, \textit{Sharia}, prohibits any interest paid on loans of money). The present study mainly contributes to the second category by examining the effect of religion on innovation.

The work of \textcite{Benabouetal2013}, to our knowledge, is the first empirical analysis of religion and innovation, while qualitative studies were present before (\cite{Kalliny07}; \cite{Preble12}). Using the World Intellectual Policy Organization (WIPO) data, they measure innovation with patents per capita. The two religiosity measures, namely, belief in God and being religious, are retrieved from the World Values Survey (WVS). Both cross-country and within-country (USA) analyses indicate a negative relationship between religiosity and innovativeness. In a follow-up paper, \textcite{Benabouetal2015} enlarge the analysis by focusing on the innovation attitudes at the individual level instead of patents per capita which is an ex-post and macro-level measure. They define eleven different innovation attitudes and five alternative measures of religiosity that are retrieved from the WVS. They find an overall negative relation across 52 model specifications. Almost in every specification, greater religiosity is significantly associated with less favourable attitudes toward innovation. Though both papers do not directly try to deal with endogeneity, the latter study decreases the influence of institutional-level confounding factors by using individual-level data.

\textcite{Perret2014} examines the relationship between religious affiliation and innovativeness in Russia. Innovation is measured by the number of patents issued by the Russian Federal Service for Intellectual Property (Rospatent). He finds that only Hindu, Buddhist, and Jewish faiths exhibit significant and negative effects on innovativeness. Christians, Muslims, Jews, and Atheists do not display any significant impact at all. 

 \textcite{Chenetal2014}\footnote{This paper follows \textcite{Kumaretal2011} and \textcite{Kumar2009} that find a positive relationship between the propensity to gamble and risky, lottery-like financial market investments.} analyse the relation between local gambling culture and corporate innovation. ``Many innovative endeavours, \textit{i.e.,} attempts to come up with new products, services and methods, represent gambles because they promise relatively small probabilities of large success and large probabilities of failure" (\cite{Adhikari:Agrawal}, pp. 229). They use religious beliefs as a measure of gambling preferences of firms' local communities. Taste for gambling is addressed by a higher Catholic-to-Protestant ratio assuming that Catholics are more likely to take risks. They find that the firms headquartered in areas with a taste for gambling tend to be more innovative.

\textcite{CinnirellaStreb2017} analyse the effect of religious tolerance on innovation in Prussia by assuming that tolerance and diversity are conducive to technological creativity and innovation. Religious tolerance is proxied by the index of the religious diversity of the population across 1278 cities in Prussia. Innovation is measured by 1740 patents issued in Prussia between 1877 and 1890. They find that higher levels of religious tolerance have a strong positive impact on innovation during the second industrial revolution. They also show that the culture of tolerance did not stem from a particular denomination but rather from the presence of different denominations and churches.

The research at the intersection of religion and innovation has begun to flourish following the aforementioned studies (\cite{Huang16}; \cite{Assouad18}; \cite{Recio19}). However, no empirical study has considered the potential endogeneity of religion so far, despite reverse causality and identification problems have been recognized in the literature (\cite{Guisoetal2003}).

The process of causation is generally considered as religion affects economic outcomes by playing an integral role in shaping culture, individual preferences, traits, and beliefs. However, the direction does not have to be from religion to economic outcomes. The secularization hypothesis suggests that economic outcomes, such as individual and country-level income and growth, may affect religious behaviour, presenting a possible reverse causality channel for our study. The core argument of secularization is that religiosity decreases as a nation gets richer. There has been a long debate on whether the secularization hypothesis holds. Many economists have analysed the main channels that might facilitate the secularization effect, yet it is unlikely to say there is a consensus in the literature. \textcite{RuiterTubergen} argue that modernization reduces the need for religious reassurance because it enables the creation of more securities ―whether it be financial, social, or political― for the population, thus decreasing the level of religiosity. \textcite{Iannaccone2008} provides valuable insights by analyzing the secularization with retrospective questions from 30 nations that stretch from the 1920s through the 1980s. She finds that secularization cannot explain the continued vitality of religion in the USA, one of the world's most modernized nations. On the contrary, she finds some favourable evidence for secularization in Britain, France, and Germany. 


Apart from religion, there is a relatively vast literature on culture and innovation that has adopted the Hofstede model (1980/\cite*{Hofstede2001}) (\cite{Shane1992}, \cite*{Shane1993}, \cite*{Shane1995}; \cite{GorodnichenkoRoland}; \cite{TaylorWilson}; \cite{Petrakisettal2015}; \cite{Chenetal2017}; \cite{Kostisetal}). Hofstede analyses the differences in national cultures using the data of business employees from more than 50 countries. He empirically identifies and validates five independent dimensions affecting economic attitudes and outcomes ―namely, power distance, uncertainty avoidance, individualism versus collectivism, masculinity versus femininity, long-term versus short-term orientation― on which national cultures differentiate \footnote{Hofstede initially defines four dimensions. In the second edition of the book (\cite*{Hofstede2001}), he adds \textit{long-term versus short-term orientation} as the fifth dimension.}. 
These five dimensions together form a model of differences among national cultures and have been pretty influential in the literature on culture and economic outcomes. \textcite{Kirkmanetal} review 180, and \textcite{Sondergaard} reviews 61 published studies that used Hofstede's dimensions, and they both confirmed their relevance to the differences in national cultures. In this study, we consider uncertainty avoidance and individualism-collectivism as potential causality channels from religion to innovation (see Section 6).

\subsection{ Innovation as a Mindset} \label{section: inno}
Innovation can be defined in diverse forms. \textcite{WangAhmet} review the literature on innovation and propose five different innovation types that jointly determine the overall innovativeness of an organisation: (1) product innovativeness (newness, novelty, originality, or uniqueness of products), (2) market innovativeness (newness of approaches that companies adopt to enter and exploit the targeted market), (3) process innovativeness (introduction of new production methods, new technology, and new management approaches that can be used to improve management and production processes), (4) strategic innovativeness (an organisation's ability to manage ambitious organisational objectives and to identify a mismatch of these ambitions and existing resources to use limited resources creatively) and (5) behavioural innovativeness (enables the formation of an innovative culture, the general internal capacity for new ideas and innovation and screens through teams, individuals, and management). Different types of innovation are expected to be related to different individual and organisational traits. \textcite{CampsMarques} name these traits as innovation enablers, \textit{i.e.,} a set of general capabilities driven by social capital that contributes to favour innovation. They associate with innovation enablers and innovation types as follows: goal alignment and knowledge enhancement for product innovation; knowledge enhancement and cooperation for process innovation; cooperation and associability for strategic innovation; associability, risk-taking and creative environment for market innovation; risk-taking, creative environment, communication and information flow for behavioural innovation. Among the innovation types, behavioural innovativeness is not only crucial for overall innovativeness but also for the other types of innovation, given that it enables the formation of an innovative culture by being present at the various levels for individuals, teams, and management (\cite{WangAhmet}). Cultivating an innovative culture through behavioural innovativeness induces other types of innovation mainly by creating an environment where new ideas are supported, failure is tolerated, creativity is promoted, and risk-taking is encouraged. 

Many studies underline the importance of innovative culture and mindset by focusing on behavioural aspects of innovation. \textcite{Goldstone1987} argues that what separated the East from the West in the early modern world was not capitalism nor rationalisation of institutions; it was the willingness to innovate, which was fostered by revolting from orthodoxy and by cultivating tolerance to internal diversity, which enhanced openness to taking risks. He compares three historical crises in the 17th-century \textemdash the Stuart king crisis in England, Celali revolts in Ottoman Turkey, and revolts against the Ming dynasty in China\textemdash that have economic roots. He argues that England fostered tolerance to diversity and openness to take risks by adopting a relatively new, risk-taking path, which resulted in a higher propensity to innovate. On the contrary, Ottoman Turkey and China followed cultural orthodoxy and suppressed alternatives, resulting in an unfavourable environment to innovate. England, therefore, reached dynamism and growth, while Ottoman Turkey and China had stagnation. Even though the analysis of Goldstone is at the macro level, the cultural traits mentioned ―tolerance to diversity, openness to taking risks, and revolting from tradition― are perfectly plausible at the individual level as well. 

\textcite{DziallasBlind} review the extensive literature on organisational innovation indicators published between 1980 and 2015. They find that \textit{innovation culture} has been one of the most used company-specific innovation indicators. Innovation culture is addressed by different indicator categories such as creativity, attitudes toward science and technology, social innovation climate and trust, support of new ideas, openness to new fields, the openness of the company towards change and innovation, resistance to change, willingness to exchange ideas, tolerance for innovation failures, et cetera. All these indicators are related to behavioural aspects of innovation. 
 
\textcite{Kahn2018} underlines that innovation should be recognized as three different things: (1)\textit{ innovation is an outcome} (product innovation, process innovation, marketing innovation, business model innovation, supply chain innovation, organizational innovation), (2) \textit{innovation is a process }(innovation process, product development process), and (3) \textit{innovation is a mindset} (individual mindset, organization culture). ``Mindset aligns employees and manifests the culture needed for innovation to happen. Encompassing a mindset that predisposes individuals and organisations to be risk-taking, cross-disciplinary, and open to varied ways of thinking helps establish the state necessary for innovation; state implies something habitual and lasting. It is about instilling and ingraining a mindset that prepares the individual and organisation for innovation so that there is proper engagement in the innovation process to achieve the desired innovation outcome." (\cite{Kahn2018}, pp. 459).

The literature mentioned above demonstrates that innovation is strongly associated with individual traits such as tolerance to diversity, risk-acceptance, creativity, and deviating from traditional paths and rules, which jointly form innovative cultures of teams, groups, and organisations. Nevertheless, previous research has generally measured innovation as an outcome, using patents, R\&D expenditure and technology improvements' expenditure at the country or firm level. There might be a couple of drawbacks in using patents as a sole measure of innovation outcome. \textcite{Moser2012} argues that almost a thousand empirical studies have used patent counts to address innovation, mostly without controlling for variation across industries and over time. However, the percentage of patents varies significantly across industries. Furthermore, relying on the historical evidence, the majority of innovations were outside of the patent system, even in countries with a developed patent system, such as the mid 19th century USA (\cite{Moser2013}). He uses the exhibition data of 8079 innovations between 1851 and 1915 and underlines that 89\% of British innovations and 85\% of American innovations were not patented. The probability of an innovation/invention being patented mainly depends on the country's patent system and law, among many other determinants. Furthermore, inventors do not have to use patents to protect their invention; they may use other property rights such as lead-time advantage and secrecy. Hence, patents as an indicator of innovation would be a biased measure in cross-country and cross-industry studies.

Another critical issue is that patents are ex-post measures of innovation that reflect country-specific institutional constraints. In other words, patents measure some portion of occurred innovation. However, they have little to say about the propensity to innovate, mainly determined by the behavioural and cultural traits mentioned above. One cannot address unhappened, \textit{i.e.,} blocked by unfavourable institutional constraints or cultural and religious traits, innovation with an ex-post measure. Many papers in this regard address culture and religion by individual measures ―belief in god, degree of religiosity, willingness to take a risk, tolerance to diversity, et cetera― but address innovation by country-level patents, which is a macro and ex-post measure. Therefore some economists ( \cite{Guisoetal2003},  \cite*{Guisoetal2006}; \cite{Estebanetal2015}, \cite*{Estebanetal2018}, \cite*{Estebanetal2019}) suggest using individual attitudes as a measure of innovation rather than aggregate macro outcomes such as patents.

\section{Estimation Strategy}\label{section: strategy}

\subsection{Data}
We use pooled cross-sectional data from the European Social Survey (ESS). The ESS is a biennial, academically-driven, and individual-level survey that includes large groups of observations about preferences, beliefs, and attitudes toward many different subjects, including immigration, religion, political choices, trust, and markets. The ESS is available for 36\footnote{Albania, Austria, Belgium, Bulgaria, Croatia, Cyprus, Czech Republic, Denmark, Estonia, Finland, France, Germany, Greece, Hungary, Iceland, Ireland, Israel, Italy, Kosovo, Latvia, Lithuania, Luxembourg, Netherlands, Norway, Poland, Portugal, Romania, Russia, Slovakia, Slovenia, Spain, Sweden, Switzerland, Turkey, Ukraine, and United Kingdom. We have dropped Israel from the sample since it is not located in mainland Europe and is the only dominantly Jewish country.} countries for nine waves from 2002 to 2018. We use the first eight waves that result in more than 380 000 observations. The survey does not homogeneously include all countries across all waves, meaning that some countries are not surveyed in certain waves and thus are represented in different proportions. 
Table A1 displays the surveyed countries across the waves. 


The main variables we are interested in are presented in different groups below. Detailed descriptions of all variables are presented in Appendix A. Descriptive statistics of the variables are presented in Table A2.

\subsection{Measuring Innovation}
 Innovation is a complex phenomenon as religion and is challenging to measure due to its many dimensions. It is important to underline that innovation mainly occurs in two phases: initiation and implementation of innovation (\cite{Glynn1996}; \cite{WilliamMcquire}). The first phase is about creating ideas which is, in general, an individual task. On the other hand, the second phase might include teams, groups, and management activities, requiring an innovative organisational culture and mindset. Most papers on innovation literature focused on the second phase by using patents as the only indicator of innovation, as underlined in Section 2.2. On the contrary, our innovation measures are mainly related to the first phase. Nonetheless, individual innovation measures are likely to stay vital in the second phase as well.
  
 Following \textcite{Guisoetal2003}, we address innovation by the individual propensity to innovate by introducing innovation attitudes that are designated under the guidance of the literature briefly reviewed in Section 2. In this regard, we exploit several variables in the ESS: \textit{\textit{creative}} (important to be creative), \textit{\textit{different}} (important to be different), \textit{\textit{free}} (important to be free), \textit{\textit{adventurous}} (important to be adventurous), \textit{\textit{traditions}} (important to follow traditions), \textit{\textit{rules}} (important to follow rules). We categorise these variables as \textit{positive} or \textit{negative innovation attitudes}. Positive innovation attitudes (PIA) are essential behavioural conditions and favourable personal traits to innovate. Negative innovation attitudes (NIA) are personal traits that contradict with an innovative mindset, thus are unfavourable to innovation. We nominate four favourable —\textit{\textit{creative}, \textit{different}, \textit{free}, \textit{adventurous}}— and two unfavourable —\textit{\textit{traditions}, \textit{rules}}— traits to innovation. 
 
We create two average measures: \textit{\textit{inposav}} and \textit{innegav}. Inposav, average positive innovation attitudes, is the mean of iprcrtiv, \textit{different}, \textit{free} and \textit{rules}; while \textit{innegav}, average negative innovation attitudes, is the mean of \textit{traditions} and \textit{rules}. \textit{Innosum} is a summary measure of innovation attitudes computed as the sum of \textit{creative}, \textit{different}, \textit{free}, \textit{adventurous} minus \textit{traditions} and \textit{rules}. 
 
\textit{Inposav} and \textit{innegav} are the primary dependent variables in our analysis, yet we also analyse each attitude that composes them. The corresponding tables of OLS and IV estimates of each attitude are presented in Appendix C and D for the sake of brevity. Detailed information on how we construct innovation variables is presented in Appendix A.

\subsection{Measuring and Instrumenting Religiosity}
The variables \textit{belonging, pbelonging, denomination, pdenomination, \textit{degree}, \textit{\textit{attendance}}, \textit{pray}} are religion-related observations in the ESS. \textit{Belonging, pbelonging} stand for respectively present and past belonging to a religion; \textit{denomination, pdenomination} present respectively current and past denomination of an individual; \textit{degree} is the degree of religiosity, \textit{attendance} is the frequency of attendance to religious activities, \textit{pray} is the frequency of praying.  \textit{Belonging, pbelonging, denomination, pdenomination} indicate religious affiliation  —belonging to a particular religion or denomination— while \textit{degree}, \textit{attendance}, \textit{pray} reveal religious intensity that we simply name religiosity. 


Different measures of religiosity are likely to have different effects on socioeconomic variables, reflecting the multi-dimensional nature of religion (see \cite{BarroMcCleary2006}). For instance, the religion market model \footnote{Developed by \textcite{FinkeStark}, \textcite{FinkeIannacone}.} suggests that if a state has a formal religion, imposing the state religion and limiting the other religions, participation in religious services is likely to decrease because the variety of services is subject to suffer (\cite{BarroMcCleary2006}). Nevertheless, people might keep firm religious beliefs without attending state-regulated formal religious services. Accordingly, low attendance may be caused by the supply side of the religion market and may not mean low religiosity. Motivated by this reasoning, we use three different measures of religious intensity (\textit{degree}, \textit{attendance}, \textit{pray}) as explanatory variables. We also create an index of general religiosity, \textit{religiosity}, that is the principal component of the three religious intensity variables.



Our IV strategy is inspired by \textcite{Estebanetal2019} where they examine the role of religiosity, along with personal liberties, in influencing the decision of labour effort. They construct an instrumental variable for religious intensity by computing the average religious intensity of people of the same sex, age bracket, and religious denomination in neighbouring countries, assuming that religiosity is a cultural trait shared by people over national borders. By doing so, they could eliminate the possible omitted variable problem for individual unobservables, such as ability, because the instrument is other people's average religiosity, which is likely to be uncorrelated with individual $i$'s personal characteristics. We partly adopt this strategy to build instruments for religious intensity variables by computing the average religiosity of people of the same sex, age range, and religious affiliation who live in countries with the same dominant religious denomination. In other words, the degree of religiosity of individual $i$ from country $x$ who belongs to a religion (does not belong to a religion) is instrumented by the average religiosity of people\textemdash who belong to a religion (do not belong to a religion), share the same age range and gender with $i$ \textemdash who live in the countries that have the same dominant religious denomination with the country $x$ has. 

Briefly, our instruments are constructed by taking into account four elements: age range, gender, religious affiliation, and dominant religious denomination. We use 11 age ranges (15-20,..., 60-65, 65+)\footnote{We have also tried 6 age ranges (15-25, 25-35...55-65, 65+). The estimation results were pretty similar to those estimated with 11 age ranges.}. Gender is a dummy variable 0= male and 1=female. Religious affiliation is based on variable \textit{belonging} which is a dummy, 0=does not belong to a religion and 1=belongs to a religion. We determine dominant religious denominations with the help of variables \textit{denomination} and \textit{country} after weighted by population size weights. We calculate the percentages of each religious denomination in each country in the sample and identify the religious denomination that has the highest percentage in a country as \textit{the dominant religious denomination}. The dominant religious denomination of each country and corresponding percentages are presented in Table A1 in Appendix. There are five dominant religious denominations: Roman Catholics, Protestants, Eastern Orthodox, Muslims, and Atheists (do not belong to a religion).

The rationale behind our IV strategy emerges from the following points. \textcite{Guisoetal2003} show that individuals who were raised in a particular religious environment possess common preferences and beliefs even though they refuse to belong to any religion as adults. The dominant religious denomination in a county is the leading actor that forms the characteristics of the religious environment in which individual preferences and beliefs are being shaped. Some individuals inherit more, some less, but the dominant religious denomination determines the rules. Thus, instrumenting individual $i$'s religiosity by the average religiosity of people who are subject to the same dominant religious denomination in their country (along with the same age range, gender, and affiliation) means that individual $i$'s average religiosity is instrumented by the average religiosity of people who live in a similar religious environment as that of $i$. Here we assume that, as Esteban et al., religiosity is not a national trait but a cultural trait that transcends national borders. People raised in a particular religious environment are likely to share a significant part of their culture with those raised in a similar religious environment regardless of national borders. However, they do not necessarily share the same institutional environment. We, therefore, use a large set of fixed effects to eliminate country-level institutional differences. Moreover, our sample consists of European countries that share many common values and customs. 

We use religious affiliation variables, along with religious intensity variables when constructing instruments. Religious affiliation generally passes down from generation to generation and is thus inherited and subject to slow change. The ESS data justify this proposition. Only 9.8\% of individuals in the sample have changed their religious denomination\footnote{Calculated by using the observations of \textit{denomination} and \textit{pdenomination}.}, meaning that religious affiliation is pretty stable throughout one's lifetime and does not significantly change along with changes in other factors, including innovation. On the other hand, religious intensity is more likely to change over one's lifetime, making it more vulnerable to reverse causality \footnote{\textcite{BotticiniEckstein} and \textcite{Guisoetal2006} underline that not only religious affiliation but also religious practices are modified only over centuries; thus any aspect of religion can largely be assumed invariant over one's lifetime.}.
Moreover, it is expected that the preferences and beliefs of an individual who belongs to a religion would be affected by the religious environment around her more than a person who does not belong to a religion. 

Another crucial point is that the average religiosity of people who belong to a particular religious denomination may inherently be more intense than those who belong to another religious denomination. For instance, Muslims may be inherently more religious than Protestants\footnote{The data support this proposition. For instance, Table A3 in Appendix A shows that Muslim countries exhibit significantly higher means for the religiosity measures.} We, therefore, use religious denomination fixed effects to be able to compare the religiosity of people from different religious denominations.

In our IV strategy, the exclusion restriction would be violated if belonging to a religion is not inherited from past generations but chosen during one's lifetime. In this case, belonging to a religion would be correlated with individual unobservables and thus cannot be considered exogenous. Nevertheless, as mentioned above, only 9.8\% of our sample have changed their religious affiliation (once-believers). As a robustness check, we drop once-believers from the main sample and reestimate the main specifications (Table 9, columns 7 and 8). The results are very similar to the full sample. Another scenario in which our instruments violate the exclusion restriction is that countries with the same dominant religious denomination are subject to correlated shocks. We use survey year fixed effects to capture global shocks along with country and country-survey year interaction fixed effects. 

It is likely that within-group observations are correlated. In our case, those groups are the five elements we use to construct instruments: age, gender, religious affiliation, denominations, and country. For instance, the religiosity of a Catholic woman at age 36 from a dominantly Catholic country could be correlated to the religiosities of other middle-aged Catholic women from dominantly Catholic countries. It is known that ignoring within-group correlations leads to understated standard errors (\cite{Sheppard96}). We thus use clustered standard errors at the instruments' level to avoid within-group correlations in error terms.

\subsection{Dealing with Endogeneity}
 Religious beliefs are embedded in social behaviour and culture, meaning that extracting the pure effect of religion can be challenging. One, therefore, must extensively consider endogeneity when working on religion. We ultimately address the possible inherent endogeneity of religion by employing the instrumental variable strategy. In addition, we consider the primary sources of endogeneity below and discuss how we overcome them. 
\medskip

\textit{\textbf{Omitted variables.}} Religion is a complex phenomenon. Therefore, it is plausible to assume that possible confounding factors and unobservable components could be added to the error term when including religion-related variables into regression, causing omitted variable bias. To minimize the effect of confounding factors, we employ a multi-way fixed effects strategy along with controls of individual characteristics.\footnote{See \textcite{GuimaraesPortugal} and \textcite{Correia} for further information on multi-way fixed effects in linear models.} We control for the following fixed effects: country, survey year, country and survey year interaction, religious denominations, occupation categories, and income level. Country and survey year fixed effects eliminate the impact of institutional features and country-level economic determinants, which are the same for the entire country. Occupation and income level fixed effects aim to minimize the differences of individual-level unobservables across respondents; religious denomination fixed effects help us compare the religiosity of people who belong to different religious denominations. Controlling for a large set of fixed effects conveys the possibility of underestimating the impact of religion, given that religion affects many determinants whose effects might be partially absorbed by fixed effects. Therefore, our estimates can be interpreted as a lower bound of the effect of religiosity on innovation attitudes.


In our model, one possible omitted variable is human capital. It is reasonable to assume that human capital has a feature somewhat unobservable, which gives rise to the omitted variable problem. Occupation and income level fixed effects can eliminate, to some degree, human capital differences among respondents, thereby reducing the possible omitted variable bias. Furthermore, we follow \textcite{BarroLee} and use educational attainment, namely competed years of education, as a measure of human capital. We also control for father's and mother's education levels, which are essential to human capital formation, especially at younger ages. \textcite{Akcigitetal} examine a large data set on innovation and inventors collected for the period between 1880 and 1940 in the USA. They uncover many micro and macro stylized facts at the individual level. One of them is that father's education is a crucial determinant of being an inventor, especially through the channel of a child's education. Hence we control for the father's and the mother's education level since the mother's education is likely to be equally important, if not more, given that mothers spend more time with children in most cultures. Another uncovered fact by Akcigit et al. is that inventors tend to migrate from their birth state to more eligible states for innovation. We, therefore, control for the born-in country of respondent.
\medskip

\textit{\textbf{Reverse causality.}} The direction of causation may go both ways when working on religion. We aim to measure the effect of religiosity on innovation. However, changes in innovation may affect one's religiosity as well, mainly through economic growth and development, which is conceptualized in the secularization hypothesis mentioned in Section 2. To deal with reverse causality in the case of culture and religion-related research, \textcite{Guisoetal2006} suggest focusing only on inherited dimensions of culture, such as religious affiliation and ethnicity, rather than accumulated ones such as social capital. We follow Guiso et al. and use religious affiliation along with religious intensity when measuring religiosity, as explained above. 

Another strategy to mitigate reverse causality is to focus on economic attitudes rather than outcomes, given that economic outcomes are partly determined by the institutional and political environment, which is rather challenging to account for (\cite{Guisoetal2003}, \cite*{Guisoetal2006}). Economic attitudes reveal \textit{individual propensity} to something; in other words, they give insights into the possibility of a particular economic outcome being real. As \textcite{Guisoetal2003} pointed out, asking somebody if she has ever cheated on taxes is different from asking somebody her opinion on cheating on taxes. 
\medskip

\textit{\textbf{Measurement error.}} Data may be measured with errors, such as coding and reporting errors. When an independent variable is measured with error, endogeneity arises since it violates the zero conditional mean assumption. Our data source is the ESS, a comparative cross-national survey that collects measures of individuals' preferences and social and political attitudes across 36 European countries. These measures may contain errors due to differences in concepts measured across the participating countries. The ESS aims to minimize such measurement errors and improve data reliability, validity, and comparability; therefore, it undertakes a range of data quality assessment activities, including the Multitrait-Multimethod (MTMM) and Survey Quality Predictor (SQP). MTMM is an experimental project in which the same respondents are asked three survey questions twice in different concepts by using different response scales each time.\footnote{The detailed test data obtained from MTMM is available online:\\
	$https://www.europeansocialsurvey.org/data/download_mtmm.html$} MTMM measures the measurement quality of individual questions, and it was implemented for all the waves of the ESS. SQP is an open-source database on which the complete set of the ESS questions is evaluated through MTMM design. Due to the rigorous approach of the ESS, we may say that it is unlikely that our independent variables suffer from measurement errors.

Furthermore, the ESS imposes a minimum target response rate of 70\% in each country to minimize non-response bias which may lead over or under-representation of some individuals with certain characteristics. To adjust for non-response bias and others, the ESS data come with three weighting variables: design weights (\textit{dweight}), post-stratification weights (\textit{pspwght}) and population size weights (\textit{pweight}). Design weights correct for sample selection bias, given that some countries use complex sampling designs and respondents have different probabilities of being a part of the sample. Post-stratification weights adjust for uneven representation of sub-groups and sub-populations with certain characteristics, correcting for sample and non-response biases. Population size weights are used when data from more than two countries are combined. Since each country has a different population size but a similar sample size, \textit{pweight} corrects for over or under-representation of countries concerning their population. We analyse more than one country and total averages of countries in some cases, hence we must use a combination of either \textit{dweight} and \textit{pweight} or \textit{pspwght} and \textit{pweight}. Given that \textit{pspwght} includes \textit{dweight}, we use the combination of \textit{pspwght} and \textit{pweight} by generating a new variable \textit{gweight= pwght $\ast$ pweight}.\footnote{For detailed information on weighting the ESS data, please refer to the guide \textit{``Weighting European Social Survey Data"}.}

\subsection{Model}
Based on the strategy above, we aim to estimate a causal relationship that goes from religion to innovation. We first estimate equation 1 with OLS. We then instrument religiosity measures, $R_{i}$, with $Riv_{i}$ and estimate the baseline model with IV/2SLS \footnote{We use ``\textit{ivreghdfe}" command of Stata which allows IV/2SLS estimation with multi-way fixed effects.}. The first stage model is specified in equation 2. All models include standard demographic controls and fixed effects. Control variables are gender (\textit{gender}), age (\textit{age}), completed years of education (\textit{education}), parent's highest education level (\textit{mothere, fathere}), born in country (\textit{bornc}), paid work status (\textit{paidwork}), whether had a paid job before (\textit{pwbefore}) whether have a partner or wife living in the house (\textit{partner}), whether have a child living in the house \textit{(child)}, subjective health status (\textit{health}). Fixed effects are country dummies (\textit{country}), survey year (\textit{essround}), country and survey year interactions (\textit{cness}), religious denominations (\textit{\textit{denomination}}), occupation categories (\textit{occupation}), and income level (\textit{income1} \& \textit{income2}).

\begin{equation}\label{eq1}
I_{i}=\alpha_{0} + \beta R_{i}+ \theta\mathbf{X}_{i} + \delta\mathbf{F} +\varepsilon_{i}
\end{equation}

\noindent where $I$ denotes innovation attitudes (\textit{creative}, \textit{different}, \textit{free}, \textit{adventurous}, \textit{traditions}, \textit{rules}, \textit{inposav}, \textit{innegav}) of individual $i$, $R$ stands for religiosity variables (\textit{degree}, \textit{attendance}, \textit{pray}, \textit{religiosity}), $\beta$ is the coefficient of interest, the vector $X$ denotes control variables, the vector $F$ represents the complete set of fixed effects absorbed into estimates, $\varepsilon$ is the unobserved error term. 

\begin{equation}\label{eq2}
 R_{i}=\alpha_{1} + \zeta Riv_{i} + \vartheta\mathbf{X}_{i} + \eta\mathbf{F} +\epsilon_{i}
\end{equation}

\noindent where $Riv$ denotes instrumental variables that we use as an exogenous predictor of each religiosity measure.

\section{Results}\label{section: result}
\subsection{OLS Estimates}
We begin by examining the relationship between religiosity and innovation attitudes estimated with OLS. All specifications include multi-level fixed effects that are fully absorbed into all specifications. We gradually control for socio-demographic indicators. Regression errors are likely correlated within groups since we expect endogeneous religiosity variables. Moreover, the data set we use has a grouped structure, \textit{i.e.,} countries. We, therefore, report robust standard errors clustered at the level of instruments, namely clustered for the country, gender, age, religious affiliation and denomination groups. We also performed OLS estimates with robust standard errors and found very little difference so that the significance levels of the coefficients did not change. However, we prefer to report the clustered version due to the unity of the analyses throughout the study.

Table 1 presents the variations in \textit{inposav} concerning different measures of religiosity. \textit{Religiosity} is the independent variable in the first three specifications. The first specification, column 1, includes control variables for gender, age, age squared, education, and the complete set of fixed effects. The insignificant correlation coefficient in column 1 endures in columns 2 and 3 after adding other control variables. Other measures of religiosity display different patterns. \textit{Degree} is statistically significant at the 99\% level in column 4 and stays robust to the further controls in column 5. On the contrary to the other religiosity measures, \textit{attendance} shows a negative relationship with \textit{inposav} in column 6. It stays robust when we add further controls in column 7. \textit{Pray} is positive and significant at the 99\% level in column 8 and stays robust to the further controls, yet the coefficient and significance level decrease in column 9. 

The detailed OLS estimates, broken down into each measure of innovation attitudes, are presented in Appendix C. Table C1 reports a stable, positive, and significant relation between \textit{importance of creativity} and the religiosity measures, except for \textit{attendance}, which has insignificant coefficients. In Table C2, all religiosity measures are positively and significantly associated with \textit{importance of being different}, including \textit{attendance}. Differently, \textit{importance of being free} is negatively related to all religiosity measures in Table C3. As Table C4 displays, all religiosity measures have insignificant coefficients for \textit{importance of being adventurous}, except \textit{attendance}, which is negative and statistically significant at the 90\% level. When compared to Table 1, the positive coefficients for \textit{degree} (0.008) and \textit{pray} (0.006) seem to be driven by \textit{importance of creativity} and \textit{importance of being different} as evidenced in Tables C1 and C2. The negative relation between \textit{attendance} and \textit{inposav} (-0.012) is apparently fostered by the negative relation between \textit{attendance} and \textit{importance of being free} in Table C3.

\begin{table}[H]
\centering
\def\sym#1{\ifmmode^{#1}\else\(^{#1}\)\fi}
\caption{OLS Estimates: Religiosity and Average Positive Innovation Attitudes} \label{tabs11}
\resizebox{0.98\columnwidth}{0.45\linewidth}{%
\begin{threeparttable}
\begin{tabular}{l*{9}{c}}
\toprule
&\multicolumn{1}{c}{(1)}&\multicolumn{1}{c}{(2)}&\multicolumn{1}{c}{(3)}&\multicolumn{1}{c}{(4)}&\multicolumn{1}{c}{(5)}&\multicolumn{1}{c}{(6)}&\multicolumn{1}{c}{(7)}&\multicolumn{1}{c}{(8)}&\multicolumn{1}{c}{(9)}\\
&\multicolumn{1}{c}{\textit{inposav}}&\multicolumn{1}{c}{\textit{inposav}}&\multicolumn{1}{c}{\textit{inposav}}&\multicolumn{1}{c}{\textit{inposav}}&\multicolumn{1}{c}{\textit{inposav}}&\multicolumn{1}{c}{\textit{inposav}}&\multicolumn{1}{c}{\textit{inposav}}&\multicolumn{1}{c}{\textit{inposav}}&\multicolumn{1}{c}{\textit{inposav}}\\
\midrule
\textit{religiosity}&0.002&0.006&0.003&&&&&&\\
&(0.004)&(0.004)&(0.004)&&&&&&\\
\addlinespace
\textit{degree}&&&&0.009\sym{***}&0.008\sym{***}&&&&\\
&&&&(0.003)&(0.003)&&&&\\
\addlinespace
\textit{attendance}&&&&&&-0.010\sym{***}&-0.012\sym{***}&&\\
&&&&&&(0.004)&(0.004)&&\\
\addlinespace
\textit{pray}&&&&&&&&0.008\sym{***}&0.006\sym{**}\\
&&&&&&&&(0.003)&(0.003)\\
\addlinespace
\textit{gender}&-0.025\sym{***}&-0.023\sym{***}&-0.023\sym{***}&-0.023\sym{***}&-0.023\sym{***}&-0.022\sym{***}&-0.023\sym{***}&-0.023\sym{***}&-0.024\sym{***}\\
&(0.002)&(0.002)&(0.002)&(0.002)&(0.002)&(0.002)&(0.002)&(0.002)&(0.002)\\
\addlinespace
\textit{age}&-0.005\sym{***}&-0.003\sym{***}&-0.002\sym{***}&-0.003\sym{***}&-0.002\sym{***}&-0.003\sym{***}&-0.002\sym{***}&-0.003\sym{***}&-0.002\sym{***}\\
&(0.000)&(0.000)&(0.000)&(0.000)&(0.000)&(0.000)&(0.000)&(0.000)&(0.000)\\
\addlinespace
\textit{age2}&0.000\sym{***}&0.000\sym{**}&0.000&0.000\sym{**}&0.000&0.000\sym{**}&0.000&0.000\sym{**}&0.000\\
&(0.000)&(0.000)&(0.000)&(0.000)&(0.000)&(0.000)&(0.000)&(0.000)&(0.000)\\
\addlinespace
\textit{education}&0.004\sym{***}&0.003\sym{***}&0.002\sym{***}&0.003\sym{***}&0.003\sym{***}&0.003\sym{***}&0.003\sym{***}&0.003\sym{***}&0.003\sym{***}\\
&(0.000)&(0.000)&(0.000)&(0.000)&(0.000)&(0.000)&(0.000)&(0.000)&(0.000)\\
\addlinespace
\textit{paidwork}&&-0.001&0.001&-0.000&0.001&-0.001&0.001&-0.000&0.001\\
&&(0.002)&(0.002)&(0.002)&(0.002)&(0.002)&(0.002)&(0.002)&(0.002)\\
\addlinespace
\textit{pwbefore}&&-0.115\sym{**}&-0.069\sym{***}&-0.115\sym{**}&-0.068\sym{***}&-0.121\sym{**}&-0.077\sym{***}&-0.115\sym{**}&-0.069\sym{***}\\
&&(0.051)&(0.013)&(0.052)&(0.011)&(0.049)&(0.013)&(0.050)&(0.012)\\
\addlinespace
\textit{partner}&&-0.019\sym{***}&-0.018\sym{***}&-0.020\sym{***}&-0.018\sym{***}&-0.019\sym{***}&-0.018\sym{***}&-0.019\sym{***}&-0.018\sym{***}\\
&&(0.002)&(0.002)&(0.002)&(0.002)&(0.002)&(0.002)&(0.002)&(0.002)\\
\addlinespace
\textit{health}&&0.017\sym{***}&0.017\sym{***}&0.017\sym{***}&0.017\sym{***}&0.017\sym{***}&0.017\sym{***}&0.017\sym{***}&0.017\sym{***}\\
&&(0.001)&(0.001)&(0.001)&(0.001)&(0.001)&(0.001)&(0.001)&(0.001)\\
\addlinespace
\textit{child}&&-0.019\sym{***}&-0.020\sym{***}&-0.019\sym{***}&-0.020\sym{***}&-0.019\sym{***}&-0.020\sym{***}&-0.019\sym{***}&-0.020\sym{***}\\
&&(0.002)&(0.002)&(0.002)&(0.002)&(0.002)&(0.002)&(0.002)&(0.002)\\
\addlinespace
\textit{bornc}&&&-0.010\sym{***}&&-0.009\sym{***}&&-0.011\sym{***}&&-0.009\sym{***}\\
&&&(0.003)&&(0.003)&&(0.003)&&(0.003)\\
\addlinespace
\textit{fathere}&&&0.005\sym{***}&&0.005\sym{***}&&0.005\sym{***}&&0.005\sym{***}\\
&&&(0.001)&&(0.001)&&(0.001)&&(0.001)\\
\addlinespace
\textit{mothere}&&&0.004\sym{***}&&0.004\sym{***}&&0.004\sym{***}&&0.004\sym{***}\\
&&&(0.001)&&(0.001)&&(0.001)&&(0.001)\\
\addlinespace
\textit{cons}&0.749\sym{***}&0.767\sym{***}&0.698\sym{***}&0.761\sym{***}&0.691\sym{***}&0.773\sym{***}&0.707\sym{***}&0.764\sym{***}&0.694\sym{***}\\
&(0.010)&(0.051)&(0.017)&(0.053)&(0.016)&(0.050)&(0.018)&(0.051)&(0.017)\\
\midrule
\(N\)&234,528&228,327&202,300&231,306&204,788&231,733&205,176&229,606&203,384\\
\textit{Adj. \(R^{2}\)}&0.135&0.146&0.150&0.147&0.151&0.147&0.151&0.147&0.150\\
\bottomrule
\end{tabular}
\begin{tablenotes}
\item Notes: OLS estimates for alternative measures of religiosity are reported. Robust standard errors clustered at the level of instruments are in parentheses. All regressions include the following fixed effects: country, survey year, country-survey year, religious denomination, occupation, and income level. * $p<0.10$; **$p<0.05$; ***$p<0.01$.
\end{tablenotes}
\end{threeparttable}
}
\end{table}

Overall, the OLS results show that higher values of \textit{degree} and \textit{pray} have a somewhat positive relationship with innovation through \textit{importance of creativity} and \textit{importance of being different}, while \textit{attendance} is negatively related to innovation. \textit{Religiosity} shows no significant relationship with \textit{inposav}, which is expected to some degree, given that \textit{religiosity} is the principal component of other three measures (\textit{degree}, \textit{attendance}, \textit{pray}) that exhibit both negative and positive relationships, yielding to cancel out each effect when the measures combined.

In contrast with Table 1, Table 2 displays a pretty stable relationship between different measures of religiosity and \textit{innegav}. Each specification is statistically significant at the 99\% level and positively related to \textit{innegav}, regardless of the gradual inclusion of controls. Another salient result is that all coefficients are substantially higher in absolute value than those in Table 1, suggesting that the religiosity measures are more strongly associated with NIA than with PIA. In other words, religiosity seems to foster individual traits that are unfavourable to innovation.

\begin{table}[H]
\centering
\def\sym#1{\ifmmode^{#1}\else\(^{#1}\)\fi}
\caption{OLS Estimates: Religiosity and Average Negative Innovation Attitudes} \label{tabs12}
\resizebox{0.98\columnwidth}{0.45\linewidth}{%
\begin{threeparttable}
\begin{tabular}{l*{9}{c}}
\toprule
&\multicolumn{1}{c}{(1)}&\multicolumn{1}{c}{(2)}&\multicolumn{1}{c}{(3)}&\multicolumn{1}{c}{(4)}&\multicolumn{1}{c}{(5)}&\multicolumn{1}{c}{(6)}&\multicolumn{1}{c}{(7)}&\multicolumn{1}{c}{(8)}&\multicolumn{1}{c}{(9)}\\
&\multicolumn{1}{c}{\textit{innegav}}&\multicolumn{1}{c}{\textit{innegav}}&\multicolumn{1}{c}{\textit{innegav}}&\multicolumn{1}{c}{\textit{innegav}}&\multicolumn{1}{c}{\textit{innegav}}&\multicolumn{1}{c}{\textit{innegav}}&\multicolumn{1}{c}{\textit{innegav}}&\multicolumn{1}{c}{\textit{innegav}}&\multicolumn{1}{c}{\textit{innegav}}\\
\midrule
\textit{religiosity}&0.180\sym{***}&0.176\sym{***}&0.175\sym{***}&&&&&&\\
&(0.005)&(0.005)&(0.005)&&&&&&\\
\addlinespace
\textit{degree}&&&&0.132\sym{***}&0.130\sym{***}&&&&\\
&&&&(0.005)&(0.005)&&&&\\
\addlinespace
\textit{attendance}&&&&&&0.129\sym{***}&0.129\sym{***}&&\\
&&&&&&(0.005)&(0.005)&&\\
\addlinespace
\textit{pray}&&&&&&&&0.077\sym{***}&0.075\sym{***}\\
&&&&&&&&(0.003)&(0.003)\\
\addlinespace
\textit{gender}&-0.009\sym{***}&-0.007\sym{***}&-0.007\sym{***}&-0.004\sym{*}&-0.004\sym{*}&0.000&0.000&-0.005\sym{**}&-0.005\sym{**}\\
&(0.003)&(0.003)&(0.003)&(0.003)&(0.003)&(0.003)&(0.003)&(0.003)&(0.003)\\
\addlinespace
\textit{age}&0.001\sym{***}&-0.000&-0.000&-0.000&-0.001&-0.000&-0.001&-0.000&-0.001\\
&(0.000)&(0.000)&(0.000)&(0.000)&(0.000)&(0.000)&(0.000)&(0.000)&(0.000)\\
\addlinespace
\textit{age2}&0.000&0.000\sym{***}&0.000\sym{***}&0.000\sym{***}&0.000\sym{***}&0.000\sym{***}&0.000\sym{***}&0.000\sym{***}&0.000\sym{***}\\
&(0.000)&(0.000)&(0.000)&(0.000)&(0.000)&(0.000)&(0.000)&(0.000)&(0.000)\\
\addlinespace
\textit{education}&-0.003\sym{***}&-0.003\sym{***}&-0.003\sym{***}&-0.003\sym{***}&-0.003\sym{***}&-0.003\sym{***}&-0.003\sym{***}&-0.003\sym{***}&-0.003\sym{***}\\
&(0.000)&(0.000)&(0.000)&(0.000)&(0.000)&(0.000)&(0.000)&(0.000)&(0.000)\\
\addlinespace
\textit{paidwork}&&0.003&0.005\sym{**}&0.002&0.004\sym{*}&0.002&0.004\sym{*}&0.003&0.004\sym{*}\\
&&(0.002)&(0.002)&(0.002)&(0.002)&(0.002)&(0.002)&(0.002)&(0.002)\\
\addlinespace
\textit{pwbefore}&&-0.244\sym{***}&-0.209\sym{***}&-0.256\sym{***}&-0.218\sym{***}&-0.253\sym{***}&-0.221\sym{***}&-0.276\sym{***}&-0.257\sym{***}\\
&&(0.034)&(0.031)&(0.037)&(0.034)&(0.032)&(0.032)&(0.027)&(0.034)\\
\addlinespace
\textit{partner}&&0.025\sym{***}&0.023\sym{***}&0.026\sym{***}&0.024\sym{***}&0.025\sym{***}&0.022\sym{***}&0.027\sym{***}&0.025\sym{***}\\
&&(0.002)&(0.002)&(0.002)&(0.002)&(0.002)&(0.002)&(0.002)&(0.002)\\
\addlinespace
\textit{health}&&0.004\sym{***}&0.004\sym{***}&0.004\sym{***}&0.004\sym{***}&0.003\sym{***}&0.003\sym{***}&0.005\sym{***}&0.005\sym{***}\\
&&(0.001)&(0.001)&(0.001)&(0.001)&(0.001)&(0.001)&(0.001)&(0.001)\\
\addlinespace
\textit{child}&&0.003&0.001&0.004\sym{**}&0.002&0.004\sym{**}&0.003&0.004\sym{**}&0.002\\
&&(0.002)&(0.002)&(0.002)&(0.002)&(0.002)&(0.002)&(0.002)&(0.002)\\
\addlinespace
\textit{bornc}&&&-0.032\sym{***}&&-0.036\sym{***}&&-0.037\sym{***}&&-0.035\sym{***}\\
&&&(0.004)&&(0.003)&&(0.003)&&(0.004)\\
\addlinespace
\textit{fathere}&&&-0.001&&-0.001&&-0.001&&-0.001\\
&&&(0.001)&&(0.001)&&(0.001)&&(0.001)\\
\addlinespace
\textit{mothere}&&&-0.004\sym{***}&&-0.004\sym{***}&&-0.004\sym{***}&&-0.004\sym{***}\\
&&&(0.001)&&(0.001)&&(0.001)&&(0.001)\\
\addlinespace
\textit{cons}&0.510\sym{***}&0.756\sym{***}&0.776\sym{***}&0.765\sym{***}&0.787\sym{***}&0.793\sym{***}&0.823\sym{***}&0.818\sym{***}&0.859\sym{***}\\
&(0.010)&(0.036)&(0.032)&(0.038)&(0.035)&(0.033)&(0.033)&(0.028)&(0.035)\\
\midrule
\(N\)&234,387&228,192&202,191&231,156&204,668&231,582&205,055&229,462&203,268\\
\textit{Adj. \(R^{2}\)}&0.239&0.242&0.246&0.236&0.240&0.231&0.236&0.231&0.235\\
\bottomrule
\end{tabular}
\begin{tablenotes}
\item Notes: OLS estimates for alternative measures of religiosity are reported. Robust standard errors clustered at the level of instruments are in parentheses. All regressions include the following fixed effects: country, survey year, country-survey year, religious denomination, occupation, and income level. * $p<0.10$; **$p<0.05$; ***$p<0.01$.
\end{tablenotes}
\end{threeparttable}
}
\end{table}

Tables C5 and C6 display detailed results for the two NIA. All religiosity measures are positively and significantly (at the 99\% level) associated with both \textit{following traditions} and \textit{following rules}, but the coefficients are almost four times higher for \textit{following traditions}.

Table 3 shows the OLS estimates broken down into various religious denominations and innovation attitudes. The dependent variables are PIA for the first four columns, NIA for columns 5 and 6, and average innovation attitudes for the last two columns. The complete set of control variables and fixed effects are included in each specification. Briefly, Roman Catholics, Protestants and Other Christians exhibit a negative relation with \textit{inposav}, while Jewish, Other Christian and Eastern religions are positively correlated. When it comes to \textit{innegav}, there is a consistent relationship across denominations, each of them is positively and significantly correlated with \textit{innegav}, meaning that all kinds of religions are highly correlated to the unfavourable individual traits to innovation. 

\begin{table}[H]
\centering
\def\sym#1{\ifmmode^{#1}\else\(^{#1}\)\fi}
\caption{OLS Estimates: Religious Denominations and Innovation Attitudes} \label{tabs13}
\resizebox{0.9\columnwidth}{0.3\linewidth}{%
\begin{threeparttable}
\begin{tabular}{l*{8}{c}}
\toprule
&\multicolumn{1}{c}{(1)}&\multicolumn{1}{c}{(2)}&\multicolumn{1}{c}{(3)}&\multicolumn{1}{c}{(4)}&\multicolumn{1}{c}{(5)}&\multicolumn{1}{c}{(6)}&\multicolumn{1}{c}{(7)}&\multicolumn{1}{c}{(8)}\\
&\textit{creative}&\textit{different}&\textit{free}&\textit{adventurous}&\textit{traditions}&\textit{rules}&\textit{inposav}&\textit{innegav}\\
\midrule
Roman Catholic&-0.006\sym{*}&-0.007\sym{**}&-0.016\sym{***}&-0.025\sym{***}&0.171\sym{***}&0.048\sym{***}&-0.014\sym{***}&0.110\sym{***}\\
&(0.004)&(0.004)&(0.004)&(0.004)&(0.004)&(0.004)&(0.003)&(0.003)\\
\addlinespace
Protestant&-0.009\sym{**}&-0.018\sym{***}&-0.017\sym{***}&-0.027\sym{***}&0.151\sym{***}&0.059\sym{***}&-0.018\sym{***}&0.105\sym{***}\\
&(0.004)&(0.004)&(0.004)&(0.003)&(0.005)&(0.005)&(0.003)&(0.005)\\
\addlinespace
Eastern Orthodox&0.008&0.005&0.003&-0.000&0.111\sym{***}&0.034\sym{***}&0.003&0.072\sym{***}\\
&(0.009)&(0.008)&(0.009)&(0.008)&(0.006)&(0.007)&(0.006)&(0.005)\\
\addlinespace
Other Christian&0.007&-0.000&-0.027\sym{***}&-0.026\sym{***}&0.111\sym{***}&0.075\sym{***}&-0.011\sym{*}&0.093\sym{***}\\
&(0.009)&(0.010)&(0.009)&(0.010)&(0.010)&(0.011)&(0.007)&(0.008)\\
\addlinespace
Jewish&0.051&0.074\sym{**}&0.011&0.073\sym{**}&0.205\sym{***}&-0.044&0.052\sym{*}&0.079\sym{***}\\
&(0.033)&(0.033)&(0.029)&(0.036)&(0.033)&(0.045)&(0.027)&(0.021)\\
\addlinespace
Muslim&-0.006&0.005&0.001&-0.013\sym{*}&0.224\sym{***}&0.090\sym{***}&-0.003&0.157\sym{***}\\
&(0.009)&(0.009)&(0.010)&(0.008)&(0.009)&(0.009)&(0.007)&(0.007)\\
\addlinespace
Eastern Religions&0.015&0.022&0.015&0.045\sym{***}&0.141\sym{***}&0.000&0.024\sym{**}&0.070\sym{***}\\
&(0.014)&(0.015)&(0.013)&(0.015)&(0.016)&(0.020)&(0.010)&(0.015)\\
\addlinespace
Other Non-Christian&0.055\sym{***}&0.013&-0.009&0.030&0.127\sym{***}&-0.025&0.021\sym{*}&0.051\sym{***}\\
&(0.016)&(0.018)&(0.016)&(0.022)&(0.018)&(0.019)&(0.012)&(0.013)\\
\addlinespace
Not declared&0.005&0.009&-0.015&-0.006&0.143\sym{***}&0.041\sym{***}&-0.002&0.092\sym{***}\\
&(0.012)&(0.009)&(0.009)&(0.011)&(0.013)&(0.010)&(0.007)&(0.010)\\
\midrule
\(N\)&204,634&204,801&204,940&204,732&204,960&204,087&205,773&205,647\\
\textit{Adj. \(R^{2}\)}&0.082&0.081&0.074&0.174&0.194&0.134&0.151&0.222\\
\bottomrule
\end{tabular}
\begin{tablenotes}
\item Notes: OLS estimates for religious denominations are reported. Robust standard errors clustered at the level of instruments are in parentheses. All regressions include controls for age, age squared, gender, education, paid work status, children, health, mother’s and father’s, education, born-in country and the following fixed effects: country, survey year, country-survey year, religious denomination, occupation and income level. * $p<0.10$; **$p<0.05$; ***$p<0.01$.
\end{tablenotes}
\end{threeparttable}
}
\end{table}

\subsection{Instrumental Variables Estimates }
The OLS estimates reported above indicate both negative and positive associations between religiosity and innovation attitudes, though the negative relation is more substantial. Nevertheless, the potential endogeneity of religion makes it challenging to interpret the results as causal links. 
Therefore, we use instruments to cope with endogeneity fundamentally. Tables 4 and 5 report the main results estimated with IV. The first stage results of each specification can be found in the corresponding columns of Tables B1 and B2 in Appendix B. The OLS version of each specification is presented in the corresponding columns of Tables 1 and 2, meaning that, for instance, column 4 in Table 1 is the OLS version of the specification in column 4 in Table 4. All specifications include the complete set of fixed effects. 

We report a couple of post-estimation diagnostic tests. Kleibergen-Paap rk LM statistic (\textit{idp}) is for under-identification. Kleibergen-Paap rk Wald F statistic (\textit{widstat}), that is F test of excluded instruments valid in non-\textit{i.i.d} case, and Cragg-Donald Wald F statistic (\textit{cdf}) are diagnostics for weak identification. \textit{Widstat} is equivalent to the first stage F statistic. Anderson-Rubin Wald F statistic (\textit{AR test}), which is weak-instrument-robust, tests if the coefficients of endogenous variables are equal to zero in the reduced form estimation. We do not need to consider the over-identification problem because there is only one instrument for each endogenous religiosity variable in all estimates; thus, our model is just-identified. All the results and statistics are efficient only for homoscedasticity since all are estimated with heteroscedasticity-robust options.

\begin{table}[H]\centering
\def\sym#1{\ifmmode^{#1}\else\(^{#1}\)\fi}
\caption{IV Estimates: Religiosity and Average Positive Innovation Attitudes} \label{tabs1iv12}
\resizebox{0.98\columnwidth}{0.45\linewidth}{%
\begin{threeparttable}
\begin{tabular}{l*{9}{c}}
\toprule
&\multicolumn{1}{c}{(1)}&\multicolumn{1}{c}{(2)}&\multicolumn{1}{c}{(3)}&\multicolumn{1}{c}{(4)}&\multicolumn{1}{c}{(5)}&\multicolumn{1}{c}{(6)}&\multicolumn{1}{c}{(7)}&\multicolumn{1}{c}{(8)}&\multicolumn{1}{c}{(9)}\\
&\multicolumn{1}{c}{\textit{inposav}}&\multicolumn{1}{c}{\textit{inposav}}&\multicolumn{1}{c}{\textit{inposav}}&\multicolumn{1}{c}{\textit{inposav}}&\multicolumn{1}{c}{\textit{inposav}}&\multicolumn{1}{c}{\textit{inposav}}&\multicolumn{1}{c}{\textit{inposav}}&\multicolumn{1}{c}{\textit{inposav}}&\multicolumn{1}{c}{\textit{inposav}}\\
\midrule
\textit{religiosity}&-0.155\sym{**}&-0.151\sym{**}&-0.198\sym{***}&&&&&&\\
&(0.070)&(0.065)&(0.062)&&&&&&\\
\addlinespace
\textit{degree}&&&&-0.099&-0.177\sym{*}&&&&\\
&&&&(0.099)&(0.096)&&&&\\
\addlinespace
\textit{attendance}&&&&&&-0.157\sym{***}&-0.185\sym{***}&&\\
&&&&&&(0.052)&(0.050)&&\\
\addlinespace
\textit{pray}&&&&&&&&-0.080\sym{**}&-0.119\sym{***}\\
&&&&&&&&(0.040)&(0.039)\\
\addlinespace
\textit{gender}&-0.015\sym{***}&-0.014\sym{***}&-0.011\sym{***}&-0.017\sym{***}&-0.012\sym{**}&-0.019\sym{***}&-0.019\sym{***}&-0.013\sym{***}&-0.009\sym{**}\\
&(0.004)&(0.004)&(0.004)&(0.006)&(0.005)&(0.003)&(0.003)&(0.005)&(0.005)\\
\addlinespace
\textit{age}&-0.005\sym{***}&-0.003\sym{***}&-0.003\sym{***}&-0.003\sym{***}&-0.003\sym{***}&-0.003\sym{***}&-0.003\sym{***}&-0.003\sym{***}&-0.003\sym{***}\\
&(0.000)&(0.000)&(0.000)&(0.000)&(0.000)&(0.000)&(0.000)&(0.000)&(0.000)\\
\addlinespace
\textit{age2}&0.000\sym{***}&0.000\sym{***}&0.000\sym{***}&0.000\sym{***}&0.000\sym{**}&0.000\sym{***}&0.000\sym{**}&0.000\sym{***}&0.000\sym{***}\\
&(0.000)&(0.000)&(0.000)&(0.000)&(0.000)&(0.000)&(0.000)&(0.000)&(0.000)\\
\addlinespace
\textit{education}&0.004\sym{***}&0.003\sym{***}&0.002\sym{***}&0.003\sym{***}&0.002\sym{***}&0.003\sym{***}&0.003\sym{***}&0.003\sym{***}&0.003\sym{***}\\
&(0.000)&(0.000)&(0.000)&(0.000)&(0.000)&(0.000)&(0.000)&(0.000)&(0.000)\\
\addlinespace
\textit{paidwork}&&-0.001&-0.000&-0.001&0.001&-0.001&0.000&-0.001&-0.000\\
&&(0.002)&(0.002)&(0.002)&(0.002)&(0.002)&(0.002)&(0.002)&(0.002)\\
\addlinespace
\textit{pwbefore}&&-0.158\sym{***}&-0.148\sym{***}&-0.146\sym{***}&-0.152\sym{***}&-0.167\sym{***}&-0.151\sym{***}&-0.135\sym{***}&-0.102\sym{***}\\
&&(0.040)&(0.037)&(0.045)&(0.055)&(0.038)&(0.034)&(0.048)&(0.019)\\
\addlinespace
\textit{partner}&&-0.017\sym{***}&-0.015\sym{***}&-0.018\sym{***}&-0.016\sym{***}&-0.016\sym{***}&-0.014\sym{***}&-0.019\sym{***}&-0.017\sym{***}\\
&&(0.002)&(0.003)&(0.003)&(0.003)&(0.003)&(0.003)&(0.002)&(0.002)\\
\addlinespace
\textit{health}&&0.017\sym{***}&0.017\sym{***}&0.018\sym{***}&0.017\sym{***}&0.019\sym{***}&0.018\sym{***}&0.017\sym{***}&0.016\sym{***}\\
&&(0.001)&(0.001)&(0.001)&(0.001)&(0.001)&(0.001)&(0.001)&(0.001)\\
\addlinespace
\textit{child}&&-0.016\sym{***}&-0.017\sym{***}&-0.017\sym{***}&-0.017\sym{***}&-0.017\sym{***}&-0.018\sym{***}&-0.017\sym{***}&-0.017\sym{***}\\
&&(0.002)&(0.002)&(0.002)&(0.002)&(0.002)&(0.002)&(0.002)&(0.002)\\
\addlinespace
\textit{bornc}&&&-0.024\sym{***}&&-0.021\sym{***}&&-0.019\sym{***}&&-0.023\sym{***}\\
&&&(0.006)&&(0.007)&&(0.004)&&(0.005)\\
\addlinespace
\textit{fathere}&&&0.005\sym{***}&&0.005\sym{***}&&0.006\sym{***}&&0.006\sym{***}\\
&&&(0.001)&&(0.001)&&(0.001)&&(0.001)\\
\addlinespace
\textit{mothere}&&&0.004\sym{***}&&0.004\sym{***}&&0.004\sym{***}&&0.004\sym{***}\\
&&&(0.001)&&(0.001)&&(0.001)&&(0.001)\\
\midrule
\textit{N}&234,528&228,327&202,300&231,306&204,788&231,733&205,176&229,606&203,384\\
\textit{idp}&0.000&0.000&0.000&0.000&0.000&0.000&0.000&0.000&0.000\\
\textit{cdf}&2,338&2,380&2,161&846&771&2,732&2,533&2,012&1,730\\
\textit{widstat}&187&194&160&84&79&141&131&192&160\\
\textit{AR test}&0.022 &0.016&0.001 & 0.297&0.052 & 0.002&0.000 & 0.046 &0.002 \\
\bottomrule
\end{tabular}
\begin{tablenotes}
\item Notes: IV estimates for alternative measures of religiosity are reported. Robust standard errors clustered at the level of instruments are in parentheses. All regressions include the following fixed effects: country, survey year, country-survey year, religious denomination, occupation, and income level. Kleibergen-Paap rk LM statistic (\textit{idp}), Kleibergen-Paap rk Wald F statistic (\textit{widstat}), Cragg-Donald Wald F statistic (\textit{cdf}), and Anderson-Rubin Wald F statistic (\textit{AR test}) are reported.
 * $p<0.10$; **$p<0.05$; ***$p<0.01$.
\end{tablenotes}
\end{threeparttable}
}
\end{table}

Table 4 displays the IV results for \textit{inposav} with respect to different measures of religiosity. The first stage results in Table B1 show that all the excluded instruments are strongly and positively correlated with the corresponding religiosity measures, with a correlation coefficient between $0.581$ and $0.784$. P-values of Kleibergen-Paap rk LM test (\textit{idp}) show that the null of under-identification is rejected, and the full rank condition is satisfied in all estimates. For all IV estimates, the first stage F statistics (\textit{widstat}) are well above Stock-Yogo critical values of weak identification. The \textit{AR test} results indicate that the instruments are relevant, except for \textit{degree}, which we will discuss later.

The IV results for \textit{inposav} exhibit a somewhat different pattern than the corresponding OLS results in Table 1. It is essential to underline that the OLS and IV results are not directly comparable since OLS estimates average treatment effect (ATE), while IV/2SLS estimates local average treatment effect (LATE) (\cite{Angrist94}). However, we emphasize the main differences to discuss the effects of vanishing endogeneity. First of all, all religiosity measures display substantial increase in the coefficients, which is also the case in other studies using a more aggregated instrument than endogenous variable (\cite{Iyer17}; \cite{Cutler96}). \textit{Religiosity} becomes significant and negative in column 3; it is positive and insignificant with OLS. \textit{Degree} gets negative and less significant in columns 5 and 6. \textit{Attendance} keeps the sign of OLS estimates, which is negative. \textit{Pray} in columns 8 and 9, changes the sign to be negative as well.

\textit{Innegav} reported in Table 5, in contrast to \textit{inposav}, displays pretty stable and consistent patterns. All religiosity measures are positively and significantly correlated to \textit{innegav}, as they are in the corresponding OLS estimates in Table 2, with the increase in the coefficients. The corresponding first stage results, reported in Table B2, show that all the excluded instruments are strongly correlated to the religiosity measures in each specification, with a correlation coefficient between $0.581$ and $0.784$. 

Appendix D provides the IV estimates broken down into each measure of innovation attitudes. Tables D1 and D2 indicate that \textit{importance of creativity} and \textit{importance of being different} are negatively affected by \textit{attendance}, while other religiosity measures are insignificant. Table D3 shows that all religiosity measures are negatively related to \textit{importance of being free}, yielding the highest coefficients among IV estimates. Religiosity does not affect \textit{importance of being adventurous} as demonstrated in Table D4. These results signal that the estimated positive associations between religiosity and PIA in OLS Tables C1 (\textit{importance of creativity}) and C2 (\textit{importance of being different}) are driven by endogeneity. Tables D5 and D6 show that religiosity measures positively affect NIA, similar to the OLS estimates. Overall, it is plausible to argue that higher degrees of religiosity are not advantageous for innovation since it fosters some personal traits unconducive to an innovative mindset. Moreover, it falls away from conductive personal traits such as freedom.

Apart from the main coefficients of interest, the control variables also provide insightful results. The effect of religiosity on \textit{importance of creativity}, \textit{importance of being adventurous}, and \textit{following rules} are stronger for men (\textit{gender}); while it is stronger for women regarding \textit{importance of being different} and \textit{following traditions}. \textit{Age} is negatively related to \textit{importance of being different}, \textit{importance of being adventurous}, and \textit{following rules} and is insignificant for \textit{importance of creativity}, \textit{importance of being free}, and \textit{following traditions}. \textit{Education} seem to foster PIA (see Tables D1-D4) and mitigate NIA (see Tables D5 and 

\begin{table}[H]\centering
\def\sym#1{\ifmmode^{#1}\else\(^{#1}\)\fi}
\caption{IV Estimates: Religiosity and Average Negative Innovation Attitudes} \label{tabs1iv2}
\resizebox{0.98\columnwidth}{0.45\linewidth}{%
\begin{threeparttable}
\begin{tabular}{l*{9}{c}}
\toprule
&\multicolumn{1}{c}{(1)}&\multicolumn{1}{c}{(2)}&\multicolumn{1}{c}{(3)}&\multicolumn{1}{c}{(4)}&\multicolumn{1}{c}{(5)}&\multicolumn{1}{c}{(6)}&\multicolumn{1}{c}{(7)}&\multicolumn{1}{c}{(8)}&\multicolumn{1}{c}{(9)}\\
&\multicolumn{1}{c}{\textit{innegav}}&\multicolumn{1}{c}{\textit{innegav}}&\multicolumn{1}{c}{\textit{innegav}}&\multicolumn{1}{c}{\textit{innegav}}&\multicolumn{1}{c}{\textit{innegav}}&\multicolumn{1}{c}{\textit{innegav}}&\multicolumn{1}{c}{\textit{innegav}}&\multicolumn{1}{c}{\textit{innegav}}&\multicolumn{1}{c}{\textit{innegav}}\\
\midrule
\textit{religiosity}&0.139\sym{***}&0.177\sym{***}&0.210\sym{***}&&&&&&\\
&(0.052)&(0.050)&(0.055)&&&&&&\\
\addlinespace
\textit{degree}&&&&0.182\sym{***}&0.223\sym{***}&&&&\\
&&&&(0.067)&(0.077)&&&&\\
\addlinespace
\textit{attendance}&&&&&&0.205\sym{***}&0.216\sym{***}&&\\
&&&&&&(0.044)&(0.047)&&\\
\addlinespace
\textit{pray}&&&&&&&&0.088\sym{**}&0.118\sym{***}\\
&&&&&&&&(0.036)&(0.040)\\
\addlinespace
\textit{gender}&-0.006&-0.007\sym{*}&-0.009\sym{**}&-0.007&-0.010\sym{*}&-0.001&-0.002&-0.007&-0.010\sym{**}\\
&(0.004)&(0.004)&(0.004)&(0.005)&(0.005)&(0.003)&(0.003)&(0.005)&(0.005)\\
\addlinespace
\textit{age}&0.001\sym{***}&-0.000&-0.000&-0.000&-0.000&-0.000&-0.000&-0.000&-0.001\\
&(0.000)&(0.000)&(0.000)&(0.000)&(0.000)&(0.000)&(0.000)&(0.000)&(0.000)\\
\addlinespace
\textit{age2}&0.000\sym{*}&0.000\sym{***}&0.000\sym{***}&0.000\sym{***}&0.000\sym{***}&0.000\sym{***}&0.000\sym{***}&0.000\sym{***}&0.000\sym{***}\\
&(0.000)&(0.000)&(0.000)&(0.000)&(0.000)&(0.000)&(0.000)&(0.000)&(0.000)\\
\addlinespace
\textit{education}&-0.003\sym{***}&-0.003\sym{***}&-0.003\sym{***}&-0.003\sym{***}&-0.003\sym{***}&-0.003\sym{***}&-0.003\sym{***}&-0.003\sym{***}&-0.003\sym{***}\\
&(0.000)&(0.000)&(0.000)&(0.000)&(0.000)&(0.000)&(0.000)&(0.000)&(0.000)\\
\addlinespace
\textit{paidwork}&&0.003&0.005\sym{**}&0.003&0.004\sym{*}&0.003&0.004\sym{*}&0.003&0.005\sym{**}\\
&&(0.002)&(0.002)&(0.002)&(0.002)&(0.002)&(0.002)&(0.002)&(0.002)\\
\addlinespace
\textit{pwbefore}&&-0.243\sym{***}&-0.195\sym{***}&-0.242\sym{***}&-0.176\sym{***}&-0.230\sym{***}&-0.184\sym{***}&-0.273\sym{***}&-0.246\sym{***}\\
&&(0.037)&(0.037)&(0.049)&(0.050)&(0.039)&(0.036)&(0.028)&(0.034)\\
\addlinespace
\textit{partner}&&0.025\sym{***}&0.022\sym{***}&0.025\sym{***}&0.022\sym{***}&0.023\sym{***}&0.020\sym{***}&0.027\sym{***}&0.024\sym{***}\\
&&(0.002)&(0.002)&(0.002)&(0.002)&(0.002)&(0.002)&(0.002)&(0.002)\\
\addlinespace
\textit{health}&&0.004\sym{***}&0.003\sym{***}&0.004\sym{***}&0.004\sym{***}&0.002\sym{**}&0.002\sym{**}&0.005\sym{***}&0.005\sym{***}\\
&&(0.001)&(0.001)&(0.001)&(0.001)&(0.001)&(0.001)&(0.001)&(0.001)\\
\addlinespace
\textit{child}&&0.003&0.000&0.003&0.000&0.003\sym{*}&0.002&0.004\sym{*}&0.001\\
&&(0.002)&(0.002)&(0.002)&(0.002)&(0.002)&(0.002)&(0.002)&(0.002)\\
\addlinespace
\textit{bornc}&&&-0.029\sym{***}&&-0.030\sym{***}&&-0.033\sym{***}&&-0.031\sym{***}\\
&&&(0.005)&&(0.006)&&(0.004)&&(0.005)\\
\addlinespace
\textit{fathere}&&&-0.001&&-0.001&&-0.001&&-0.001\\
&&&(0.001)&&(0.001)&&(0.001)&&(0.001)\\
\addlinespace
\textit{mothere}&&&-0.004\sym{***}&&-0.004\sym{***}&&-0.004\sym{***}&&-0.004\sym{***}\\
&&&(0.001)&&(0.001)&&(0.001)&&(0.001)\\
\midrule
\textit{N}&234,387&228,192&202,191&231,156&204,668&231,582&205,055&229,462&203,268\\
\textit{idp}&0.000&0.000&0.000&0.000&0.000&0.000&0.000&0.000&0.000\\
\textit{cdf}&2,327&2,369&2,147&839&763&2,729&2,525&2,001&1,722\\
\textit{widstat}&186&193&159&82&77&141&130&190&159\\
\textit{AR test}& 0.001& 0.000 & 0.000 &0.007 & 0.004 & 0.000 &0.000  & 0.012 & 0.002\\
\bottomrule
\end{tabular}
\begin{tablenotes}
\item Notes: IV estimates for alternative measures of religiosity are reported. Robust standard errors clustered at the level of instruments are in parentheses. All regressions include the following fixed effects: country, survey year, country-survey year, religious denomination, occupation, and income level. Kleibergen-Paap rk LM statistic (\textit{idp}), Kleibergen-Paap rk Wald F statistic (\textit{widstat}), Cragg-Donald Wald F statistic (\textit{cdf}), and Anderson-Rubin Wald F statistic (\textit{AR test}) are reported. * $p<0.10$; **$p<0.05$; ***$p<0.01$.
\end{tablenotes}
\end{threeparttable}
}
\end{table}

\noindent D6). Having paid work in the last seven days (\textit{paidwork}) is mostly insignificant. Having paid work before (\textit{pwbefore}) yields the highest coefficients among control variables, implying that individuals with work experience are less affected by each measure of religiosity. Individuals living with a partner are less affected by religiosity with respect to PIA (except for \textit{importance of creativity}), while they are more affected for NIA. \textit{Health} and born-in country (\textit{bornc}) do not introduce heterogeneous effects for PIA and NIA. \textit{Health} is positively related to all innovation attitudes. \textit{Bornc} is negative in all specifications, indicating the fact that being born in a different country is a factor that increases the effect of religiosity on innovation attitudes. Interestingly, individuals living with a child (\textit{child}) are less affected by religiosity regarding PIA, while it does not matter for NIA. Both mother's (\textit{mothere}) and father's education (\textit{fathere}) are positively related to PIA, while only (\textit{mothere}) significantly and negatively relates with NIA.

Overall, the negative effect of religiosity on innovativeness is in line with the findings of prior studies (\cite{Benabouetal2013}, \cite*{Benabouetal2015}). Using OLS estimates, \textcite{Benabouetal2015} find that religiosity is positively related to creativity, which they describe as  ``puzzling". We contribute to this literature by uncovering that the positive relationship between religiosity and creativity may be driven by endogeneity. We find a negative relationship between religiosity and creativity once we rule out endogeneity by applying the IV method.

The question is that, compared to OLS estimates, why does \textit{inposav} change its behaviour with respect to the religiosity measures while \textit{innegav} stays stable? The answer is hidden in the relationship between religious belonging (\textit{belonging}) and other religiosity measures, which are summarized in Tables C7 and C8 in Appendix C. 

In the ESS, a respondent answers the questions regarding \textit{degree}, \textit{attendance} and \textit{pray} regardless of being belong to a religion. In other words, even though the respondent does not believe in a religion, she still answers the questions such as ``How religious are you?", ``How frequently do you pray?", ``How frequently do you attend religious services?" Interestingly, the answers are generally not zero. Table C7 displays the mean values of all religiosity measures, broken down into religious belonging. For instance, the mean value of \textit{religiosity} is $0.51$ for believers —people who belong to a religion at present (\textit{belonging}=1 \& \textit{belongingp}=1)— , while it is $0.14$ for never-believers —people who have never belonged to a religion (\textit{belonging}=0 \& \textit{belongingp}=0 )— , and is $0.18$ for once-believers —people who do not belong to a religion at present but used to belong at past (\textit{belonging}=0 \& \textit{belongingp}=1).

Table C8 shows disaggregated OLS results for religious belonging. For each sub-sample, all religiosity measures are positively and significantly associated with \textit{innegav}. This pattern disappears when we look at \textit{inposav}. For the sub-sample of believers, only \textit{attendance} is significantly and negatively related to \textit{inposav}, meaning that other religiosity measures of believers are not associated with \textit{inposav}. On the contrary, all religiosity measures of never-believers and once-believers are positively and significantly related to \textit{inposav}. Under the light of this disaggregation, one can see that the OLS results in Table 1, namely, the positive correlations between \textit{degree}, \textit{pray}, and \textit{inposav} are seemed to be driven by the sub-sample of non-believers, raising the concerns for a spurious correlation. We, therefore, consider religious belonging in our instrumental variables strategy and construct the instruments by taking it into account. We instrument the religiosity of a believer with the average religiosity of believers, along with other measures explained in Section 3. By doing so, we address the differences between the religiosities of believers and non-believers. Thus, the IV results for \textit{inposav} in Table 4 display different signs than the overall OLS results.

\begin{sidewaystable}\centering
\def\sym#1{\ifmmode^{#1}\else\(^{#1}\)\fi}
\caption{Disaggregate IV Results for Age and Gender} \label{tabs1age}
\resizebox{0.95\columnwidth}{!}{%
\begin{threeparttable}
\begin{tabular}{l*{16}{c}}
\toprule
\addlinespace
\textit{inposav}	&\multicolumn{2}{c}{(15-25)}&\multicolumn{2}{c}{(25-35)}&\multicolumn{2}{c}{(35-45)}&\multicolumn{2}{c}{(45-55)}&\multicolumn{2}{c}{(55-65)}&\multicolumn{2}{c}{(65+)} &\multicolumn{2}{c}{(female)}&\multicolumn{2}{c}{(male)} \\
&	\multicolumn{1}{c}{(OLS)}&\multicolumn{1}{c}{(IV)}&\multicolumn{1}{c}{(OLS)}&\multicolumn{1}{c}{(IV)}&\multicolumn{1}{c}{(OLS)}&\multicolumn{1}{c}{(IV)}&\multicolumn{1}{c}{(OLS)}&\multicolumn{1}{c}{(IV)}&\multicolumn{1}{c}{(OLS)}&\multicolumn{1}{c}{(IV)}&\multicolumn{1}{c}{(OLS)}&\multicolumn{1}{c}{(IV)}&\multicolumn{1}{c}{(OLS)}&\multicolumn{1}{c}{(IV)}&\multicolumn{1}{c}{(OLS)}&\multicolumn{1}{c}{(IV)}    \\
\toprule
&	\multicolumn{1}{c}{(1)}&\multicolumn{1}{c}{(2)}&\multicolumn{1}{c}{(3)}&\multicolumn{1}{c}{(4)}&\multicolumn{1}{c}{(5)}&\multicolumn{1}{c}{(6)}&\multicolumn{1}{c}{(7)}&\multicolumn{1}{c}{(8)}&\multicolumn{1}{c}{(9)}&\multicolumn{1}{c}{(10)}&\multicolumn{1}{c}{(11)}&\multicolumn{1}{c}{(12)}&\multicolumn{1}{c}{(13)}&\multicolumn{1}{c}{(14)}&\multicolumn{1}{c}{(15)}&\multicolumn{1}{c}{(16)}    \\
\midrule
\textit{religiosity}& -0.014& -0.087& -0.004&-0.178&-0.004  &-0.191&0.027***&-0.241** &0.012&-0.243**&-0.004& -0.192&0.000&-0.154*& 0.010**&-0.213   \\
&(0.013)& (0.143)& (0.009)& (0.173)&(0.008) &(0.133) &(0.008) &(0.100)&(0.008)& (0.097)  & (0.008)&  (0.119)  &(0.005)&(0.084) &(0.005)&(0.149)    \\
&16,883 &16,898& 33,172&33,200&38,457 & 38,493&38,165 &38,199&35,236&35,269& 40,387 &   40,404  &104,447 &104,447  & 98,016& 98,016     \\

&0.065&0.542& 0.094&0.312&0.086&0.127 &0.72&0.010&0.096&0.011& 0.154 &0.076 & 0.160  & 0.057 & 0.130 & 0.115       \\

\addlinespace
\textit{degree}& -0.000&0.020&0.002&1.110&0.006&0.458&0.024***&-0.158&0.018**&-0.282**& -0.003&-0.160  &0.006&-0.148&0.012***&-0.054       \\
&(0.010)&(0.122)& (0.007)&(3.310)&(0.007)&(0.338)&(0.007)&(0.191)&(0.007)& (0.122)& (0.007)&    (0.135)  &(0.004)&(0.111) &(0.004)&(0.158)           \\
& 17,031&17,047&33,607&33,636&38,944&38,982&38,641&38,675&35,695& 35,729& 40,870& 40,887 & 105,725  &105,725   & 99,231 &  99,231       \\

&0.067&0.869& 0.093&0.464 &0.086&0.035&0.073&0.411&0.097&0.053& 0.155&  0.198 & 0.161  & 0.146 & 0.131& 0.727           \\

\addlinespace
\textit{\textit{attendance}} &-0.025**&-0.145& -0.011&-0.111&-0.026***&-0.329*&-0.001&-0.250** &-0.012&  -0.181**&  -0.008&-0.266* &-0.017***&-0.172**&-0.005&-0.304**    \\
& (0.012)&(0.113)& (0.008)&(0.125)&(0.008) &(0.177)&(0.009)& (0.099)&(0.008)& (0.091)&  (0.007)&   (0.101) &(0.005)&(0.080) &(0.005)&(0.147)        \\
& 17,063&17,079&33,621&33,651&39,005&39,043&38,721& 38,756& 35,799& 35,833&  40,967&    40,984  & 105,912 & 105,912 & 99,434 & 99,434         \\

&0.068&0.120&0.094& 0.308&0.087&0.013&0.074&0.002&0.097&0.029& 0.154& 0.030  & 0.161  & 0.026  & 0.130 & 0.019  \\

\addlinespace
\textit{pray}&-0.001&0.059& -0.001&-0.195**&0.006&-0.107&0.024***&-0.144**&0.011**& -0.229**&  0.000&  -0.108 &0.007**&-0.011  &0.009***&-0.176      \\
&(0.008)&(0.182)&(0.005)&(0.097)&(0.005)&(0.089)&(0.005)&(0.059)&(0.005)& (0.149)&  (0.005)    &(0.113) &(0.003)&(0.076)&(0.003)&(0.109)           \\
&16,970& 16,985&33,335&33,365& 38,643&38,679&38,369&38,404&35,448& 35,481&  40,619&  40,636 & 104,992 & 104,992  &98,558  & 98,558    \\     

&0.066&0.741&0.093&0.065&0.085&0.219&0.073& 0.009&0.095&0.007& 0.155& 0.313 & 0.161 &0.887   &  0.130& 237      \\  

\addlinespace
\bottomrule 
\toprule
\addlinespace
&\multicolumn{2}{c}{(15-25)}&\multicolumn{2}{c}{(25-35)}&\multicolumn{2}{c}{(35-45)}&\multicolumn{2}{c}{(45-55)}&\multicolumn{2}{c}{(55-65)}&\multicolumn{2}{c}{(65+)} &\multicolumn{2}{c}{(female)}&\multicolumn{2}{c}{(male)} \\
\textit{innegav}&	\multicolumn{1}{c}{(OLS)}&\multicolumn{1}{c}{(IV)}&\multicolumn{1}{c}{(OLS)}&\multicolumn{1}{c}{(IV)}&\multicolumn{1}{c}{(OLS)}&\multicolumn{1}{c}{(IV)}&\multicolumn{1}{c}{(OLS)}&\multicolumn{1}{c}{(IV)}&\multicolumn{1}{c}{(OLS)}&\multicolumn{1}{c}{(IV)}&\multicolumn{1}{c}{(OLS)}&\multicolumn{1}{c}{(IV)}&\multicolumn{1}{c}{(OLS)}&\multicolumn{1}{c}{(IV)}&\multicolumn{1}{c}{(OLS)}&\multicolumn{1}{c}{(IV)}    \\
\toprule
&	\multicolumn{1}{c}{(1)}&\multicolumn{1}{c}{(2)}&\multicolumn{1}{c}{(3)}&\multicolumn{1}{c}{(4)}&\multicolumn{1}{c}{(5)}&\multicolumn{1}{c}{(6)}&\multicolumn{1}{c}{(7)}&\multicolumn{1}{c}{(8)}&\multicolumn{1}{c}{(9)}&\multicolumn{1}{c}{(10)}&\multicolumn{1}{c}{(11)}&\multicolumn{1}{c}{(12)}&\multicolumn{1}{c}{(13)}&\multicolumn{1}{c}{(14)}&\multicolumn{1}{c}{(15)}&\multicolumn{1}{c}{(16)}    \\
\midrule
\addlinespace
\textit{religiosity}	&0.220*** &-0.244* &0.216*** &0.323** &0.184*** &0.647*** &0.167***& 0.354***& 0.153***&0.378** &0.151*** &0.609** &0.170***&0.292***&0.182***&0.189 \\
&(0.015) &(0.144) & (0.011)&(0.163) &(0.009) &(0.161) & (0.009)&(0.128) &(0.009) &(0.154)&(0.008)  &(0.286)  &(0.006)&(0.113)&(0.006)&(0.138) \\
& 16,879&16,894&33,154&33,182 &38,443 &38,479&38,149 &38,183&35,208 &35,241& 40,358 & 40,375 &104,396 & 104,396 &97,958  &97 958   \\
& 0.240& 0.073& 0.236&0.029&0.240&0.000 & 0.242&0.001 & 0.220&0.001&0.167 &0.003& 0.248 & 0.007& 0.248 &0.174  \\

\addlinespace	
\textit{degree}& 0.168***&0.033 & 0.141***&2.274 & 0.134***&0.860*&0.118*** & 0.352&0.116*** &0.477*&0.123***&0.546***&0.122***&0.142&0.136***&0.305**  \\
& (0.012)&(0.155) &(0.009) &(3.875) & (0.008)&(0.462) &(0.008) &(0.223) &(0.008) &(0.248) &(0.007)& (0.192) &(0.005)&(0.113)&(0.005)&(0.154)  \\
& 17,027& 17,043&33,587&33,616 & 38,926& 38,964&38,623 &38,657 &35,665&35,699 &40,840& 40,857 & 105,671 &  105,671  & 99,165 & 99,165  \\
& 0.231&0.830&0.229 & 0.115&0.232 &0.001& 0.236& 0.110&0.213&0.011&0.163& 0.002& 0.241 & 0.197& 0.243 &0.058\\
\addlinespace
\textit{\textit{attendance}}& 0.143***& -0.236*&0.168*** & 0.221**& 0.140***&0.649*** &0.136*** & 0.323**&0.125*** & 0.332**&0.093***& 0.622*&0.133***&0.332*** &0.132***&0.198 \\
& (0.015)& (0.125)&(0.010) &(0.108) & (0.009)&(0.232) & (0.009)&(0.126) &(0.009) &(0.139) &(0.007)&(0.317) &(0.005)&(0.091)&(0.006)&(0.130)  \\
&17,058 &17,074 & 33,601&33,631&38,987 & 39,025& 38,704& 38,739&35,770&35,804 &40,935& 40,952 & 105,858 & 105,858&99,367  & 99,367  \\
&0.218&0.021& 0.227&0.006 & 0.227&0.000 &0.236& 0.000&0.212 &0.000 &0.155&0.001& 0.240 &  0.000 & 0.236 &0.123 \\

\addlinespace
\textit{pray}&0.103*** &-0.371* & 0.101***& 0.081&0.079*** & 0.317***&0.066*** &0.175** &0.057*** &0.264* &0.071***& 0.435&0.068***&0.268** &0.084***&0.119\\
& (0.010)&(0.217) &(0.007) &(0.111) &(0.006) &(0.094) &(0.005) &(0.074) &(0.006) & (0.141)&(0.005)& (0.347)&(0.003)&(0.121) &(0.004)&(0.113) \\
& 16,965&16,980 &33,317&33,347 &38,627 &38,663 &38,352 & 38,387&35,420 &35,453 &40,587&40,604 &  104,939&   104,939& 98,495 & 98,495  \\
& 0.222& 0.028&0.226 &0.455 & 0.228&0.000 &0.231 &0.009&0.209&0.021& 0.159 &0.051& 0.237 & 0.011& 0.240 & 0.279  \\

\addlinespace
\bottomrule 
\end{tabular}
\begin{tablenotes}
\item Notes: Disaggregated OLS and IV estimates for age and gender are reported. Robust standard errors clustered at the level of instruments are in parentheses. The observation numbers are below standard errors, followed by adjusted R squared for OLS and Anderson-Rubin Wald F statistic (\textit{AR test}) for IV estimates. All regressions include controls for age, gender, education, paid work status, partner, children, health, born in country, mother’s and father’s, education and the following fixed effects: country, survey year, country-survey year, religious denomination, occupation and income level. * $p<0.10$; **$p<0.05$; ***$p<0.01$.
\end{tablenotes}
\end{threeparttable}
}
\end{sidewaystable}

\subsection{Heterogeneity of Religiosity: Gender and Age}
We reestimate the baseline specifications to examine whether the main findings hold for different sub-samples of age and gender. Table 6 presents the results. Columns $1$-$12$ show OLS and IV estimates disaggregated by six age brackets\footnote{The instruments are built with eleven age brackets. Here we only consider six age brackets for brevity. The model is estimated also with eleven age brackets, but the overall results do not change significantly.}. The last four columns ($13$-$16$) display the baseline OLS and IV estimates disaggregated by gender. The dependent variable is \textit{inposav} on the upper part of the table and \textit{innegav} on the lower part. Each coefficient comes from a different regression and includes the complete set of control variables and fixed effects which are not reported due to space constraints. The number of observations is reported below standard errors, followed by adjusted $R^{2}$ for OLS and AR test for IV estimates. 

Considering age, OLS results show that the religiosity measures, when statistically significant, are negatively related to \textit{inposav} in early and lower-middle ages, \textit{i.e.,} 15-25 and 35-45, but are positively related in middle and upper-middle ages, \textit{i.e.,} 45-55 and 55-65. On the other hand, IV results are always negatively related to \textit{inposav}, regardless of age bracket. As mentioned above, this alteration probably stems from the vanishment of spurious correlations thanks to the indirect inclusion of religious belonging via the instruments. 

The OLS and IV results in the lower part of Table 6 show that the religiosity measures are always positively associated with \textit{innegav} for each age bracket except for the 15-25, which is the only negative association between religiosity measures and \textit{innegav} throughout this paper. One explanation is that young adults are generally open to questioning the values and beliefs they inherited from their parents as well as from the culture in which they live, assuming that in the early ages of life, people make up their minds, see the world from their own eyes, form their own beliefs and values. Moreover, fulfilling religious duties —such as going to church or mosque— to meet the expectations of parents and society is not uncommon in the early ages, yet, being exposed to religion by doing so might make them question traditions and rules more, which can explain the negative association between \textit{attendance} and \textit{innegav}. 


Regarding gender, the OLS results for \textit{inposav} show that \textit{attendance} is negatively and \textit{pray} is positively correlated for women, while higher religiosity favour \textit{inposav} for men. In the IV results, only \textit{attendance} is significant for both genders, introducing a higher negative effect for men. 


The overall picture changes when we consider \textit{innegav} as the dependent variable. Considering OLS estimates, each measure of religiosity is positively and significantly correlated to \textit{innegav} with relatively high coefficients for both genders. IV results suggest that this pattern stays the same for women except for \textit{degree}, while \textit{degree} is the only significant effect for men. These findings are consistent with the existing literature on gender. Religions, generally, have different approaches and prescribed behavioural rules for men and women, imposing various restrictions on women's liberty and rights. In contrast, men face very little of them, if not none. Therefore, the results are reasonable, suggesting that higher values of \textit{attendance} and \textit{pray} causes women to follow traditions and established rules more.

\section {Sensitivity Analysis}\label{section: robust}

\subsection{Alternative Dependent Variable}
So far, we have analysed innovation attitudes separately, concluding that higher religiosity is negatively related to PIA and positively correlated to NIA, which means that being religious is not a catalyser for innovation-related traits and is possibly an obstacle. In order to assess

\begin{table}[H]\centering
\def\sym#1{\ifmmode^{#1}\else\(^{#1}\)\fi}
\caption{Religiosity and Summary Innovation Attitudes} \label{tabs1iv89}
\resizebox{0.98\columnwidth}{0.45\linewidth}{%
\begin{threeparttable}
\begin{tabular}{l*{8}{c}}
\toprule
&\multicolumn{1}{c}{(1)}&\multicolumn{1}{c}{(2)}&\multicolumn{1}{c}{(3)}&\multicolumn{1}{c}{(4)}&\multicolumn{1}{c}{(5)}&\multicolumn{1}{c}{(6)}&\multicolumn{1}{c}{(7)}&\multicolumn{1}{c}{(8)}\\
&\multicolumn{1}{c}{\textit{innosum}}&\multicolumn{1}{c}{\textit{innosum}}&\multicolumn{1}{c}{\textit{innosum}}&\multicolumn{1}{c}{\textit{innosum}}&\multicolumn{1}{c}{\textit{innosum}}&\multicolumn{1}{c}{\textit{innosum}}&\multicolumn{1}{c}{\textit{innosum}}&\multicolumn{1}{c}{\textit{innosum}}\\
\midrule
\textit{religiosity}&-0.056\sym{***}&&&&-0.193\sym{***}&&&\\
&(0.003)&&&&(0.042)&&&\\
\addlinespace
\textit{degree}&&-0.038\sym{***}&&&&-0.176\sym{***}&&\\
&&(0.003)&&&&(0.065)&&\\
\addlinespace
\textit{attendance}&&&-0.051\sym{***}&&&&-0.185\sym{***}&\\
&&&(0.003)&&&&(0.038)&\\
\addlinespace
\textit{pray}&&&&-0.021\sym{***}&&&&-0.112\sym{***}\\
&&&&(0.002)&&&&(0.027)\\
\addlinespace
\textit{gender}&-0.012\sym{***}&-0.014\sym{***}&-0.015\sym{***}&-0.013\sym{***}&-0.005&-0.006&-0.012\sym{***}&-0.003\\
&(0.002)&(0.002)&(0.002)&(0.002)&(0.003)&(0.004)&(0.002)&(0.003)\\
\addlinespace
\textit{age}&-0.002\sym{***}&-0.002\sym{***}&-0.002\sym{***}&-0.001\sym{***}&-0.002\sym{***}&-0.002\sym{***}&-0.002\sym{***}&-0.002\sym{***}\\
&(0.000)&(0.000)&(0.000)&(0.000)&(0.000)&(0.000)&(0.000)&(0.000)\\
\addlinespace
\textit{age2}&-0.000&-0.000&-0.000&-0.000&0.000&0.000&0.000&0.000\\
&(0.000)&(0.000)&(0.000)&(0.000)&(0.000)&(0.000)&(0.000)&(0.000)\\
\addlinespace
\textit{education}&0.003\sym{***}&0.003\sym{***}&0.003\sym{***}&0.003\sym{***}&0.003\sym{***}&0.002\sym{***}&0.003\sym{***}&0.003\sym{***}\\
&(0.000)&(0.000)&(0.000)&(0.000)&(0.000)&(0.000)&(0.000)&(0.000)\\
\addlinespace
\textit{paidwork}&-0.000&0.000&-0.000&-0.000&-0.001&-0.000&-0.001&-0.001\\
&(0.002)&(0.002)&(0.002)&(0.002)&(0.002)&(0.002)&(0.002)&(0.002)\\
\addlinespace
\textit{pwbefore}&0.022\sym{*}&0.026\sym{**}&0.021\sym{*}&0.039\sym{***}&-0.031&-0.036&-0.036&0.015\\
&(0.012)&(0.013)&(0.012)&(0.011)&(0.025)&(0.039)&(0.024)&(0.012)\\
\addlinespace
\textit{partner}&-0.020\sym{***}&-0.021\sym{***}&-0.020\sym{***}&-0.021\sym{***}&-0.018\sym{***}&-0.019\sym{***}&-0.017\sym{***}&-0.020\sym{***}\\
&(0.002)&(0.001)&(0.001)&(0.002)&(0.002)&(0.002)&(0.002)&(0.002)\\
\addlinespace
\textit{health}&0.010\sym{***}&0.010\sym{***}&0.010\sym{***}&0.009\sym{***}&0.010\sym{***}&0.010\sym{***}&0.011\sym{***}&0.009\sym{***}\\
&(0.001)&(0.001)&(0.001)&(0.001)&(0.001)&(0.001)&(0.001)&(0.001)\\
\addlinespace
\textit{child}&-0.014\sym{***}&-0.014\sym{***}&-0.014\sym{***}&-0.014\sym{***}&-0.011\sym{***}&-0.012\sym{***}&-0.012\sym{***}&-0.012\sym{***}\\
&(0.001)&(0.001)&(0.001)&(0.001)&(0.002)&(0.002)&(0.002)&(0.002)\\
\addlinespace
\textit{bornc}&0.004\sym{*}&0.006\sym{***}&0.005\sym{**}&0.006\sym{**}&-0.005&-0.003&-0.001&-0.004\\
&(0.002)&(0.002)&(0.002)&(0.002)&(0.004)&(0.005)&(0.003)&(0.004)\\
\addlinespace
\textit{fathere}&0.004\sym{***}&0.004\sym{***}&0.004\sym{***}&0.004\sym{***}&0.004\sym{***}&0.004\sym{***}&0.004\sym{***}&0.004\sym{***}\\
&(0.001)&(0.001)&(0.001)&(0.001)&(0.001)&(0.001)&(0.001)&(0.001)\\
\addlinespace
\textit{mothere}&0.004\sym{***}&0.004\sym{***}&0.004\sym{***}&0.004\sym{***}&0.004\sym{***}&0.004\sym{***}&0.004\sym{***}&0.004\sym{***}\\
&(0.001)&(0.001)&(0.001)&(0.001)&(0.001)&(0.001)&(0.001)&(0.001)\\
\addlinespace
\textit{cons}&0.546\sym{***}&0.538\sym{***}&0.536\sym{***}&0.516\sym{***}&&&&\\
&(0.014)&(0.015)&(0.014)&(0.013)&&&&\\
\midrule
\textit{N}&198,145&200,453&200,808&199,128&198,145&200,453&200,808&199,128\\
\textit{Adj. \(R^{2}\)}&0.242 &0.241 &0.242 &0.239 &&&&\\
\textit{idp}&&&&&0.000&0.000&0.000&0.000\\
\textit{cdf}&&&&&2,080&716&2,426&1,711\\
\textit{widstat}&&&&&158&68&130&167\\
\textit{AR test} &&&&& 0.000&0.003&0.000& 0.000       \\
\bottomrule
\end{tabular}
\begin{tablenotes}
\item Notes: OLS and IV estimates for alternative measures of religiosity are reported. Robust standard errors clustered at the level of instruments are in parentheses. All regressions include the following fixed effects: country, survey year, country-survey year, religious denomination, occupation, and income level. Kleibergen-Paap rk LM statistic (\textit{idp}), Kleibergen-Paap rk Wald F statistic (\textit{widstat}), Cragg-Donald Wald F statistic (\textit{cdf}), and Anderson-Rubin Wald F statistic (\textit{AR test}) are reported.
 * $p<0.10$; **$p<0.05$; ***$p<0.01$.
\end{tablenotes}
\end{threeparttable}
}
\end{table}

\noindent the sensitivity of this proposition, we introduce an overall, summary measure of innovation attitudes: \textit{\textit{innosum}= \textit{creative} + \textit{free} + \textit{different} + \textit{adventurous} - \textit{traditions} - \textit{rules}}, which is inspired by \textcite{Tabellini2005}\footnote{\textcite{Tabellini2005} introduces a summary variable of cultural traits which is the sum of the three positive beliefs (control, respect, trust) minus the negative belief (obedience).}. \textit{Innosum} is rescaled to be between 0 (low propensity to innovate) and 1 (high propensity to innovate); thus, higher values favour innovation. 

We re-estimate the baseline specifications with OLS and IV by identifying \textit{\textit{innosum}} as the dependent variable. The results are presented in Table 7. Columns 1-4 are estimated with OLS, while columns 5-8 are estimated with IV. All specifications include the complete set of control variables and fixed effects. The OLS results favour the negative association between religiosity and innovation attitudes through each measure of religiosity. The IV results are also in line with the previous results.

\begin{table}[H]\centering
\def\sym#1{\ifmmode^{#1}\else\(^{#1}\)\fi}
\caption{Alternative Model Specifications}  \label{tabs1iv8}
\resizebox{0.6\columnwidth}{0.47\linewidth}{%
\begin{threeparttable}
\begin{tabular}{l*{5}{c}}
\toprule
&\multicolumn{1}{c}{(1)}&\multicolumn{1}{c}{(2)}&\multicolumn{1}{c}{(3)}&\multicolumn{1}{c}{(4)}&\multicolumn{1}{c}{(5)}\\
&\multicolumn{1}{c}{\textit{inposav}}&\multicolumn{1}{c}{\textit{inposav}}&\multicolumn{1}{c}{\textit{inposav}}&\multicolumn{1}{c}{\textit{innosum}}&\multicolumn{1}{c}{\textit{inposav}}\\
\midrule
\textit{religiosity}&-0.204\sym{***}&-0.204\sym{***}&-0.204\sym{***}&-0.201\sym{***}&-0.201\sym{***}\\
&(0.058)&(0.058)&(0.058)&(0.041)&(0.059)\\
\addlinespace
\textit{gender}&-0.011\sym{***}&-0.011\sym{***}&-0.011\sym{***}&-0.004&-0.011\sym{***}\\
&(0.004)&(0.004)&(0.004)&(0.003)&(0.004)\\
\addlinespace
\textit{age}&-0.003\sym{***}&-0.003\sym{***}&-0.003\sym{***}&-0.002\sym{***}&-0.003\sym{***}\\
&(0.000)&(0.000)&(0.000)&(0.000)&(0.000)\\
\addlinespace
\textit{age2}&0.000\sym{***}&0.000\sym{***}&0.000\sym{***}&0.000&0.000\sym{***}\\
&(0.000)&(0.000)&(0.000)&(0.000)&(0.000)\\
\addlinespace
\textit{education}&0.002\sym{***}&0.002\sym{***}&0.002\sym{***}&0.003\sym{***}&0.002\sym{***}\\
&(0.000)&(0.000)&(0.000)&(0.000)&(0.000)\\
\addlinespace
\textit{paidwork}&-0.000&-0.000&-0.000&-0.001&-0.000\\
&(0.002)&(0.002)&(0.002)&(0.002)&(0.002)\\
\addlinespace
\textit{pwbefore}&-0.151\sym{***}&-0.151\sym{***}&-0.151\sym{***}&-0.035&-0.149\sym{***}\\
&(0.036)&(0.036)&(0.036)&(0.025)&(0.036)\\
\addlinespace
\textit{partner}&-0.015\sym{***}&-0.015\sym{***}&-0.015\sym{***}&-0.018\sym{***}&-0.015\sym{***}\\
&(0.002)&(0.002)&(0.002)&(0.002)&(0.002)\\
\addlinespace
\textit{health}&0.017\sym{***}&0.017\sym{***}&0.017\sym{***}&0.010\sym{***}&0.017\sym{***}\\
&(0.001)&(0.001)&(0.001)&(0.001)&(0.001)\\
\addlinespace
\textit{child}&-0.017\sym{***}&-0.017\sym{***}&-0.017\sym{***}&-0.011\sym{***}&-0.017\sym{***}\\
&(0.002)&(0.002)&(0.002)&(0.002)&(0.002)\\
\addlinespace
\textit{bornc}&-0.024\sym{***}&-0.024\sym{***}&-0.024\sym{***}&-0.006&-0.024\sym{***}\\
&(0.005)&(0.005)&(0.005)&(0.004)&(0.005)\\
\addlinespace
\textit{fathere}&0.006\sym{***}&0.006\sym{***}&0.006\sym{***}&0.004\sym{***}&0.005\sym{***}\\
&(0.001)&(0.001)&(0.001)&(0.001)&(0.001)\\
\addlinespace
\textit{mothere}&0.004\sym{***}&0.004\sym{***}&0.004\sym{***}&0.004\sym{***}&0.004\sym{***}\\
&(0.001)&(0.001)&(0.001)&(0.001)&(0.001)\\
\midrule
\textit{N}&202,300&202,300&202,300&198,145&202,300\\
\textit{idp}&0.000&0.000&0.000&0.000&0.000\\
\textit{cdf}&1,100&1,100&1,100&1,064&738\\
\textit{widstat}&83&83&83&82&58\\
\textit{AR test}&0.000&0.000&0.000&0.000&0.000\\
\textit{Hansen J} &0.480&0.480&0.480&0.195&0.612\\
\bottomrule
\end{tabular}
\begin{tablenotes}
\item Notes: Robust standard errors clustered at the level of instruments are in parentheses. All regressions include the following fixed effects: country, survey year, country-survey year, religious denomination, occupation, and income level. Kleibergen-Paap rk LM statistic (\textit{idp}), Kleibergen-Paap rk Wald F statistic (\textit{widstat}), Cragg-Donald Wald F statistic (\textit{cdf}), Anderson-Rubin Wald F statistic (\textit{AR test}), and Sargan-Hansen test \textit{Hansen J} are reported. * $p<0.10$; **$p<0.05$; ***$p<0.01$.
\end{tablenotes}
\end{threeparttable}
}
\end{table}

\subsection{Alternative Model Specifications}
\textit{Religiosity} is defined as the principal component of three measures: \textit{degree}, \textit{attendance}, and \textit{pray}. However, \textit{degree} differs from the other two in one important dimension. \textit{Pray} and \textit{attendance} are measures of an activity, an action, while \textit{degree} is self-evaluation of a belief, a value. It is plausible to think that one cannot easily overestimate or underestimate the frequency of an activity but can do so to evaluate a belief. Therefore an interpersonal comparison of \textit{degree} might be biased since some individuals may overvalue or undervalue their religiosity. This point is especially crucial to our instrumental variables, given that we instrument individual $i$'s \textit{degree} by the average \textit{degree} of other people. If many respondents tend to misvalue their \textit{degree}, then this self-report bias will lead to weak instruments. Indeed, the diagnostic tests' results for \textit{degree} in Tables 4 and 5 support this evaluation. Furthermore, \textit{degree} is a component of \textit{religiosity}; it thus endangers the instrument for \textit{religiosity} as well. In order to see if the results are sensitive to the possible self-report bias of \textit{degree}, we modify the IV strategy. Table 8 displays the results. 

The first column is estimated with 2SLS and \textit{religiosity} is instrumented with \textit{attendance} and \textit{pray}. Columns 2 and 3 follow the specification of column 1 but are estimated with \textit{k}-class estimators. Prior studies suggest that Limited Information Maximum Likelihood Method (LIML) and Fuller edited LIML have a better finite-sample performance than 2SLS in the presence of weak instruments (\cite{Baum07}). Accordingly, in case of strong instruments, the estimates of 2SLS, LIML and Fuller should yield very similar results. Column 2 is estimated with LIML, and column 3 is with Fuller edited LIML with Fuller parameter $k=1$. Column 4 is estimated with 2SLS, but the dependent variable is \textit{\textit{innosum}} and the instruments are \textit{attendance} and \textit{pray}. In specification 5, we add \textit{iv\_degree} as the third instrument for \textit{religiosity}. The results of 2SLS, LIML, and Fuller estimates are the same, suggesting that our instruments are strong. The coefficients of \textit{religiosity} in columns 1, 2, and 3 are qualitatively and quantitatively similar to the corresponding coefficient of \textit{religiosity} (-0.198) in column 3 in Table 4, indicating that the initial instrument of \textit{religiosity} (\textit{iv\_religiosity}) is not significantly affected by the potential self-report bias of \textit{degree}. 

\subsection{Reduced Sample Estimates}
Table 9 displays the estimates from four different sub-samples. Each coefficient comes from a different IV estimate. The number of observations is below standard errors, which is followed by Cragg-Donald Wald F statistic (\textit{cdf}), Kleibergen-Paap rk Wald F statistic (\textit{widstat}), and Anderson-Rubin Wald F statistic (\textit{AR test}) results. 

First, we drop Muslim countries. Turkey, Albania, and Kosova are the only dominantly Muslim counties in the ESS. The rest of the sample consists of dominantly Christian countries, if not dominantly Atheist\footnote{The second dominant religion is Christianity in each dominantly Atheist countries.}. Despite controlling for country and denomination fixed effects, combining Muslim and Christian countries may complicate the interpretation since we consider inter-religion religiosity regardless of denomination. Moreover, Muslim countries exhibit outlier values for religiosity, especially for \textit{degree}, such as $0.7$-$0.73$ on average (see Table A3), which is pretty high given that the overall cross-country average of \textit{degree} is $0.46$. Therefore, we exclude Muslim countries from the sample and re-estimate the baseline specification as a robustness check. Columns 1 and 2 report the results. The overall negative effect of \textit{attendance} increases compared to the full sample results; from $-0.185$ to $-0.256$ on \textit{inposav} and from $0.216$ to $0.281$ on \textit{innegav}. On the other hand, the effect of \textit{degree} on \textit{innegav} becomes less significant, which is not surprising since the Muslim countries exhibit outlier values of \textit{degree}. All in all, the overall picture stays robust.

\begin{table}[H]\centering
\def\sym#1{\ifmmode^{#1}\else\(^{#1}\)\fi}
\caption{Reduced Sample Estimates} \label{tabs1iv10}
\resizebox{0.95\columnwidth}{0.315\linewidth}{%
\begin{threeparttable}
\begin{tabular}{l*{10}{c}}
\toprule
&\multicolumn{1}{c}{(1)}&\multicolumn{1}{c}{(2)}&\multicolumn{1}{c}{(3)}&\multicolumn{1}{c}{(4)}&\multicolumn{1}{c}{(5)}&\multicolumn{1}{c}{(6)}&\multicolumn{1}{c}{(7)}&\multicolumn{1}{c}{(8)} &\multicolumn{1}{c}{(9)}&\multicolumn{1}{c}{(10)} \\
&\multicolumn{1}{c}{\textit{inposav}}&\multicolumn{1}{c}{\textit{innegav}}&\multicolumn{1}{c}{\textit{inposav}}&\multicolumn{1}{c}{\textit{innegav}}&\multicolumn{1}{c}{\textit{inposav}}&\multicolumn{1}{c}{\textit{innegav}}&\multicolumn{1}{c}{\textit{inposav}}&\multicolumn{1}{c}{\textit{innegav}}&\multicolumn{1}{c}{\textit{inposav}}&\multicolumn{1}{c}{\textit{innegav}} \\
\midrule
\textit{religiosity}&-0.214\sym{***}&0.199\sym{***}&-0.235***&0.212***&-0.259***&0.192***&-0.162***&0.200***&-0.153**&0.206***\\
&(0.065)&(0.058)&(0.056)&(0.062)&(0.059)&(0.065)&(0.061)&(0.055)& (0.063)& (0.057)\\
&199,429  & 199,324 & 184,551 & 184,458 & 181,680 &  181,591 &181,332  &184,230&174,626&174,533 \\
&2,041  & 2,028 & 1,632 & 1,624 & 1,530 & 1,523 & 1,999 &1,987 &1,916& 1,903\\
& 151  & 150 &105  & 105 &99  & 98 &155  &156 & 133&133\\
&0.000  &0.000  & 0.000 & 0.000 & 0.000 &0.003  & 0.004 & 0.001&0.012&0.000 \\
\addlinespace
\textit{degree}&-0.178*&0.164**&-0.251***&0.332***:&-0.259***&0.220**&-0.157*&0.228***&-0.125 &0.236***\\
&(0.093)&(0.074)&(0.091)&(0.107)&(0.056)&(0.093)&(0.088)&(0.074)&(0.106) & (0.081)\\
&201,906  & 201,790 &186,578  &186,474  & 183,696 & 183,596 &183,690  &183,577&176,822&176,718\\
& 877 & 868 & 358 & 354 &415  &412  & 866   &857&615&609\\
&101 & 99 &45 & 44 & 57 & 56 &98   &97&57&57\\
&0.043  &0.031  & 0.004 & 0.001 &0.002 &0.019  &0.061  &0.002&0.228&0.006  \\
\addlinespace
\textit{\textit{attendance}}&-0.256\sym{***}&0.281\sym{***}&-0.196***&0.192***&-0.287***&0.256***&-0.140***&0.194***&-0.161***&0.192***\\
&(0.066)&(0.060)&(0.048)&(0.048)&(0.062)&(0.064)&(0.051) &(0.047)&(0.052) &(0.049) \\
& 202,287 &202,170  &187,023  & 186,920 &184,134  & 184, 035 & 184,024 & 183,910&177,171&177,067 \\
&1,653  & 1,646 & 2,165 & 2,159 &1,328  &1,322  & 2,248 & 2,243& 2,262&2,253\\
& 113 &113  &100  &99  & 81 & 81 & 120 & 120&108&108 \\
& 0.000 &0.000  & 0.000 & 0.000 & 0.000 &0.000  & 0.003 &0.000&0.001&0.000  \\
\addlinespace
\textit{pray}&-0.117\sym{***}&0.104***&-0.148***   &0.105** &-0.147***&0.084* &-0.099**&0.121***&-0.072*&0.123*** \\
&(0.039)&(0.040)&(0.039)&(0.044)&(0.038)&(0.044)&(0.039)&(0.041)&(0.040)&(0.042) \\
& 200,499 & 200,387 &185,498  &185,399  &  182,613& 182,518 & 182,319 &182,210& 175,562&175,462 \\
& 1,776 &1,769  &1,411  & 1,406 & 1,457 &1,452  & 1,586 &1,580 & 1,645&1,638\\
& 162 &161  & 113 &113  &115 & 115&152  &152& 156&155 \\
&0.002  &0.008  & 0.000 & 0.014 & 0.000 & 0.049 & 0.008 & 0.002&0.067 &0.002 \\
\addlinespace
\bottomrule
\end{tabular}
\begin{tablenotes}
\item Notes: Robust standard errors clustered at the level of instruments are in parentheses and followed by the number of observations, Cragg-Donald Wald F statistic (\textit{cdf}), Kleibergen-Paap rk Wald F statistic (\textit{widstat}), and Anderson-Rubin Wald F statistic (\textit{AR test}). All specifications include the full set of control variables and fixed effects. * $p<0.10$; **$p<0.05$; ***$p<0.01$.
\end{tablenotes}
\end{threeparttable}
}
\end{table}

Second, we exclude Russia and Norway from the main sample because the percentages of the first and the second dominant denominations among respondents are pretty close, less than 1\%, for those countries. The majority of respondents from Norway are Atheists with $47.02\%$; Protestants come second with $46.36\%$. For Russia, the majority is Eastern Orthodox with $46.03\%$; Atheists come second with $45.83\%$. Since the dominant denomination is one of the elements we construct instruments, any misidentification would cause bias in the instruments of individuals who share the misidentified dominant denomination. We thus drop Russia and Norway and re-estimate the baseline specifications to assess the difference. Columns 3 and 4 in Table 9 report the results. Column 3 shows that the coefficient of each religiosity measure increases for \textit{inposav} and \textit{degree} becomes significant at the 99\% level compared to the full sample results in Table 4. 

Third, we exclude Muslim countries along with Russia and Norway to obtain a relatively homogeneous sample. Columns 5 and 6 display the results. Coefficients of all religiosity measures substantially increase for \textit{inposav} and \textit{degree} becomes more significant compared to Table 4. For \textit{innegav}, \textit{degree} and \textit{pray} become less significant compared to the full sample.

Fourth, we exclude once-believers. Non-believers are split into two categories among themselves: never-believers and once-believers. Table C7 shows that once-believers exhibit slightly higher religiosity means. The reason behind is that religiosity tends to be an absorbed value. Even though a person has stopped believing in a religion, there might be persistent effects, let alone the fact that personal values and traits might have been mostly grown when the person was a believer. Thus, to examine whether the higher religiosity of once-believers substantially affects the overall results, we drop once-believers from the full sample and leave the respondents who have been either an always-believer or never-believer. The corresponding results are presented in columns 7 and 8 in Table 9. Column 8 shows that the overall results for \textit{innegav} do not change substantially compared to Table 5. However, the coefficient of each religiosity measure somewhat decreases in column 7 compared to Table 4.

Lastly, we exclude second-generation immigrants. Immigrants\footnote{Note that immigrants are controlled for in the baseline specifications with variable \textit{bornc}: born-in country.} and their descendants might possess peculiar traits due to different cultural and institutional environments of their origin country. Including second-generation immigrants in the sample might introduce some unobservable effects regarding innovation attitudes. Columns 9 and 10 report the results. The coefficients for \textit{inposav} slightly decrease and \textit{degree} turns to be insignificant compared to Table 4. On the contrary, the results of \textit{innegav} are pretty stable and similar to Table 5. 

\subsection{Possible Violations of Exogeneity}
The dominant denomination is one of the factors by which we construct the instruments.As Table A1 indicates, a very high percentage of the population belongs to the dominant religion in many countries, while some countries have relatively low percentages of the dominant religion. For instance, the prevailing denomination is atheism in Switzerland (34\%) and Germany (38\%), yet they comprise roughly one-third of the population. Apparently, there are other religions in those countries with a significant presence. However, the significant presence of religions that are different from the dominant religion of a country, and their combination, might introduce some unobservable traits related to innovation attitudes. Suppose the instruments for religiosity are computed on a group of individuals that exhibit some peculiar values of the unobservable traits (that stem from the presence of different religion combinations in the country) related to innovation attitudes. In that case, the exogeneity condition may not be met. Since the instrument is the average religiosity of a group of countries sharing the same dominant denomination, it is unlikely that such unobservable traits prevail in every country in the group.

\begin{table}[H]\centering
\def\sym#1{\ifmmode^{#1}\else\(^{#1}\)\fi}
\caption{IV Estimates with Fixed Effects for Religion Combinations} \label{tabs1ivrelpair2}
\resizebox{0.98\columnwidth}{0.45\linewidth}{%
\begin{threeparttable}
\begin{tabular}{l*{8}{c}}
\toprule
&\multicolumn{1}{c}{(1)}&\multicolumn{1}{c}{(2)}&\multicolumn{1}{c}{(3)}&\multicolumn{1}{c}{(4)}&\multicolumn{1}{c}{(5)}&\multicolumn{1}{c}{(6)}&\multicolumn{1}{c}{(7)}&\multicolumn{1}{c}{(8)}\\
&\multicolumn{1}{c}{\textit{inposav}}&\multicolumn{1}{c}{\textit{inposav}}&\multicolumn{1}{c}{\textit{inposav}}&\multicolumn{1}{c}{\textit{inposav}}&\multicolumn{1}{c}{\textit{innegav}}&\multicolumn{1}{c}{\textit{innegav}}&\multicolumn{1}{c}{\textit{innegav}}&\multicolumn{1}{c}{\textit{innegav}}\\
\midrule
\textit{religiosity}&-0.193\sym{***}&&&&0.180\sym{***}&&&\\
&(0.074)&&&&(0.066)&&&\\
\addlinespace
\textit{degree}&&-0.227\sym{*}&&&&0.205\sym{**}&&\\
&&(0.123)&&&&(0.093)&&\\
\addlinespace
\textit{attendance}&&&-0.140\sym{**}&&&&0.161\sym{***}&\\
&&&(0.055)&&&&(0.054)&\\
\addlinespace
\textit{pray}&&&&-0.118\sym{***}&&&&0.110\sym{**}\\
&&&&(0.043)&&&&(0.045)\\
\addlinespace
\textit{gender}&-0.011\sym{**}&-0.009&-0.019\sym{***}&-0.009\sym{*}&-0.008\sym{*}&-0.009&-0.001&-0.009\sym{*}\\
&(0.004)&(0.007)&(0.002)&(0.005)&(0.005)&(0.006)&(0.003)&(0.006)\\
\addlinespace
\textit{age}&-0.003\sym{***}&-0.003\sym{***}&-0.003\sym{***}&-0.003\sym{***}&-0.000&-0.000&-0.001&-0.001\\
&(0.000)&(0.000)&(0.000)&(0.000)&(0.000)&(0.000)&(0.000)&(0.000)\\
\addlinespace
\textit{age2}&0.000\sym{***}&0.000\sym{***}&0.000\sym{**}&0.000\sym{**}&0.000\sym{***}&0.000\sym{***}&0.000\sym{***}&0.000\sym{***}\\
&(0.000)&(0.000)&(0.000)&(0.000)&(0.000)&(0.000)&(0.000)&(0.000)\\
\addlinespace
\textit{education}&0.002\sym{***}&0.002\sym{***}&0.003\sym{***}&0.003\sym{***}&-0.003\sym{***}&-0.003\sym{***}&-0.003\sym{***}&-0.003\sym{***}\\
&(0.000)&(0.000)&(0.000)&(0.000)&(0.000)&(0.000)&(0.000)&(0.000)\\
\addlinespace
\textit{paidwork}&-0.000&0.001&0.000&-0.000&0.005\sym{**}&0.004\sym{*}&0.004\sym{*}&0.005\sym{**}\\
&(0.002)&(0.002)&(0.002)&(0.002)&(0.002)&(0.002)&(0.002)&(0.002)\\
\addlinespace
\textit{pwbefore}&-0.139\sym{***}&-0.171\sym{**}&-0.127\sym{***}&-0.093\sym{***}&-0.224\sym{***}&-0.198\sym{***}&-0.223\sym{***}&-0.267\sym{***}\\
&(0.042)&(0.070)&(0.035)&(0.022)&(0.046)&(0.060)&(0.044)&(0.040)\\
\addlinespace
\textit{partner}&-0.015\sym{***}&-0.015\sym{***}&-0.015\sym{***}&-0.017\sym{***}&0.022\sym{***}&0.022\sym{***}&0.021\sym{***}&0.024\sym{***}\\
&(0.003)&(0.003)&(0.003)&(0.002)&(0.002)&(0.002)&(0.002)&(0.002)\\
\addlinespace
\textit{health}&0.017\sym{***}&0.017\sym{***}&0.018\sym{***}&0.016\sym{***}&0.004\sym{***}&0.004\sym{***}&0.003\sym{**}&0.005\sym{***}\\
&(0.001)&(0.001)&(0.001)&(0.001)&(0.001)&(0.001)&(0.001)&(0.001)\\
\addlinespace
\textit{child}&-0.017\sym{***}&-0.017\sym{***}&-0.018\sym{***}&-0.017\sym{***}&0.001&0.001&0.002&0.001\\
&(0.002)&(0.003)&(0.002)&(0.002)&(0.002)&(0.003)&(0.002)&(0.002)\\
\addlinespace
\textit{bornc}&-0.023\sym{***}&-0.023\sym{***}&-0.017\sym{***}&-0.023\sym{***}&-0.029\sym{***}&-0.029\sym{***}&-0.033\sym{***}&-0.029\sym{***}\\
&(0.006)&(0.007)&(0.004)&(0.005)&(0.005)&(0.006)&(0.004)&(0.005)\\
\addlinespace
\textit{fathere}&0.006\sym{***}&0.005\sym{***}&0.006\sym{***}&0.006\sym{***}&-0.001&-0.001&-0.001&-0.001\\
&(0.001)&(0.001)&(0.001)&(0.001)&(0.001)&(0.001)&(0.001)&(0.001)\\
\addlinespace
\textit{mothere}&0.004\sym{***}&0.004\sym{***}&0.005\sym{***}&0.004\sym{***}&-0.004\sym{***}&-0.004\sym{***}&-0.004\sym{***}&-0.004\sym{***}\\
&(0.001)&(0.001)&(0.001)&(0.001)&(0.001)&(0.001)&(0.001)&(0.001)\\
\midrule
\textit{N}&202,298&204,787&205,175&203,383&202,189&204,667&205,054&203,267\\
\textit{idp}&0.000&0.000&0.000&0.000&0.000&0.000&0.000&0.000\\
\textit{cdf}&1,768&584&2,356&1,536&1,755&575&2348&1528\\
\textit{widstat}&126&51&114&139&125&50&113&138\\
\textit{AR test} &0.003&0.027&0.003&0.004&0.008&0.101&0.008& 0.010       \\
\bottomrule
\end{tabular}
\begin{tablenotes}
\item Notes: IV estimates for alternative measures of religiosity are reported. Robust standard errors clustered at the level of instruments are in parentheses. All regressions include the following fixed effects: country, survey year, country-survey year, religious denomination, occupation, and income level. Kleibergen-Paap rk LM statistic (\textit{idp}), Kleibergen-Paap rk Wald F statistic (\textit{widstat}), Cragg-Donald Wald F statistic (\textit{cdf}), and Anderson-Rubin Wald F statistic (\textit{AR test}) are reported.
 * $p<0.10$; **$p<0.05$; ***$p<0.01$.
\end{tablenotes}
\end{threeparttable}
}
\end{table}

Nevertheless, as a robustness check, we include fixed effects for each pair of religion-the dominant religion to account for diverse religious environments resulting from different combinations of religions in a country. The results are presented in Table 10. Despite moderate decreases in the coefficients, the results are pretty similar to the baseline specifications reported in Tables 4 and 5, suggesting that the instruments are relevant and representative of various religious environments.

\begin{table}[H]\centering
\def\sym#1{\ifmmode^{#1}\else\(^{#1}\)\fi}
\caption{IV Estimates with Instruments of Non-Neighbouring Countries } \label{tabs1ivrelpair}
\resizebox{0.98\columnwidth}{0.45\linewidth}{%
\begin{threeparttable}
\begin{tabular}{l*{8}{c}}
\toprule
&\multicolumn{1}{c}{(1)}&\multicolumn{1}{c}{(2)}&\multicolumn{1}{c}{(3)}&\multicolumn{1}{c}{(4)}&\multicolumn{1}{c}{(5)}&\multicolumn{1}{c}{(6)}&\multicolumn{1}{c}{(7)}&\multicolumn{1}{c}{(8)}\\
&\multicolumn{1}{c}{\textit{inposav}}&\multicolumn{1}{c}{\textit{inposav}}&\multicolumn{1}{c}{\textit{inposav}}&\multicolumn{1}{c}{\textit{inposav}}&\multicolumn{1}{c}{\textit{innegav}}&\multicolumn{1}{c}{\textit{innegav}}&\multicolumn{1}{c}{\textit{innegav}}&\multicolumn{1}{c}{\textit{innegav}}\\
\midrule
\textit{religiosity}&-0.253\sym{***}&&&&0.270\sym{***}&&&\\
&(0.071)&&&&(0.062)&&&\\
\addlinespace
\textit{degree}&&-0.271\sym{**}&&&&0.324\sym{***}&&\\
&&(0.133)&&&&(0.098)&&\\
\addlinespace
\textit{attendance}&&&-0.251\sym{***}&&&&0.259\sym{***}&\\
&&&(0.065)&&&&(0.057)&\\
\addlinespace
\textit{pray}&&&&-0.127\sym{***}&&&&0.160\sym{***}\\
&&&&(0.042)&&&&(0.043)\\
\addlinespace
\textit{gender}&-0.008\sym{*}&-0.007&-0.017\sym{***}&-0.008\sym{*}&-0.013\sym{***}&-0.016\sym{**}&-0.003&-0.015\sym{***}\\
&(0.004)&(0.007)&(0.003)&(0.005)&(0.004)&(0.006)&(0.003)&(0.006)\\
\addlinespace
\textit{age}&-0.003\sym{***}&-0.003\sym{***}&-0.003\sym{***}&-0.003\sym{***}&-0.000&-0.000&-0.000&-0.000\\
&(0.000)&(0.000)&(0.000)&(0.000)&(0.000)&(0.000)&(0.000)&(0.000)\\
\addlinespace
\textit{age2}&0.000\sym{***}&0.000\sym{***}&0.000\sym{**}&0.000\sym{***}&0.000\sym{***}&0.000\sym{***}&0.000\sym{***}&0.000\sym{***}\\
&(0.000)&(0.000)&(0.000)&(0.000)&(0.000)&(0.000)&(0.000)&(0.000)\\
\addlinespace
\textit{education}&0.002\sym{***}&0.002\sym{***}&0.003\sym{***}&0.003\sym{***}&-0.003\sym{***}&-0.003\sym{***}&-0.003\sym{***}&-0.003\sym{***}\\
&(0.000)&(0.000)&(0.000)&(0.000)&(0.000)&(0.000)&(0.000)&(0.000)\\
\addlinespace
\textit{paidwork}&-0.001&0.000&-0.000&-0.000&0.005\sym{**}&0.004\sym{*}&0.005\sym{**}&0.005\sym{**}\\
&(0.002)&(0.002)&(0.002)&(0.002)&(0.002)&(0.002)&(0.002)&(0.002)\\
\addlinespace
\textit{pwbefore}&-0.170\sym{***}&-0.194\sym{***}&-0.180\sym{***}&-0.104\sym{***}&-0.171\sym{***}&-0.131\sym{**}&-0.165\sym{***}&-0.235\sym{***}\\
&(0.043)&(0.075)&(0.042)&(0.019)&(0.038)&(0.061)&(0.039)&(0.033)\\
\addlinespace
\textit{partner}&-0.014\sym{***}&-0.014\sym{***}&-0.013\sym{***}&-0.017\sym{***}&0.021\sym{***}&0.021\sym{***}&0.020\sym{***}&0.024\sym{***}\\
&(0.003)&(0.003)&(0.003)&(0.002)&(0.002)&(0.003)&(0.002)&(0.002)\\
\addlinespace
\textit{health}&0.017\sym{***}&0.017\sym{***}&0.019\sym{***}&0.016\sym{***}&0.003\sym{***}&0.003\sym{***}&0.002&0.005\sym{***}\\
&(0.001)&(0.001)&(0.001)&(0.001)&(0.001)&(0.001)&(0.001)&(0.001)\\
\addlinespace
\textit{child}&-0.016\sym{***}&-0.016\sym{***}&-0.017\sym{***}&-0.017\sym{***}&-0.001&-0.001&0.001&0.000\\
&(0.002)&(0.003)&(0.002)&(0.002)&(0.002)&(0.003)&(0.002)&(0.002)\\
\addlinespace
\textit{bornc}&-0.027\sym{***}&-0.026\sym{***}&-0.022\sym{***}&-0.023\sym{***}&-0.025\sym{***}&-0.024\sym{***}&-0.031\sym{***}&-0.026\sym{***}\\
&(0.006)&(0.009)&(0.005)&(0.006)&(0.005)&(0.007)&(0.004)&(0.006)\\
\addlinespace
\textit{fathere}&0.006\sym{***}&0.005\sym{***}&0.006\sym{***}&0.006\sym{***}&-0.001&-0.000&-0.001&-0.001\\
&(0.001)&(0.001)&(0.001)&(0.001)&(0.001)&(0.001)&(0.001)&(0.001)\\
\addlinespace
\textit{mothere}&0.004\sym{***}&0.004\sym{***}&0.005\sym{***}&0.004\sym{***}&-0.004\sym{***}&-0.004\sym{***}&-0.004\sym{***}&-0.003\sym{***}\\
&(0.001)&(0.001)&(0.001)&(0.001)&(0.001)&(0.001)&(0.001)&(0.001)\\
\midrule
\textit{N}&202,300&204,788&205,176&203,384&202,191&204,668&205,055&203,268\\
\textit{idp}&0.000&0.000&0.000&0.000&0.000&0.000&0.000&0.000\\
\textit{cdf}&1,621&469&1,823&1,408&1,610&463&1,815&1,401\\
\textit{widstat}&121&42&95&144&120&42&94&144\\
\textit{AR test}&0.000 &0.024&0.000&0.002&0.000&0.001&0.000&0.000\\
\bottomrule
\end{tabular}
\begin{tablenotes}
\item Notes: IV estimates for alternative measures of religiosity are reported. Robust standard errors clustered at the level of instruments are in parentheses. All regressions include the following fixed effects: country, survey year, country-survey year, religious denomination, occupation, and income level. Kleibergen-Paap rk LM statistic (\textit{idp}), Kleibergen-Paap rk Wald F statistic (\textit{widstat}), Cragg-Donald Wald F statistic (\textit{cdf}), and Anderson-Rubin Wald F statistic (\textit{AR test}) are reported.
 * $p<0.10$; **$p<0.05$; ***$p<0.01$.
\end{tablenotes}
\end{threeparttable}
}
\end{table}

Another possible violation of exogeneity might stem from using neighbouring countries when computing the instruments. We argue that religiosity is a transnational trait; therefore, one's religiosity can be instrumented by other people's religiosity who live in a similar religious environment. However, suppose there are substantial similarities in the institutional environments of countries that share the same dominant denomination. In that case, individual innovation attitudes could be affected by these similarities in a particular way, violating the exogeneity assumption. We, therefore, use a large set of fixed effects to eliminate the institutional level factors. However, using neighbouring countries when computing instruments might introduce another channel of institutional similarity, given that sharing a border and spatial proximity can reinforce a similar institutional environment. Motivated by this reasoning, we change the computation of instruments as a robustness check. We build instruments for religious intensity variables by computing the average religiosity of people of the same sex, age range, and religious affiliation who live in countries with the same dominant religious denomination that are not neighbours, \textit{i.e.,} do not share a border. Table 11 presents the results. All coefficients for \textit{inposav} and \textit{innegav} are increased compared to the main results in Tables 4 and 5. The negative effect of religion on innovation attitudes persists.

\section{Possible Channels of Causality}\label{section: chan}
\subsection{Time Allocation}
Starting from Gary Becker's pioneering work \textit{``A Theory of the Allocation of Time"} (\cite{Becker1965}), numerous papers have analysed agents' time allocation among various activities. The work of \textcite{AzziEhrenberg} is the first paper that considers religious participation in the context of household members' time allocation, presenting the first model of consumer choice in religious markets. They use a microeconomic approach to examine the demand of religion, introducing a multi-period utility-maximising model of household behaviour. In the model, religious participation is a part of household members' utility function and is a time-consuming activity as well as an investment in after-life consumption. Household members are men and women who have different opportunity costs of time due to different wages they are subject to in the labour market. They allocate their time by considering not only their wage but also the marginal utilities of after-life and in-life consumption. Suppose household members allocate more time for religious participation, favouring after-life consumption. In that case, they will have less time for productive activities for in-life consumption, resulting in lower total household production. 

The point we are interested in the model of Azzi and Ehrenberg is the opportunity cost of time spent on religious activities, given that religious participation is a time-consuming activity and time is scarce. The concept of \textit{opportunity cost of religious participation} can also be considered for other relevant time-consuming activities such as human capital formation, which requires many years of education, a lifelong habit of reading, and attending scientific and intellectual activities. All these activities demand time, and time is limited. If an individual is religious and allocates a certain amount of time to religious activities, then there will be a decrease in the maximum amount of potential time that could be devoted to human capital formation. It is well-known that human capital is one of the most critical drivers of innovation. Thus, any unfavourable activity to human capital formation can be plausibly assumed unfavourable to innovation. 

On the contrary, the social capital approach is a perspective that contrasts with the negative effect of time spent on religious activities. The main idea is that participating in religious services is a form of networking and helps the attendant build social capital. \textcite{BarroMcCleary2006} underline that social capital formation through religious participation is favourable for the economy only if it fosters economically relevant individual traits such as work ethic, thrifty, honesty, and trust. 

"The social capital and cultural aspects of religion (communal services, rituals, religious schools) are significant only to the extent that they influence beliefs and, hence, behaviour. In fact, for given beliefs, more time spent on communal activities would tend to be an economic drag, at least as measured by market output (GDP)." (\cite{BarroMcCleary2006}, pp. 51). 

Consequently, the net effect of religious participation, positive or negative, on economic outcomes would be determined by taking into account the comparative consequences of the two approaches mentioned: the opportunity cost of time spent on religious activities and the social capital formation through religious participation. In our case, it is somewhat arguable to claim that social capital formation fosters favourable individual traits for innovation. As mentioned in Section 2.2, taking risks, being different, adventurous, and creative are the main individual traits that are considered favourable to innovation. However, as shown by our analyses, religious participation does not seem to foster any of them. Moreover, it is likely to impose traditional values and rules. The results throughout the present paper support this argument. \textit{Attendance} has been the most robust religiosity measure across different specifications, and it negatively affects innovation attitudes.

\subsection{The Fear of Uncertainty}

As mentioned in Section 2, Hofstede (1980/\cite*{Hofstede2001}) identifies uncertainty avoidance as one of the five dimensions on which national cultures differ. Innovation requires newness and risk tolerance, thus is related to uncertainty which hints at unpredictability and a lack of structure and information (\cite{Rogers1983}). Individuals or the decision-making unit involved in the first stages of the innovation process would be naturally unsure of the new idea's results and face the inherent uncertainty of newness. Therefore, an individual's propensity to innovate is expected to be positively related to the ability to deal with uncertainty\footnote{Uncertainty and risk are initially different concepts. Risk is the perceived probability that a particular event will happen. Uncertainty is about ambiguity. Any event may happen, and there is no probability attached to it (\cite{HofstedeMinkov2010}). However, for simplicity, we use uncertainty as inclusive of risk since any degree of uncertainty contains risk, whether with a known or unknown probability. Hence, we use the term ``the fear of uncertainty" instead of ``uncertainty avoidance" to distinguish it from Hofstede's concept.}. Low levels of fear of uncertainty refer to more openness toward new ideas and change, more willingness to take risks, and less fear of novelty. 

\textcite{Shane1993} shows that uncertainty avoidance is one of the values that have a negative impact on a country's overall innovativeness, and it explains, to some degree, the variation in national rates of innovativeness across countries. \textcite{Shane1995} uses the data of 4405 individuals in 43 organizations from 68 different countries and examines the relationship between uncertainty avoidance and individual preferences for four innovation championing roles: the organizational maverick, the network facilitator, the transformational leader, and the organizational buffer. He shows that individuals from uncertainty-avoiding cultures are less likely to prefer championing roles. ``One might argue that uncertainty-accepting societies are more innovative (\cite{Shane1993}) because championing roles which overcome organizational inertia to innovation are more likely to be accepted in those societies." (\cite{Shane1995}, pp. 64).

\textcite{Chenetal2014} find that higher levels of uncertainty avoidance have negative effects on corporate innovation. Firms located in countries with higher levels of uncertainty avoidance generate fewer and less critical patents, and their R\&D expenditures are less efficient. \textcite{WilliamMcquire} show that uncertainty avoidance has a negative effect on economic creativity which facilitates innovation implementation.

Religion is one of the ways to cope with uncertainty (Hofstede \cite*{Hofstede2001}). It can be argued that religious people tend to fear uncertainty more than irreligious people. The first empirical study that used risk analysis in the context of religiosity is that of \textcite{MillarHoffman1995}. They consider religious acceptance a risk-averse behaviour and the rejection of religious beliefs as risk-taking behaviour.  
One salient feature of religion is to provide a sort of protection, both materially and mentally. Religion mitigates the uncertainties and risks of daily life by providing spiritually rewarding networks of welfare activities for the community, such as charity for the poor, assistance for individuals who experience personal disasters, elder care, medical assistance, orphanages, and education (\cite{GillLundsgaarde}; \cite{BarroMcCleary2006}; \cite{ScheveStasavage}). Those activities are provided by religious organizations (\textit{e.g.,} churches, temples, mosques, or synagogues) and may be crucial in encouraging individuals to attend religious activities and be a part of religious organizations. Religion also helps one deal with uncertainty and fear mentally by introducing the salvation motive and after-life rewards. Therefore, religious people are used to being protected by divine power and religious organizations. They have preset answers for the unknown, meaning they have less experience dealing with uncertainty than irreligious people. In order to test this hypothesis, we examine the relationship between religiosity and the fear of uncertainty as follows.

We nominate five indicators from the ESS that are related to the fear of uncertainty; then, we examine their relation with religiosity measures. Each indicator corresponds to one or more differences between weak and strong uncertainty avoidance societies in Hofstede's model. For instance, societies with strong uncertainty avoidance exhibit the following behaviours: \textit{``The uncertainty inherent in life is felt like a continuous threat that must be fought."}, \textit{``Higher stress, emotionality, anxiety, neuroticism."}, \textit{``Staying in jobs even if disliked."} (\cite{Hofstede2011}, pp.10). More broadly, Hofstede defines three components of uncertainty avoidance: rule orientation, employment stability, and stress. The variables \textit{\textit{unemployed}} (ever been unemployed and seeking job more than 3 months), \textit{\textit{securejob}} (important when choosing job: secure job) refer to employment stability; \textit{\textit{trust}} (most people can be trusted ), \textit{\textit{safe}} (important to live in secure surroundings ), \textit{\textit{government}} (important: government ensures safety) refer to stress. Apart from these measures, \textit{following rules}, one of the NIA, also refers to rule orientation.

\begin{table}[H]\centering
\def\sym#1{\ifmmode^{#1}\else\(^{#1}\)\fi}
\caption{IV Estimates: The Fear of Uncertainty} \label{s1indcollec}
\resizebox{0.62\columnwidth}{0.28\linewidth}{%
\begin{threeparttable}
\begin{tabular}{l*{5}{c}}
\toprule
&\multicolumn{1}{c}{(1)}&\multicolumn{1}{c}{(2)}&\multicolumn{1}{c}{(3)}&\multicolumn{1}{c}{(4)}&\multicolumn{1}{c}{(5)}\\
&\multicolumn{1}{c}{\textit{unemployed}}&\multicolumn{1}{c}{\textit{securejob}}&\multicolumn{1}{c}{\textit{trust}}&\multicolumn{1}{c}{\textit{safe}}&\multicolumn{1}{c}{\textit{government}}\\
\midrule
\textit{religiosity}&-0.128\sym{***}&0.028\sym{**}&0.027\sym{***}&0.109\sym{***}&0.098\sym{***}\\
&(0.024)&(0.012)&(0.009)&(0.011)&(0.010)\\
&206,589  & 43,307 & 206,639 & 201,571 & 200,322 \\
& 0.000 &  0.021& 0.002&0.000& 0.000\\
\addlinespace
\textit{degree}&-0.100\sym{***}&0.022\sym{**}&0.027\sym{***}&0.102\sym{***}&0.090\sym{***}\\
&(0.023)&(0.011)&(0.008)&(0.010)&(0.009)\\
& 209,106 & 43,773  & 209,164  & 204,032 & 202,752 \\
& 0.000& 0.045&0.001& 0.000& 0.000\\
\addlinespace
\textit{\textit{attendance}}&-0.189\sym{***}&0.030\sym{**}&0.036\sym{***}&0.135\sym{***}&0.125\sym{***}\\
&(0.030)&(0.015)&(0.011)&(0.013)&(0.012)\\
& 209,522 &43,901   & 209,566  & 204,416 & 203,132 \\
&0.000&0.041&0.004& 0.000&0.000 \\
\addlinespace
\textit{pray}&-0.100\sym{***}&0.022\sym{**}&0.021\sym{***}&0.088\sym{***}&0.080\sym{***}\\
&(0.020)&(0.010)&(0.007)&(0.009)&(0.008)\\
&207,701  &43,547   & 207,743  & 202,632 &201,377  \\
& 0.000&0.030& 0.000&0.000&0.000 \\
\bottomrule
\end{tabular}
\begin{tablenotes}
\item Notes: Robust standard errors clustered at the level of instruments are in parentheses, followed by sample size and Anderson-Rubin Wald F statistic (\textit{AR test}). All specifications include the full set of control variables and fixed effects.  * $p<0.10$; **$p<0.05$; ***$p<0.01$.
\end{tablenotes}
\end{threeparttable}
}
\end{table}

The results are presented in Table 12. \textit{Unemployed} is negatively, \textit{securejob} is positively related to the religiosity measures. It can be said that religiosity and employment stability are associated. The variables referring to stress are positively and significantly correlated with religiosity measures, suggesting that religious people do not necessarily feel stressed. The IV results for \textit{following rules} (Table D6) are positively and significantly correlated with all religiosity measures. Thus, rule orientation relates positively to religiosity. Overall, there are signals that religiosity and the fear of uncertainty are related, but further evidence is needed. 
 
\subsection{Roles Reinforced by Religion}
Roles reinforced by religion can be read through various concepts. Yet, we consider two of them: individualism-collectivism and conventional gender roles\footnote{Masculinity-femininity is one of the five dimensions of Hofstede's model, and it is a combination of conventional gender roles and the degree of orientation to material achievement (\cite{WilliamMcquire}). However, there is no consensus in the literature that masculinity (or femininity) favours innovation. We, therefore, simplify this dimension and take only conventional gender roles into account.}.

In Hofstede's work, individualism-collectivism is one of the five dimensions on which national cultures differ and has been widely used in the empirical literature related to culture (\cite{Sondergaard}). In individualist societies, the interests of individuals come before the interests of the group. On the other hand, in collectivist societies, the group's interests come first. Individuals work for the group, not for themselves. Therefore, complying with the group's rules and staying loyal to the group are crucial behaviours in collectivist societies that require one to follow the established rules and traditions, leaving limited space for reformist and creative endeavours. Nevertheless, creativity, which is one of the main ingredients of innovation, is said to be the act of an individual, at times in contradiction with the norms and values of the group (\cite{Amabile1996}; \cite{WilliamMcquire}).

Individualism may facilitate innovation by fostering a tendency to accept novelty (\cite{Steenkampetal}), giving courage to individuals to defend new ideas in the face of resistance, enabling the emergence of champion roles (\cite{Shane1995}). Collectivism, on the other hand, may damage innovation by fostering the ideas that are acceptable to all interested parties only (\cite{WilliamMcquire}). 

\textcite{WilliamMcquire} show that individualism positively influences economic creativity, which facilitates innovation implementation. \textcite{Chenetal2017} show that higher levels of individualism foster corporate innovation by generating more and higher impact patents and by being more efficient in converting R\&D into innovative output. \textcite{TaylorWilson} find that most measures of individualism have a strong and positive effect on national innovation rates. They also find that certain types of collectivism, such as patriotism and nationalism, may also foster innovation at the country level, while other types of collectivism, such as familism and localism, harm national innovation rates. 

\textcite{GorodnichenkoRoland} examine the effect of individualism and collectivism on the long-run economic growth. They find that individualism has a strong and positive effect on economic growth, mainly because higher degrees of individualism lead to more innovation, given that innovation is associated with higher personal and social rewards in an individualistic culture. Individualist countries, in general, are more prosperous than collectivist societies (\cite{HofstedeMinkov2010}). 

Religion is likely to favour collectivist culture since following rules and staying loyal to the group, church, and god are inherent to religion. One cannot modify religion to best self-interest, as an individualistic approach would require. Some denominations are said to be more individualistic than others, such as Protestantism, yet we do not compare denominations and only discriminate between being religious and irreligious. 

We examine the relationship between individualism-collectivism and religiosity based on the approaches above. We define four indicators from the ESS data that are somewhat related to individualism and collectivism. \textit{Devote} (important to devote himself to people close to him), \textit{\textit{family}} (family should be priority in life), \textit{\textit{success}} (important: being successful and recognised achievements), \textit{\textit{lookafter}} (everyone should look after himself). \textit{Lookafter} is related to personal freedom and self-care, \textit{success} is related to personal achievement, which both are stronger in individualistic cultures than they are in collectivist cultures (\cite{GorodnichenkoRoland}). \textit{Devote} and \textit{family} refer to \textit{we consciousness} and superiority of group over individual, thus expected to be stronger in collectivist cultures.

\begin{table}[H]\centering
\def\sym#1{\ifmmode^{#1}\else\(^{#1}\)\fi}
\caption{IV Estimates: Individualism-Collectivism} \label{s1indcollec2}
\resizebox{0.56\columnwidth}{0.25\linewidth}{%
\begin{threeparttable}
\begin{tabular}{l*{4}{c}}
\toprule
&\multicolumn{1}{c}{(1)}&\multicolumn{1}{c}{(2)}&\multicolumn{1}{c}{(3)}&\multicolumn{1}{c}{(4)}  \\
\multicolumn{1}{c}{}&\multicolumn{1}{c}{\textit{devote}}&\multicolumn{1}{c}{\textit{success}}&\multicolumn{1}{c}{\textit{lookafter}}&\multicolumn{1}{c}{\textit{family}} \\
\midrule
\textit{religiosity}&0.035\sym{***}&0.043\sym{***}&-0.035&0.120\sym{***}\\
&(0.007)&(0.012)&(0.021)&(0.016)\\
  &  201,446 &201,068   &21,621  &21,699   \\
  & 0.000&0.000&0.101&0.000  \\
\addlinespace
\textit{degree}&0.032\sym{***}&0.037\sym{***}&-0.039\sym{*}&0.112\sym{***}\\
&(0.007)&(0.011)&(0.020)&(0.015)\\
& 203,904  &  203,497 &21,798   & 21,874  \\
& 0.000&0.001&0.054& 0.000 \\
\addlinespace
\textit{attendance}&0.045\sym{***}&0.058\sym{***}&-0.040&0.147\sym{***}\\
&(0.008)&(0.014)&(0.027)&(0.019)\\
 & 204,293  &203,871   &  21,858 & 21,937  \\
 &0.000&0.001& 0.135&0.000 \\
\addlinespace
\textit{pray}&0.028\sym{***}&0.033\sym{***}&-0.030\sym{*}&0.097\sym{***}\\
&(0.006)&(0.010)&(0.017)&(0.014)\\
 & 202,510  & 202,119  & 21,715  & 21,793  \\
 & 0.000&0.001& 0.085& 0.000\\
\bottomrule
\end{tabular}
\begin{tablenotes}
\item Notes: Robust standard errors clustered at the level of instruments are in parentheses, followed by sample size and Anderson-Rubin Wald F statistic (\textit{AR test}). All specifications include the full set of control variables and fixed effects.  * $p<0.10$; **$p<0.05$; ***$p<0.01$.
\end{tablenotes}
\end{threeparttable}
}
\end{table}

Table 13 presents the results. \textit{Devote} and \textit{family} are positively and significantly related to the religiosity measures, suggesting that higher religiosity favours collectivist culture. On the other hand, \textit{success} and \textit{lookafter} show contrasting results, leaving the question open whether religiosity and individualism negatively relate.

When it comes to gender differences and traditional roles, there is a vast literature and a lot to say, yet we only discuss the matter non-exhaustively for the sake of clarity. 

Higher degrees of religiosity may cause individuals to be more submissive, obedient, and passive regardless of gender. Women are said to be more inclined toward those characteristics than men, mainly because of two reasons. Firstly, religions can foster gender differences since many major denominations impose strict rules on women, such as wifely submission, veiling, et cetera. Therefore, women are taught to be homemakers, fulfil childcare duties, and be committed to men and family, rather than being encouraged to work outside the home. Secondly, women are generally more religious than men because women have lower labour force participation rates and greater responsibility for housework and childcare, leading them to greater (time-wise) involvement in religion\footnote{See \textcite{MillerStark}, \textcite{DeVausMcAllister} for gender differences and religiosity.} (\cite{MillarHoffman1995}). 

Being taught to be submissive and obedient is expected to restrain novelty, creativity, and innovative endeavours, thus a negative channel from religion to innovation for both genders, yet expected to be more influential for women. Conventional gender roles empowered by religion are a significant drawback for innovation and the economy, allegedly a major reason for the lower female labour supply. If nothing, half of a given population is being treated differently regarding access to education, liberty, and social and legal rights means that up to $50\%$ of the capacity to create and produce cannot be properly used.

\begin{table}[H]\centering
\def\sym#1{\ifmmode^{#1}\else\(^{#1}\)\fi}
\caption{IV Estimates: Gender Roles} \label{s1indcollec3}
\resizebox{0.5\columnwidth}{0.25\linewidth}{%
\begin{threeparttable}
\begin{tabular}{l*{3}{c}}
\toprule
&\multicolumn{1}{c}{(1)}&\multicolumn{1}{c}{(2)}&\multicolumn{1}{c}{(3)}\\
&\multicolumn{1}{c}{\textit{lgbt}}&\multicolumn{1}{c}{\textit{womenwork}}&\multicolumn{1}{c}{\textit{menwork}}\\
\midrule
\textit{religiosity}&-0.196\sym{***}&0.187\sym{***}&0.136\sym{***}\\
&(0.011)&(0.018)&(0.020)\\
&200,719   &79,002  &106,126   \\
&0.000&0.000&0.000\\
\addlinespace
\textit{degree}&-0.176\sym{***}&0.175\sym{***}&0.128\sym{***}\\
&(0.011)&(0.016)&(0.019)\\
& 202,984  & 79,994  & 107,436  \\
&0.000&0.000&0.000 \\
\addlinespace
\textit{\textit{attendance}}&-0.249\sym{***}&0.236\sym{***}&0.166\sym{***}\\
&(0.014)&(0.021)&(0.025)\\
& 203,336  & 80,200  & 107,666  \\
&0.000&0.000&0.000 \\
\addlinespace
\textit{pray}&-0.158\sym{***}&0.151\sym{***}&0.113\sym{***}\\
&(0.009)&(0.015)&(0.017)\\
&  201,706 &79,491   & 106,709  \\
&0.000&0.000&0.000 \\
\bottomrule
\end{tabular}
\begin{tablenotes}
\item Notes: Robust standard errors clustered at the level of instruments are in parentheses, followed by sample size and Anderson-Rubin Wald F statistic (\textit{AR test}). All specifications include the full set of control variables and fixed effects.  * $p<0.10$; **$p<0.05$; ***$p<0.01$.
\end{tablenotes}
\end{threeparttable}
}
\end{table}

We define three variables from the ESS that could reflect the strong connection between traditional gender roles and religiosity. \textit{\textit{lgbt}} (gay men and lesbians should be free), \textit{\textit{womenwork}} (a woman should be prepared to cut down on her paid work for the sake of her family), \textit{\textit{menwork}} (when jobs are scarce, men should have more right to a job than women). Table 14 displays the results. \textit{lgbt} is negatively and significantly correlated, while \textit{womenwork} and \textit{menwork} are positively correlated with all measures of religiosity, supporting the hypothesis that religiosity favours conventional gender roles.

\section{Conclusion} \label{section: conc}
Religion is a multi-dimensional and complex phenomenon whose effect is embedded in social and economic behaviour. Hence, using religion in regression without considering its multi-dimensionality and embeddedness would yield spurious correlations. More formally, the potential sources of the endogeneity of religion, including omitted variables, reverse causality, and measurement error should be considered to reach an accountable and causal effect of religion on economic outcomes; otherwise, estimates will be unreliable.

The present study provides the first attempt to tackle the inherent endogeneity of religion with respect to innovation. We use different measures of religiosity, adopt an individual-level approach to innovation, use multi-way fixed effects, and employ the instrumental variables method. The results show that higher religiosity negatively affects positive innovation attitudes and positively affects negative innovation attitudes, implying that being religious is not a catalyser for innovation-related traits and is possibly an obstacle. This finding is in line with the literature (\cite{Benabouetal2013}, \cite*{Benabouetal2015}). Using OLS estimates, \textcite{Benabouetal2015} find that religiosity is positively related to creativity, which they describe as ``puzzling". We contribute to this literature by uncovering that the positive relationship between religiosity and creativity is driven by endogeneity. We find a negative relationship between religiosity and creativity once we rule out endogeneity by applying the IV method.

As possible channels from religion to innovation, we empirically examine and discuss three approaches: time allocation, the fear of uncertainty, and roles reinforced by religion, such as traditional roles and gender roles. 

Although the instrumental variables method is seen as effective in dealing with endogeneity and the diagnostic tests suggest non-weak instruments, the results should be approached with caution since we use observational survey data. As \textcite{Iannaccone1998} points out, nothing less than a genuine experiment that is probably unattainable will demonstrate the true causal effect of religion. In this study, we aim to approximate the causal effect of religion on innovation, yet further research is needed. First, the negative effect of religion on innovation should be tested with different instruments and sophisticated estimation methods. Second, different data sources, especially not self-reported attendance data, can be used to rule out the potential self-report bias fundamentally. Third, the causal channels from religion to innovation and their underlying mechanisms should more deeply be examined since we discuss them non-exhaustively.

\pagebreak
\section*{Appendix} 
\subsection*{A. Variables Index and Data Summary }\label{section: appa}

\textbf{Innovation Variables}
\vspace{2mm}

All the innovation variables below are based on self-reported assessments on the following statements, where the options are; \textit{very much like me} (1), \textit{like me} (2), \textit{somewhat like me }(3), \textit{a little like me} (4), \textit{not like me} (5), \textit{not like me at all }(6). All variables are recoded as to be increasing in a scale from 0 (\textit{not like me at all}) to 1 (\textit{very much like me}) for the sake of easier interpretation. 

We distinguish innovation attitudes as \textit{positive innovation attitudes} (PIA) and \textit{negative innovation attitudes} (NIA). PIA include creativity, being different, being free, and being adventurous, while NIA include following rules and following traditions.
\vspace{2mm}

\textit{\textit{adventurous}}: being adventurous, based on the statement \textit{``He looks for adventures and
	likes to take risks. He wants to have an exciting life."}
\vspace{2mm}

\textit{\textit{creative}}:  creativity, based on the statement \textit{``Thinking up new ideas and
	being creative is important to him. He likes to do things in his own original way."}
\vspace{2mm}

\textit{different}: being different, based on the statement
\textit{``He likes surprises and is always looking for new things to do. He thinks it is important to do lots of different things in life."}.

\textit{free}: being free, based on the statement  \textit{``It is important to him to make his own decisions about what he does. He likes to be free and not depend on others."}
\vspace{2mm}

\textit{innegav}: average negative innovation attitudes. Computed as the simple mean of \textit{\textit{rules}, \textit{traditions}}.
\vspace{2mm}

\textit{inposav}: average positive innovation attitudes. Computed as the simple mean of \textit{\textit{creative}, \textit{different}, \textit{free}, \textit{adventurous}}.. 
\vspace{2mm}

\textit{innosum}: summary innovation attitudes. Computed as the sum of positive innovation attitudes minus the sum of negative innovation attitudes: insum= (\textit{creative} + \textit{different} + \textit{free} + \textit{adventurous}) - (\textit{traditions} + \textit{rules}). Rescaled to be between 0 (low propensity to innovate) and 1 (high propensity to innovate). 
\vspace{2mm}

\textit{\textit{rules}}: following rules, based on the statement \textit{``He believes that people should do what they're told. He thinks people should follow rules at all times, even when no-one is watching."}

\textit{\textit{traditions}}: following traditions, based on the statement \textit{``Tradition is important to him.  He tries to follow the customs handed down by his religion or his family."}
\vspace{2mm}
\vspace{2mm}

\textbf{Religion Variables}
\vspace{2mm}
\textit{attendance}: frequency of attendance to religious activities, based on the question \textit{``Apart from special occasions such as weddings and funerals, about how often do you attend religious services nowadays?"} where the options are;  \textit{every day} (1), \textit{more than once a week} (2), \textit{once a week} (3), \textit{at least once a month} (4), \textit{only on special holy days} (5),\textit{ less often} (6), \textit{never} (7). The variable is rescaled to be between 0 (\textit{never})  and 1 (\textit{every day}) to ease the interpretation.
\vspace{2mm}

\textit{belonging}: religious affiliation, belonging to a religion or denomination, a dummy variable based on the question  \textit{``Do you consider yourself as belonging to any particular religion or denomination?"} 0 is \textit{no}, and 1 is \textit{yes}. 
\vspace{2mm}

\textit{belongingp}: past belonging to a religion or domination, a dummy variable based on the question \textit{``Have you ever considered yourself as belonging to any particular religion or denomination?"} 0 is \textit{no}, and 1 is \textit{yes}. 
\vspace{2mm}

\textit{degree}: the degree of being religious, based on the question \textit{``Regardless of whether you belong to
	a particular religion, how religious would you say you are?"} where 0 is \textit{not religious at all}, 10 is \textit{very religious}. The variable is rescaled to be between 0 (\textit{not religious at all})  and 1 (\textit{very religious}) to ease the interpretation.
\vspace{2mm}

\textit{denomination}: belonging to a particular denomination, based on the question  \textit{``If you consider yourself as belonging to any particular religion or denomination, which one is it?"} where the options are; \textit{Roman Catholic} (1), \textit{Protestant} (2), \textit{Eastern Orthodox} (3), \textit{Other Christian denominations} (4), \textit{Jewish} (5), \textit{Islamic} (6), \textit{Eastern religions }(7), \textit{Other non-Christian religions} (8), \textit{not declare which religion} (9), \textit{not belong to a religion} (10).
\vspace{2mm}

\textit{denominationp}: past belonging to a particular denomination, based on the question \textit{``If you have ever considered yourself as belonging to any particular religion or denomination, which one was it?"} where the options are the same as those of \textit{\textit{denomination}}.
\vspace{2mm}

\textit{pray}: frequency of praying, based on the question \textit{``Apart from when you are at religious
	services, how often, if at all, do you pray?"} where the options are the same as those of \textit{attendance}. The variable is rescaled to be between 0 (\textit{never})  and 1 (\textit{every day}) to ease the interpretation.
\vspace{2mm}

\textit{religiosity}: religiosity index, computed as the principal component of \textit{degree}, \textit{attendance}, and \textit{pray}. Rescaled to be between 0 \textit{not religious on average}, and 1 \textit{very religious on average}.
\vspace{2mm}
\vspace{2mm}

\textbf{Instrumental Variables}
\vspace{2mm}

\textit{iv\_attendance}: instrumental variable for \textit{attendance}. Frequency of attendance to religious activities of individual $i$, instrumented by the average frequency of attendance to religious activities of people who have the same age, gender, and religious affiliation with $i$ and live in a country with the same dominant denomination as the country of $i$. Takes values between 0 (\textit{low frequency of attendance to religious activities}) and 1 (\textit{high frequency of attendance to religious activities}). 
\vspace{2mm}

\textit{iv\_degree}: instrumental variable for \textit{degree}. The degree of religiosity of individual $i$, instrumented by the average degree of religiosity of people who have the same age, gender, and religious affiliation with $i$ and live in a country with the same dominant denomination as the country of $i$. Takes values between 0 (\textit{not religious}) and 1 (\textit{very religious}). 
\vspace{2mm}

\textit{iv\_pray}: instrumental variable for \textit{pray}. Frequency of praying of individual $i$, instrumented by the average frequency of praying of people who have the same age, gender, and religious affiliation with $i$ and live in a country that has the same dominant denomination as the country of $i$. Takes values between 0 (\textit{low frequency of pray}) and 1 (\textit{high frequency of pray}). 
\vspace{2mm}

\textit{iv\_religiosity}: instrumental variable for \textit{religiosity}. religiosity index of individual $i$, instrumented by the average of religiosity index of people who have the same age, gender and religious affiliation with $i$ and live in a country with the same dominant denomination as the country of $i$. Takes values between 0 (\textit{low religiosity}) and 1 (\textit{high religiosity}). 
\vspace{2mm}
\vspace{2mm}

\textbf{Control Variables}
\vspace{2mm}

\textit{age}: age of respondent.
\vspace{2mm}

\textit{age2}: square of age of respondent.
\vspace{2mm}

\textit{bornc}: born-in country of respondent. A standard dummy variable, 0 is \textit{no} which means that the respondent was born in a different country than the country where she takes the survey; and 1 is \textit{yes}, meaning that the respondent was born in the same country where she takes the survey. 
\vspace{2mm}

\textit{child}: whether respondent lives with children at home or not. A standard dummy variable with 0 is \textit{no} which means that the respondent does not live with children, and 1 is\textit{yes} meaning that the respondent lives with children.
\vspace{2mm}

\textit{education}: completed years of education of respondent, based on the question \textit{``About how many years of education have you completed, whether full-time or part-time? Please report these in full-time equivalents and include compulsory years of schooling."}
\vspace{2mm}

\textit{fathere}: completed level of education of respondent's father. The same classification problem, as of \textit{mothere}, stand for \textit{fathere} as well. The variable is, therefore, constructed by grouping two variables (\textit{escedf, edulvlfa}) from ESS data. The values are the same as those of \textit{mothere}.
\vspace{2mm}

\textit{gender}: a standard dummy variable for the gender of respondent, 0 is \textit{male}, and 1 is \textit{female}. 
\vspace{2mm}

\textit{health}: subjective general health of respondent where the values are:\textit{ very bad} (1), \textit{bad} (2), \textit{fair} (3), \textit{good} (4), \textit{very good} (5).
\vspace{2mm}

\textit{mothere}: completed level of education of respondent's mother. The variable is constructed by grouping two variables (\textit{escedm, edulvlma}) from ESS data. The variable \textit{edulvlma} is classified by ISCED 1997 and only exists for ESS rounds 1,2,3,4; while \textit{escedm} is classified by ISCED 2011 and only exists for ESS rounds 4,5,6,7,8. \textit{mothere} thus a grouped version\footnote{For detailed information about ISCED classifications please refer to the guide \textit{International Standard Classification of Education (ISCED)} by Eurostat. The related link: 
	\\
	\url{https://ec.europa.eu/eurostat/statistics-explained/pdfscache/44322.pdf }} of these two variables and the values are: \textit{less than lower secondary education completed} (1), \textit{lower secondary education completed} (2), \textit{upper secondary education completed} (3), \textit{post-secondary non-tertiary education completed }(4), \textit{tertiary education completed} (5). 
\vspace{2mm}

\textit{paidwork}: paid work status of respondent, based on the question \textit{``Have paid work in last seven days?"} where 0 is \textit{not marked} which means respondent does not have a paid job, and 1 is \textit{marked} meaning that respondent has a paid job. The variable is controlled with the help of the variable \textit{crpdwk} (control for paid work). 
\vspace{2mm}

\textit{partner}: whether respondent lives with a husband/wife/partner at home or not. A standard dummy variable with 0 is \textit{no} which means that the respondent does not live with a partner, and 1 is\textit{yes} meaning that the respondent lives with a partner.
\vspace{2mm}

\textit{pwbefore}: whether respondent has ever had a paid job or not. A standard dummy variable with 0 is \textit{no} which means that respondent has never had a paid job before; 1 is \textit{yes} meaning that respondent has had a paid job before. 
\vspace{2mm}
\vspace{2mm}

\textbf{Fixed Effects}
\vspace{2mm}

\textit{country}: country of the respondent. The ESS has observations for 36 European countries, but we drop Israel since it is not located in continental Europe. 
\vspace{2mm}

\textit{denomination}: please refer to Religion Variables.
\vspace{2mm}

\textit{essround}: survey dummies for eight rounds (2002, 2004, 2006, 2008, 2010, 2012, 2014, 2016) of the ESS.
\vspace{2mm}

\textit{income1 $\&$ income2}: both variables stand for net income of household from all sources. \textit{income1} only exists for the ESS rounds 1,2,3; while \textit{income2} exists for the ESS rounds 4,5,6,7,8 due to a change of classification.  But they cannot be grouped because \textit{income1} takes values with (12) ranges, while \textit{income2} with (10) deciles. Thus we use them separately.

\textit{occupation}: occupation of respondent. Constructed by grouping the two occupation variables with different classifications (\textit{iscoco} and \textit{isco08}) from the ESS data. \textit{iscoco} exists for the ESS rounds 1,2,3,4,5 and takes values of \textit{International Standard Classification of Occupations 1988} (ISCO-88); while \textit{isco08} only exists for the ESS rounds 6,7,8 and takes values of \textit{International Standard Classification of Occupations 2008 }(ISCO-08). The values of \textit{occupation} are relabeled in accordance with ISCO-08\footnote{The correspondences between the classifications of ISCO-88 and ISCO-08 are taken from the guide: \textit{International Standard Classification of Occupations: ISCO-08.} (2012) Volume I: Structure Group Definitions and Correspondence Tables by International Labour Organization.} classification. Occupation variable includes 582 different values (between 0 and 9999) that refer to different occupation labels. We consider only the main categories, which are 9, as fixed effects; armed forces occupations (0-1000), managers (1000-2000), professionals (2000-3000), technicians, and associate prof. (3000-4000), clerical support workers  (4000-5000), services and sales workers (5000-6000), skilled agricultural, forestry, and fishery workers (6000-7000), plant and machine operators and assemblers (7000-8000), elementary occupations (8000-9000), no answer (9000-9999).
\vspace{2mm}
\vspace{2mm}

\textbf{Other Variables}
\vspace{2mm}

\textit{\textit{devote}}: Based on the statement \textit{``It is important to him to be loyal to his friends. He wants to devote himself to people close to him."} Takes the same values as \textit{\textit{safe}}.
\vspace{2mm}

\textit{\textit{family}}: Based on the statement  \textit{``A person's family ought to be his or her main
priority in life."} Takes the same values as \textit{\textit{lookafter}}.
\vspace{2mm}

\textit{\textit{government}}: Based on the statement \textit{``It is important to him that the government ensures his safety against all threats. He wants the state to be strong so it can defend its citizens."} Takes the same values as \textit{\textit{safe}}.
\vspace{2mm}

\textit{\textit{lgbt}}: Based on the statement \textit{``Gay men and lesbians should be free to live their own life as they wish."} Takes the same values as \textit{\textit{lookafter}}.
\vspace{2mm}

\textit{\textit{lookafter}}: Based on the statement  \textit{``Society would be better off if everyone just looked after themselves."} where the options are \textit{ agree strongly} (1), \textit{agree} (2), \textit{neither agree or disagree} (3), \textit{disagree} (4), \textit{disagree strongly} (5). Recoded as to be increasing in a scale from 0 (\textit{disagree strongly}) to 1 (\textit{agree strongly}) for the sake of easier interpretation. 
\vspace{2mm}

\textit{ \textit{menwork}}: Based on the statement \textit{``When jobs are scarce men should have more right to a job than women."} Takes the same values as \textit{\textit{lookafter}}.
\vspace{2mm}

\textit{\textit{safe}}: Based on the statement \textit{``It is important to him to live in secure surroundings. He avoids anything that might endanger his safety."} where the options are \textit{very much like me} (1), \textit{like me} (2), \textit{somewhat like me }(3), \textit{a little like me} (4), \textit{not like me} (5), \textit{not like me at all }(6). Recoded as to be increasing in a scale from 0 (\textit{not like me at all}) to 1 (\textit{very much like me}) for the sake of easier interpretation. 
\vspace{2mm}

\textit{\textit{securejob}}: Based on the statement \textit{``How important do you think each of the following would be if you were choosing a job: a secure job"} where the options are \textit{not important at all} (1), \textit{not important} (2), \textit{neither important nor unimportant} (3), \textit{important} (4), \textit{very important} (5). Recoded as to be increasing in a scale from 0 (\textit{not important at all}) to 1 (\textit{very important}) for the sake of easier interpretation.
\vspace{2mm}

\textit{\textit{success}}: Based on the statement \textit{``Being very successful is important to him. He hopes people will recognize his achievements."} Takes the same values as \textit{\textit{safe}}.
\vspace{2mm}

\textit{\textit{trust}}: Based on the statement \textit{``Using this card, generally speaking, would you say that most people can be trusted, or that you cannot be too careful in dealing with people? Please tell me on a score of 0 to 10, where 0 means you cannot be too careful and 10 means that most people can be trusted."} Normalized to be between 0 (\textit{cannot be too careful}) and 1 (\textit{most people can be trusted})
\vspace{2mm}

\textit{\textit{unemployed}}: A dummy based on the question \textit{``Have you ever been unemployed and seeking work for a period of more than three months?"} 0 is \textit{no}, and 1 is \textit{yes}. 
\vspace{2mm}

\textit{\textit{womenwork}}: Based on the statement  \textit{``A woman should be prepared to cut down on
her paid work for the sake of her family."} Takes the same values as \textit{\textit{lookafter}}.
\vspace{2mm}

\pagebreak

\begin{table}[H]
	\tiny
	\caption*{\emph{Table A1. The ESS Data, Countries and Dominant Denominations}}	
	\scriptsize
	\centering
	{\fontsize{7}{7}\selectfont
	\begin{threeparttable}
		\begin{tabular}{l cccccccc}	
			\hline \hline 
			\scshape	Country	& Ess 1  & Ess 2 & Ess  3 & Ess 4 & Ess 5 & Ess 6 & Ess 7 & Ess 8 \\ 	
			\hline
			Albania (AL)&&&&&& $ \bullet$  &&       \\
			 Muslim (57\%) \\
			\hline
			Austria	 (AT)&$ \bullet$ &$ \bullet$ & $ \bullet$ & && & $\bullet$ & $\bullet $\\ 
			Catholic (62\%)	&  \\ 
			\hline
			Belgium (BE)	 &$ \bullet$ &$ \bullet$ & $\bullet$ &$ \bullet$ &$ \bullet$ &$ \bullet$ & $\bullet$ & $\bullet $\\
			Not belong (56\%)	&   \\ 
			\hline
			Bulgaria (BG)& & & $\bullet$ &$ \bullet$ &$ \bullet$ &$ \bullet$ &  & \\
			Orthodox (51\%)	& &&  & &&  \\
			\hline
			Croatia (HR)& &&  &$ \bullet$ &$ \bullet$ & & & \\
			Catholic (76\%)&  & &   & & \\ 
			\hline
			Cyprus (CY)& & & $\bullet$ &$ \bullet$ &$ \bullet$ &$ \bullet$ &  & \\
			Orthodox (75\%)& &&   &&  \\ 
			\hline
			Czech Republic (CZ)&$ \bullet$ &$ \bullet$ &  &$ \bullet$ &$ \bullet$ &$ \bullet$ & $\bullet$ & $\bullet $\\
			Not belong (78\%)& \\ 
			\hline
			Denmark (DK)&$ \bullet$ &$ \bullet$ & $\bullet$ &$ \bullet$ &$ \bullet$ &$ \bullet$ & $ \bullet$  &$ \bullet$  \\
			Protestant (52\%)	&  \\ 
			\hline
			Estonia (EE)& &$ \bullet$ & $\bullet$ &$ \bullet$ &$ \bullet$ &$ \bullet$ & $\bullet$ & $\bullet $\\
			Not belong (74\%)&  &    \\ 
			\hline
			Finland (FI) &$ \bullet$ &$ \bullet$ & $\bullet$ &$ \bullet$ &$ \bullet$ &$ \bullet$ & $\bullet$ & $\bullet $\\ 
			Protestant (55\%)&  &  \\ 
			\hline
			France (FR)&$ \bullet$ &$ \bullet$ & $\bullet$ &$ \bullet$ &$ \bullet$ &$ \bullet$ & $\bullet$ & $\bullet $\\
			Not belong (51\%)	&  \\ 
			\hline
			Germany (DE)&$ \bullet$ &$ \bullet$ & $\bullet$ &$ \bullet$ &$ \bullet$ &$ \bullet$ & $\bullet$ & $\bullet $\\
			Not belong (38\%)	&& \\ 
			\hline
			Great Britain (GE)&$ \bullet$ &$ \bullet$ & $\bullet$ &$ \bullet$ &$ \bullet$ &$ \bullet$ & $\bullet$ & $\bullet $\\
			Not belong (53\%)&  \\ 
			\hline
			Greece (GR)&$ \bullet$ &$ \bullet$ &  &$ \bullet$ &$ \bullet$ & & & \\
			Orthodox (89\%)& && & &&&& \\ 
			\hline
			Hungary (HU)&$ \bullet$ &$ \bullet$ & $\bullet$ &$ \bullet$ &$ \bullet$ &$ \bullet$ & $\bullet$ & $\bullet $\\
			Not belong (45\%)&  &&&  \\ 
			\hline
			Iceland (IS)& &$ \bullet$ & & & & $\bullet$ &  & $\bullet $\\
			Not belong (57\%)&   \\ 
			\hline
			Ireland (IE)&$ \bullet$ &$ \bullet$ & $\bullet$ &$ \bullet$ &$ \bullet$ &$ \bullet$ & $\bullet$ & $\bullet $\\
			Catholic (72\%)&  \\ 
			\hline
			Italy (IT)&$\bullet$ &&&&&$ \bullet$ &  & $\bullet $\\
			Catholic (71\%)& \\ 
			\hline
			Kosovo (XK)& &&&&& $ \bullet$ &&    \\
			Muslim (88\%)  \\
			\hline
			Lithuania (LT)&&& &&$ \bullet$ &$ \bullet$ & $\bullet$ & $\bullet $\\
			Catholic (79\%)&  &  &  &    \\ 
				\hline
			Latvia (LV)&&& &$ \bullet$ && & & \\
			Not belong (53\%)&  &  &  &    \\ 
			\hline
			Luxemburg (LU)&$ \bullet$ &$ \bullet$ &  & & &&  & \\
			Catholic (53\%)&  &&&&& \\
			\hline
			Netherlands (NL) &$ \bullet$ &$ \bullet$ & $\bullet$ &$ \bullet$ &$ \bullet$ &$ \bullet$ & $\bullet$ & $\bullet $\\
			Not belong (61\%)&   \\ 
			\hline
			Norway (NO)&$ \bullet$ &$ \bullet$ & $\bullet$ &$ \bullet$ &$ \bullet$ &$ \bullet$ & $\bullet$ & $\bullet $\\
			Not belong (47\%)&   \\ 
			\hline
			Poland (PL)&$ \bullet$ &$ \bullet$ & $\bullet$ &$ \bullet$ &$ \bullet$ &$ \bullet$ & $\bullet$ & $\bullet $\\
			Catholic (90\%)&  & &  \\ 
			\hline
			Portugal (PT)&$ \bullet$ &$ \bullet$ & $\bullet$ &$ \bullet$ &$ \bullet$ &$ \bullet$ & $\bullet$ & $\bullet $\\
			Catholic (80\%)& \\ 
				\hline
			Romania (RO)& && &$ \bullet$ & && & \\
			Orthodox (81\%)&  &  &&  \\ 
			\hline
			Russia (RU)& && $\bullet$ &$ \bullet$ &$ \bullet$ &$ \bullet$ & & $\bullet $\\
			Orthodox (46\%)&  &  &&  \\ 
			\hline
			Slovakia(SK)&&$ \bullet$ & $\bullet$ &$ \bullet$ &$ \bullet$ &$ \bullet$ &  & \\
			Catholic (63\%)&  &&  \\ 
			\hline
			Slovenia (SI)&$ \bullet$ &$ \bullet$ & $\bullet$ &$ \bullet$ &$ \bullet$ &$ \bullet$ & $\bullet$ & $\bullet $\\
			Catholic 50\%& \\ 
			\hline
			Spain (ES)&$ \bullet$ &$ \bullet$ & $\bullet$ &$ \bullet$ &$ \bullet$ &$ \bullet$ & $\bullet$ & $\bullet $\\
			Catholic (65\%) &   \\ 
			\hline
			Sweden (SE)&$ \bullet$ &$ \bullet$ & $\bullet$ &$ \bullet$ &$ \bullet$ &$ \bullet$ & $\bullet$ & $\bullet $\\
			Not belong (68\%)& \\ 
			\hline
			Switzerland (CH)	&$ \bullet$ &$ \bullet$ & $\bullet$ &$ \bullet$ &$ \bullet$ &$ \bullet$ & $\bullet$ & $\bullet $\\
			Not belong (34\%)		&   \\  
			\hline
			Turkey(TR)&&$ \bullet$ &  &$ \bullet$ & & & & \\
			Muslim	(96\%)&  & &   & &  \\ 
			\hline
			Ukraine(UA) &&$ \bullet$ & $\bullet$ &$ \bullet$ &$ \bullet$ &$ \bullet$ &  &\\
			Orthodox (58\%)& &  &  \\ 
			\hline \hline
		\end{tabular}
		\label {table: A1}
\begin{tablenotes}
\item Notes: Surveyed countries across the ESS waves are reported. First column displays countries and their alphabetical codes. Dominant denomination and its percentage among respondents are presented below each country.
\end{tablenotes}	
	\end{threeparttable}
}
\end{table}
\pagebreak[5]

\begin{table}[H]\centering
\def\sym#1{\ifmmode^{#1}\else\(^{#1}\)\fi}
\caption*{\emph{Table A2. Summary Statistics}} \label{tabsum}
\resizebox{!}{0.7\linewidth}{%
\begin{threeparttable}
\begin{tabular}{l*{1}{ccccc}}
\toprule
&obs.&mean&sd&min&max\\
\midrule
\textit{creative}& 284,142&0.68& 0.24& 0& 1\\
\textit{different}&284,142& 0.60&  0.27& 0& 1\\
\textit{free}&284,142& 0.76& 0.22& 0& 1\\
\textit{adventurous}& 284,142& 0.42& 0.29& 0& 1\\
\textit{traditions}&284,142& 0.67& 0.27& 0& 1\\
\textit{rules}&284,142&0.58& 0.27& 0& 1\\
\textit{inposav}&284,142&0.62&0.18&  0&  1\\
\textit{innegav}&284,142&0.62&0.22&  0&  1\\
\textit{innosum}&284,142& 0.54& 0.14& 0& 1\\
\textit{belonging}&284,142&0.62&0.48&  0&  1\\
\textit{belonginge}&107,060&0.25&0.43&  0&  1\\
\textit{denomination}&284,142&5.11&4.06&  1& 10\\
\textit{denominatione}&25,829&1.66& 1.12&  1& 8\\
\textit{religiosity}& 284,142& 0.37& 0.27& 0& 1\\
\textit{iv\_religiosity}& 284,142& 0.47& 0.23& 0& 1\\
\textit{degree}&284,142& 0.48&  0.30& 0& 1\\
\textit{iv\_degree}&284,142& 0.48&  0.18& 0& 1\\
\textit{\textit{attendance}}&284,142& 0.27& 0.25& 0& 1\\
\textit{\textit{iv\_attendance}}&284,142& 0.35& 0.19& 0& 1\\
\textit{pray}&284,142& 0.40& 0.41& 0& 1\\
\textit{iv\_pray}&284,142& 0.40& 0.24& 0& 1\\
\textit{age}&284,142& 47.63& 18.29&15&  114\\
 \textit{age2}&284,142 &   2603&    1818&    225    &  12996\\
\textit{education}&282,293& 12.42& 4.09& 0&56\\
\textit{gender}&284,087&0.53&0.50&  0&  1\\
\textit{health}&284,142&  3.78&  0.91& 1& 5\\
\textit{paidwork}&284,142&  0.55& 0.50& 0& 1\\
\textit{pwbefore}&284,142& 0.90& 0.29& 0& 1\\
\textit{child}&283,749& 0.38& 0.49& 0& 1\\
\textit{partner}&284,142& 0.53&  0.50& 0& 1\\
\textit{bornc}&283,964& 0.92& 0.27& 0& 1\\
\textit{fathere}&284,142& 2.55& 1.44& 1& 5\\
\textit{mothere}&284,142& 2.37& 1.37& 1& 5\\
\textit{essround}&284,142& 4.58&2.17&  1&  8\\
\textit{occupation}&253,195&4.74&2.50&  0&  9\\
\textit{income1}&68.04&6.17&2.65&  1& 12\\
\textit{income2}&150,443&5.30&2.78&  1& 10\\
\textit{lgbt}&273,645&0.69&0.30&  0&  1\\
\textit{womenwork}& 114,915&0.55&0.29&  0&  1\\
\textit{menwork}&148,614&0.35&0.31&  0&  1\\
\textit{devote}& 283,564&0.81&0.18&  0&  1\\
\textit{success}& 283,209&0.57& 0.27&  0&  1\\
\textit{lookafter}& 32,966& 0.36&0.28&  0&  1\\
\textit{family}& 33,114&0.80& 0.19&  0&  1\\
\textit{unemployed} & 283,170&0.26& 0.44&  0&  1\\
\textit{trust} &283,516&0.50&0.25&  0&  1\\
\textit{safe} & 283,522&0.74&0.24&  0&  1\\
\textit{securejob}&62,677&0.84&0.20&  0&  1\\
\textit{government} & 282,017&4.69&1.18&  1&  6\\

\midrule
\(N\)&284,142&&&&\\
\bottomrule
\end{tabular}
\begin{tablenotes}
\item Notes: Summary statistics for all variables are reported. All are weighted by gweight.
\end{tablenotes}
\end{threeparttable}
}
\end{table}

\begin{table}[H]\centering
	\tiny
\caption*{\emph{A3. Means by Country}} \label{tabmeans}
\resizebox{\columnwidth}{0.4\linewidth}{%
\begin{threeparttable}
\begin{tabular}{l*{14}{c}}
	\toprule
  &  \textit{religiosity} &   \textit{degree}   &\textit{attendance}    &  \textit{pray}  & \textit{creative} &  \textit{different} &  \textit{free} &  \textit{advent.}  & \textit{trad.} &  \textit{rules}& \textit{inposav}&\textit{innegav}&\textit{innosum}\\
\toprule
AL  & 0.49  &    0.73   &   0.24 &     0.50   &  0.70&       0.69 &      0.76 &      0.42 &      0.77      & 0.67  &     0.64 &      0.72 &      0.53\\
AT  &   0.41   &   0.49  &    0.30    &  0.43   &    0.71    &    0.60   &    0.79  &     0.44      & 0.65   &    0.52  &     0.64    &   0.59  &     0.56\\
BE& 0.32   &   0.47 &     0.17 &     0.30 &      0.67   &    0.63    &   0.77    &   0.43    &   0.66   &    0.56   &   0.62  &     0.61   &    0.55\\
BG & 0.34 &     0.43 &     0.27  &    0.32&       0.59    &   0.60     &  0.71    &   0.47   &    0.76   &     0.67&       0.59     &  0.71  &      0.49\\
CH&  0.40  &     0.51  &     0.25  &     0.45   &    0.74   &     0.64    &    0.84  &      0.42   &     0.63    &    0.50  &      0.66  &     0.57   &     0.59\\
CY&  0.60 &      0.69   &    0.44   &    0.70   &     0.78   &     0.63   &      0.80    &    0.47   &     0.81   &    0.55  &      0.67  &      0.68  &      0.55\\
CZ& 0.18 &      0.24&       0.14   &    0.15   &     0.68    &    0.58   &     0.73   &     0.41     &   0.64    &    0.63   &      0.60   &     0.64   &     0.52\\
DE&  0.34&       0.44  &     0.23   &    0.36  &   0.70    &     0.60     &   0.80  &     0.35    &   0.62  &     0.51 &      0.61  &     0.57  &     0.55\\
DK&  0.27   &    0.41 &     0.19    &  0.22   &   0.73   &    0.58    &   0.77   &    0.48  &     0.66  &    0.61    &   0.64   &    0.63     &  0.55\\
EE   &   0.24 &     0.35  &    0.19  &    0.19   &  0.60      & 0.59   &   0.75    &   0.41   &    0.59 &      0.53  &     0.59   &    0.56    &   0.54\\
ES  &    0.36   &   0.44    &  0.25 &     0.39  &  0.71  &     0.61&       0.78    &    0.40   &    0.66  &    0.58 &      0.62  &     0.62   &   0.54\\
FI   &   0.37  &    0.51   &   0.21 &     0.38    &   0.67  &     0.63  &     0.76  &     0.43    &   0.59  &      0.60  &     0.62&        0.60   &    0.55\\
FR   &  0.28   &   0.41  &    0.17  &    0.26   &    0.68    &   0.61   &    0.69   &    0.36   &    0.54  &     0.43   &    0.59  &     0.49   &   0.56\\
GB   &0.31     & 0.40  &    0.20  &    0.33  &   0.69  &     0.62   &    0.76  &     0.46   &   0.62  &     0.53   &    0.63   &   0.58   &  0.56\\
GR &    0.60 &     0.68  &    0.42 &     0.70   &   0.74   &   0.65  &    0.79    &   0.48   &   0.81  &    0.64 &      0.67  &   0.73   &    0.54\\
HR &  0.55   &   0.62  &    0.42   &   0.62 &      0.64    &   0.51  &    0.72  &  0.33 &      0.74  &    0.55 &      0.55  &   0.64   &    0.49\\
HU&0.31  &    0.41 &     0.20      &0.33 &     0.71  &     0.65  &    0.79   &    0.42  &    0.70   &    0.51     &  0.64  &    0.61    &   0.56\\
IE & 0.52    &  0.52 &     0.43   &   0.63  &    0.71 &      0.64  &     0.77    &   0.49  &     0.67 &    0.57   &  0.65   &    0.62    &  0.56\\
IS &    0.36 &     0.53 &     0.18   &   0.36 &   0.66  &    0.58   &    0.76   &    0.50   &    0.52    &   0.49   &   0.62  &    0.51    &   0.58\\
IT &  0.50   &   0.59  &    0.37    &  0.54  &    0.69  &     0.66&       0.75  &    0.39   &    0.75  &     0.65    &   0.62    &   0.70   &   0.52\\
LT   & 0.43   &   0.54  &    0.35    &  0.40 &     0.58   &    0.52  &    0.69   &    0.45   &   0.67  &   0.52  &    0.56     &  0.60   &    0.51\\
LU   &    0.33 &     0.43  &    0.25   &   0.31  &    0.70 &      0.66  &     0.74   &     0.40    &   0.64   &     0.60&       0.62   &    0.62    &   0.54\\
LV    & 0.31    &  0.38 &     0.23 &     0.31   &   0.68     &  0.68      & 0.82  &     0.52    &   0.71  &     0.53    &   0.67 &      0.62   &    0.58\\
NL      &0.33 &     0.46  &    0.19   &   0.33 &     0.71     &  0.63   &    0.79      & 0.47 &      0.63    &    0.60    &   0.65     &  0.62    &   0.56\\
NO &     0.28   &   0.38 &     0.19  &    0.25   &    0.69  &     0.54     &   0.70     &  0.45 &      0.61  &     0.64    &    0.60   &    0.63   &    0.52\\
PL &      0.62     & 0.64   &   0.53    &  0.71  &    0.63   &    0.62  &     0.77    &   0.41    &   0.77   &    0.71   &    0.61    &   0.74    &    0.50 \\
PT &      0.48    &  0.54  &    0.34  &    0.56 &      0.65    &   0.57    &    0.70      &  0.40  &     0.64     &  0.55   &    0.58    &   0.59     &  0.52\\
RO &    0.58     & 0.68  &    0.39   &   0.67  &    0.66    &   0.61  &     0.67      & 0.43     &  0.74     &   0.70     &  0.59     &  0.72   &    0.49\\
RU &   0.32  &    0.44   &   0.21   &   0.31 &    0.63    &   0.55  &     0.73     &  0.44        &0.70     &  0.58    &   0.59     &  0.64    &   0.51\\
SE   & 0.24 &     0.33   &   0.18   &   0.20 &      0.70    &   0.57    &   0.73       &0.43       &0.57  &     0.52     &  0.61    &   0.54    &   0.56\\
SI &  0.36      &0.47  &    0.29   &   0.32 &      0.71      &  0.70  &     0.81  &     0.46       &0.69   &    0.58    &   0.67    &   0.64    &   0.57\\
SK &   0.49 &     0.59 &     0.37   &   0.52 &    0.67    &   0.62   &    0.73 &       0.40    &   0.71    &   0.67&        0.60   &    0.69    &   0.51\\
TR  & 0.66  &    0.70  &    0.40 &     0.87      &0.71        &0.68        &0.76        &0.48        &0.81        &0.75        &0.65       &0.78        &0.52\\
UA &  0.43   &   0.50 &     0.32  &    0.49   &0.55       &0.53        &0.69        &0.37        &0.69        &0.61        &0.53        &0.65       &0.48\\
XK  &0.56  &    0.73 &     0.33    &  0.62     &  0.82      &0.77        &0.83        &0.52        &0.85        &0.73       &0.73        &0.79        &0.57\\
\midrule
Total &    0.37   &   0.48 &     0.27  &    0.40      &0.68        &0.60        &0.76        &0.42        &0.67        &0.58       &0.62       &0.62       &0.54\\
\bottomrule
\end{tabular}
\begin{tablenotes}
\item Notes: Country means of dependent and independent variables are reported. All are weighted by gweight. See Appendix A for definitions.
\end{tablenotes}
\end{threeparttable}
}
\end{table}
\pagebreak

\subsection*{B. First Stage Results}\label{section: appb}

\begin{table}[H]\centering
	\tiny
	\def\sym#1{\ifmmode^{#1}\else\(^{#1}\)\fi}	
\caption*{\emph{Table B1. Religiosity and Innovation Attitudes: First Stage Results of Table 4}} \label{tabfirst}
\resizebox{\columnwidth}{0.3\linewidth}{%
\begin{threeparttable}
\begin{tabular}{l*{9}{c}}
\toprule
&\multicolumn{1}{c}{(1)}&\multicolumn{1}{c}{(2)}&\multicolumn{1}{c}{(3)}&\multicolumn{1}{c}{(4)}&\multicolumn{1}{c}{(5)}&\multicolumn{1}{c}{(6)}&\multicolumn{1}{c}{(7)}&\multicolumn{1}{c}{(8)}&\multicolumn{1}{c}{(9)}\\
&\multicolumn{1}{c}{\textit{religiosity}} &\multicolumn{1}{c}{\textit{religiosity}}&\multicolumn{1}{c}{\textit{religiosity}}&\multicolumn{1}{c}{\textit{degree}}&\multicolumn{1}{c}{\textit{degree}}&\multicolumn{1}{c}{\textit{attendance}}&\multicolumn{1}{c}{\textit{attendance}}&\multicolumn{1}{c}{\textit{pray}}& \multicolumn{1}{c}{\textit{pray}}  \\
\toprule	
\textit{iv\_religiosity}&0.582*** &0.596***&0.600***&& &&&&\\
&(0.043) &(0.043)&(0.047)&& & &&&\\
\addlinespace
\textit{iv\_degree}&& &&0.644***&0.648***&& &&\\
&&&&(0.070)&(0.073)&&&&\\
\addlinespace
\textit{iv\_attendance}&&& &&&0.775***&0.784***&&\\
&&&&&&(0.065)&(0.068)&&\\
\addlinespace
\textit{iv\_pray}&&&&&&&&0.719***&0.707***\\
&&&&&&&&(0.052)&(0.056)\\
\addlinespace

\textit{gender}&0.007&0.005&0.004 &0.013 &0.012 &-0.005 &-0.007&0.022\sym{***}&0.022\sym{**}\\
&(0.004)&(0.004)&(0.004)&(0.005)&(0.005)&(0.004)&(0.004)&(0.007)&(0.008)\\
\addlinespace

\textit{age}&-0.000 &-0.002\sym{***} &-0.002\sym{***} &-0.002\sym{***} &-0.002\sym{***} &-0.001\sym{**}&-0.002\sym{***} &-0.002\sym{***} &-0.003\sym{***} \\
&(0.000)&(0.000)&(0.000)&(0.000)&(0.001)&(0.000)&(0.000)&(0.001)&(0.001)\\
\addlinespace

\textit{age2}&0.000**&0.000\sym{***}&0.000\sym{***}&0.000\sym{***} &0.000\sym{***} &0.000\sym{**}  &0.000\sym{**}&0.000\sym{***}&0.000\sym{***}\\
&(0.000)&(0.000)&(0.000)&(0.000)&(0.000)&(0.000)&(0.000)&(0.000)&(0.000)\\
\addlinespace

\textit{education}&-0.000&0.000&-0.000&-0.001\sym{***} &-0.002\sym{***} &0.001&0.001 &0.001&0.001\\\
&(0.000)&(0.000)&(0.000)&(0.000)&(0.000)&(0.000)&(0.000)&(0.001)&(0.001)\\
\addlinespace

\textit{paidwork}&&-0.003&-0.002&-0.002 &0.000 &-0.001  &-0.001&-0.007&-0.005\\
&&(0.002)&(0.002)&(0.002)&(0.002)&(0.002)&(0.002)&(0.004)&(0.004)\\
\addlinespace

\textit{pwbefore}&&-0.276**&-0.384\sym{***}  &-0.280  &-0.450\sym{***} &-0.315  &-0.424\sym{***}&-0.218\sym{***}&-0.241\sym{***} \\
&&(0.133)&(0.085)&(0.204)&(0.132)&(0.131)&(0.084)&(0.062)&(0.064)\\
\addlinespace

\textit{partner}&&0.018\sym{***}&0.017\sym{***}&0.016\sym{***} &0.015\sym{***} &0.024\sym{***}&0.023\sym{***}  &0.011&0.010\\
&&(0.002)&(0.002)&(0.002)&(0.002)&(0.002)&(0.002)&(0.004)&(0.004)\\
\addlinespace
 
\textit{health}&&0.003*** &0.002**&0.003\sym{*} &0.002  &0.009\sym{***}  &0.009\sym{***}  &-0.006\sym{***}&-0.008\sym{***}\\
&&(0.001)&(0.001)&(0.001)&(0.001)&(0.001)&(0.001)&(0.002)&(0.002)\\
\addlinespace

\textit{child}&&0.015\sym{***}&0.015\sym{***}&0.013\sym{***}&0.014\sym{***} &0.011\sym{***}&0.010\sym{***}&0.023\sym{***}&0.023\sym{***}\\
&&(0.002)&(0.002)&(0.002)&(0.002)&(0.002)&(0.002)&(0.003)&(0.003)\\
\addlinespace

\textit{bornc}&&&-0.069\sym{***}&&-0.061\sym{***}&&-0.049\sym{***}&&-0.108\sym{***}\\
&&&(0.004)&&(0.004)&&(0.005)&&(0.007)\\
\addlinespace

\textit{fathere}&&&0.000&&-0.001&&0.001&&0.002\\
&&&(0.001)&&(0.001)&&(0.001)&&(0.001)\\
\addlinespace

\textit{mothere}&&&-0.001&&0.001&&0.000&&-0.005\sym{**}\\
&&&(0.001)&&(0.001)&&(0.001)&&(0.001)\\
\addlinespace
\midrule
\textit{N}&234,528 &228,327  & 202,300 &231,306 &204,788 & 231,733& 205,176 & 229,606 & 203,384\\
\midrule
\end{tabular}
\begin{tablenotes}
\item Notes: The first stage results are reported. Robust standard errors clustered at the level of instruments are in parentheses. All regressions include the following fixed effects: country, survey year, country-survey year, religious denomination, occupation, and income level. * $p<0.10$; **$p<0.05$; ***$p<0.01$.
\end{tablenotes}
\end{threeparttable}
}
\end{table}

\begin{table}[H]\centering
	\tiny
	\def\sym#1{\ifmmode^{#1}\else\(^{#1}\)\fi}	
	\caption*{\emph{Table B2. Religiosity and Innovation Attitudes: First Stage Results of Table 5}} \label{tabfsecond}
\resizebox{\columnwidth}{0.3\linewidth}{%
\begin{threeparttable}
	\begin{tabular}{l*{9}{c}}
		\toprule
&\multicolumn{1}{c}{(1)}&\multicolumn{1}{c}{(2)}&\multicolumn{1}{c}{(3)}&\multicolumn{1}{c}{(4)}&\multicolumn{1}{c}{(5)}&\multicolumn{1}{c}{(6)}&\multicolumn{1}{c}{(7)}&\multicolumn{1}{c}{(8)}&\multicolumn{1}{c}{(9)}\\
&\multicolumn{1}{c}{\textit{religiosity}} &\multicolumn{1}{c}{\textit{religiosity}}&\multicolumn{1}{c}{\textit{religiosity}}&\multicolumn{1}{c}{\textit{degree}}&\multicolumn{1}{c}{\textit{degree}}&\multicolumn{1}{c}{\textit{attendance}}&\multicolumn{1}{c}{\textit{attendance}}&\multicolumn{1}{c}{\textit{pray}}& \multicolumn{1}{c}{\textit{pray}}  \\

		\toprule	
		\textit{iv\_religiosity}&0.581*** &0.595***&0.600***&& &&&&\\
		&(0.043) &(0.043)&(0.047)&& & &&&\\
		\addlinespace
		\textit{iv\_degree}&& &&0.644***&0.648***&& &&\\
		&&&&(0.070)&(0.073)&&&&\\
		\addlinespace
		\textit{iv\_attendance}&&& &&&0.775***&0.784***&&\\
		&&&&&&(0.065)&(0.068)&&\\
		\addlinespace
		\textit{iv\_pray}&&&&&&&&0.718***&0.707***\\
		&&&&&&&&(0.052)&(0.056)\\

\addlinespace		
\textit{gender}&0.007&0.005&0.004 &0.013 &0.012 &-0.005 &-0.007&0.022\sym{**}&0.022\sym{**}\\
&(0.004)&(0.004)&(0.004)&(0.005)&(0.005)&(0.004)&(0.004)&(0.007)&(0.008)\\
\addlinespace

\textit{age}&-0.000 &-0.002\sym{***} &-0.002\sym{***} &-0.002\sym{***} &-0.002\sym{***} &-0.001\sym{**}&-0.002\sym{***} &-0.002\sym{***} &-0.003\sym{***} \\
&(0.000)&(0.000)&(0.000)&(0.000)&(0.000)&(0.000)&(0.000)&(0.001)&(0.001)\\
\addlinespace

\textit{age2}&0.000&0.000\sym{***}&0.000\sym{***}&0.000\sym{***} &0.000\sym{***} &0.000\sym{**}  &0.000\sym{**}&0.000\sym{***}&0.000\sym{***}\\
&(0.000)&(0.000)&(0.000)&(0.000)&(0.000)&(0.000)&(0.000)&(0.000)&(0.000)\\
\addlinespace

\textit{education}&-0.000&0.000&-0.000&-0.001\sym{***} &-0.002\sym{***} &0.001&0.001 &0.001&0.001\\\
&(0.000)&(0.000)&(0.000)&(0.000)&(0.000)&(0.000)&(0.000)&(0.001)&(0.001)\\
\addlinespace

\textit{paidwork}&&-0.003&-0.003&-0.002 &0.000 &-0.001  &-0.001&-0.007&-0.005\\
&&(0.002)&(0.002)&(0.002)&(0.002)&(0.002)&(0.002)&(0.004)&(0.004)\\
\addlinespace

\textit{pwbefore}&&-0.276**&-0.384\sym{***}  &-0.280  &-0.450\sym{***} &-0.314  &-0.424\sym{***}&-0.218\sym{***}&-0.241\sym{***} \\
&&(0.133)&(0.085)&(0.204)&(0.132)&(0.131)&(0.084)&(0.062)&(0.064)\\
\addlinespace

\textit{partner}&&0.018\sym{***}&0.016\sym{***}&0.017\sym{***} &0.014\sym{***} &0.024\sym{***}&0.023\sym{***}  &0.011&0.010\\
&&(0.002)&(0.002)&(0.002)&(0.002)&(0.002)&(0.002)&(0.004)&(0.004)\\
\addlinespace
 
\textit{health}&&0.002 &0.003***&0.002\sym{**} &0.002  &0.009\sym{***}  &0.009\sym{***}  &-0.006\sym{***}&-0.008\sym{***}\\
&&(0.001)&(0.001)&(0.001)&(0.001)&(0.001)&(0.001)&(0.002)&(0.002)\\
\addlinespace

\textit{child}&&0.015\sym{***}&0.016\sym{***}&0.015\sym{***}&0.014\sym{***} &0.012\sym{***}  &0.010\sym{***}&0.023\sym{***}&0.023\sym{***}\\
&&(0.002)&(0.002)&(0.002)&(0.002)&(0.002)&(0.002)&(0.003)&(0.003)\\
\addlinespace

\textit{bornc}&&&-0.069\sym{***}&&-0.069\sym{***}&&-0.049\sym{***}&&-0.108\sym{***}\\
&&&(0.004)&&(0.004)&&(0.005)&&(0.007)\\
\addlinespace

\textit{fathere}&&&0.000&&-0.001&&0.001&&0.002\\
&&&(0.001)&&(0.001)&&(0.001)&&(0.001)\\
\addlinespace

\textit{mothere}&&&-0.001&&-0.001&&0.000&&-0.005\sym{**}\\
&&&(0.001)&&(0.001)&&(0.001)&&(0.001)\\
\addlinespace		
\midrule		
\addlinespace	
\textit{N}&234,387 &228,192  & 202,191&231,156 &204,668  &231,582  &205,055  &229,462  &203,268  \\
\midrule
\end{tabular}
\begin{tablenotes}
\item Notes: The first stage results are reported. Robust standard errors clustered at the level of instruments are in parentheses. All regressions include the following fixed effects: country, survey year, country-survey year, religious denomination, occupation, and income level. * $p<0.10$; **$p<0.05$; ***$p<0.01$.
\end{tablenotes}
\end{threeparttable}
}
\end{table}	

\pagebreak[5]	


\subsection*{C. Detailed OLS Results}\label{section: appc}

\begin{table}[H]\centering
\def\sym#1{\ifmmode^{#1}\else\(^{#1}\)\fi}
\caption*{\emph{Table C1. OLS Estimates: Religiosity and Creativity}} \label{creative}
\resizebox{\columnwidth}{0.305\linewidth}{%
\begin{threeparttable}
\begin{tabular}{l*{9}{c}}
\toprule
\addlinespace
\textit{creative}&\multicolumn{1}{c}{(1)}&\multicolumn{1}{c}{(2)}&\multicolumn{1}{c}{(3)}&\multicolumn{1}{c}{(4)}&\multicolumn{1}{c}{(5)}&\multicolumn{1}{c}{(6)}&\multicolumn{1}{c}{(7)}&\multicolumn{1}{c}{(8)}&\multicolumn{1}{c}{(9)}\\
\addlinespace
\midrule
\textit{religiosity}&0.035\sym{***}&0.035\sym{***}&0.033\sym{***}&&&&&&\\
&(0.005)&(0.005)&(0.005)&&&&&&\\
\textit{degree}&&&&0.032\sym{***}&0.031\sym{***}&&&&\\
&&&&(0.004)&(0.004)&&&&\\
\textit{attendance}&&&&&&0.005&0.002&&\\
&&&&&&(0.005)&(0.005)&&\\
\textit{pray}&&&&&&&&0.025\sym{***}&0.024\sym{***}\\
&&&&&&&&(0.004)&(0.004)\\
\textit{gender}&-0.015\sym{***}&-0.014\sym{***}&-0.015\sym{***}&-0.013\sym{***}&-0.015\sym{***}&-0.011\sym{***}&-0.013\sym{***}&-0.014\sym{***}&-0.016\sym{***}\\
&(0.003)&(0.003)&(0.003)&(0.003)&(0.003)&(0.003)&(0.003)&(0.003)&(0.003)\\
\textit{age}&-0.001&-0.000&0.001&-0.000&0.001&-0.000&0.001&-0.000&0.001\\
&(0.001)&(0.001)&(0.001)&(0.001)&(0.001)&(0.001)&(0.001)&(0.001)&(0.001)\\
\textit{age2}&-0.000&-0.000\sym{*}&-0.000\sym{***}&-0.000&-0.000\sym{***}&-0.000&-0.000\sym{**}&-0.000\sym{*}&-0.000\sym{***}\\
&(0.000)&(0.000)&(0.000)&(0.000)&(0.000)&(0.000)&(0.000)&(0.000)&(0.000)\\
\textit{education}&0.005\sym{***}&0.005\sym{***}&0.004\sym{***}&0.005\sym{***}&0.004\sym{***}&0.005\sym{***}&0.004\sym{***}&0.005\sym{***}&0.004\sym{***}\\
&(0.000)&(0.000)&(0.000)&(0.000)&(0.000)&(0.000)&(0.000)&(0.000)&(0.000)\\
\textit{paidwork}&&-0.000&-0.000&-0.000&0.000&-0.000&-0.000&-0.000&-0.000\\
&&(0.003)&(0.003)&(0.003)&(0.003)&(0.003)&(0.003)&(0.003)&(0.003)\\
\textit{pwbefore}&&-0.252\sym{**}&-0.193\sym{*}&-0.253\sym{**}&-0.192\sym{*}&-0.261\sym{***}&-0.205\sym{*}&-0.257\sym{**}&-0.200\sym{*}\\
&&(0.103)&(0.106)&(0.105)&(0.108)&(0.101)&(0.105)&(0.101)&(0.104)\\
\textit{partner}&&0.004\sym{*}&0.005\sym{*}&0.005\sym{*}&0.005\sym{**}&0.005\sym{*}&0.005\sym{**}&0.005\sym{*}&0.005\sym{**}\\
&&(0.003)&(0.003)&(0.003)&(0.003)&(0.002)&(0.003)&(0.003)&(0.003)\\
\textit{health}&&0.014\sym{***}&0.014\sym{***}&0.014\sym{***}&0.014\sym{***}&0.014\sym{***}&0.014\sym{***}&0.014\sym{***}&0.014\sym{***}\\
&&(0.002)&(0.002)&(0.002)&(0.002)&(0.002)&(0.001)&(0.002)&(0.001)\\
\textit{child}&&-0.010\sym{***}&-0.013\sym{***}&-0.010\sym{***}&-0.013\sym{***}&-0.010\sym{***}&-0.012\sym{***}&-0.011\sym{***}&-0.013\sym{***}\\
&&(0.002)&(0.002)&(0.002)&(0.002)&(0.002)&(0.002)&(0.002)&(0.002)\\
\textit{bornc}&&&-0.011\sym{**}&&-0.011\sym{***}&&-0.013\sym{***}&&-0.011\sym{**}\\
&&&(0.004)&&(0.004)&&(0.004)&&(0.004)\\
\textit{fathere}&&&0.007\sym{***}&&0.008\sym{***}&&0.008\sym{***}&&0.007\sym{***}\\
&&&(0.001)&&(0.001)&&(0.001)&&(0.001)\\
\textit{mothere}&&&0.004\sym{***}&&0.004\sym{***}&&0.004\sym{***}&&0.004\sym{***}\\
&&&(0.001)&&(0.001)&&(0.001)&&(0.001)\\
\textit{cons}&0.654\sym{***}&0.843\sym{***}&0.757\sym{***}&0.839\sym{***}&0.750\sym{***}&0.859\sym{***}&0.779\sym{***}&0.848\sym{***}&0.764\sym{***}\\
&(0.013)&(0.104)&(0.107)&(0.106)&(0.109)&(0.101)&(0.106)&(0.101)&(0.106)\\
\midrule
\(N\)&233,137&226,965&201,219&229,884&203,663&230,311&204,054&228,212&202,280\\
\textit{Adj. \(R^{2}\)}&0.076&0.079&0.081&0.080&0.082&0.080&0.082&0.079&0.081\\
\bottomrule
\end{tabular}
\begin{tablenotes}
\item Notes: Robust standard errors clustered at the level of instruments are in parentheses. All regressions include the full set of fixed effects. * $p<0.10$; **$p<0.05$; ***$p<0.01$.
\end{tablenotes}
\end{threeparttable}
}
\end{table}


\begin{table}[H]\centering
\def\sym#1{\ifmmode^{#1}\else\(^{#1}\)\fi}
\caption*{\emph{Table C2. OLS Estimates: Religiosity and Being Different}} \label{different}
\resizebox{\columnwidth}{0.305\linewidth}{%
\begin{threeparttable}
\begin{tabular}{l*{9}{c}}
\toprule
\addlinespace
\textit{different}&\multicolumn{1}{c}{(1)}&\multicolumn{1}{c}{(2)}&\multicolumn{1}{c}{(3)}&\multicolumn{1}{c}{(4)}&\multicolumn{1}{c}{(5)}&\multicolumn{1}{c}{(6)}&\multicolumn{1}{c}{(7)}&\multicolumn{1}{c}{(8)}&\multicolumn{1}{c}{(9)}\\
\addlinespace
\midrule
\textit{religiosity}&0.034\sym{***}&0.036\sym{***}&0.032\sym{***}&&&&&&\\
&(0.006)&(0.006)&(0.006)&&&&&&\\
\textit{degree}&&&&0.032\sym{***}&0.029\sym{***}&&&&\\
&&&&(0.004)&(0.005)&&&&\\
\textit{attendance}&&&&&&0.015\sym{***}&0.013\sym{**}&&\\
&&&&&&(0.005)&(0.006)&&\\
\textit{pray}&&&&&&&&0.021\sym{***}&0.018\sym{***}\\
&&&&&&&&(0.004)&(0.004)\\
\textit{gender}&0.002&0.006\sym{*}&0.007\sym{**}&0.006\sym{*}&0.007\sym{**}&0.008\sym{**}&0.008\sym{***}&0.006\sym{*}&0.007\sym{**}\\
&(0.003)&(0.003)&(0.003)&(0.003)&(0.003)&(0.003)&(0.003)&(0.003)&(0.003)\\
\textit{age}&-0.006\sym{***}&-0.004\sym{***}&-0.004\sym{***}&-0.004\sym{***}&-0.004\sym{***}&-0.004\sym{***}&-0.004\sym{***}&-0.004\sym{***}&-0.004\sym{***}\\
&(0.001)&(0.001)&(0.001)&(0.001)&(0.001)&(0.001)&(0.001)&(0.001)&(0.001)\\
\textit{age2}&0.000\sym{***}&0.000\sym{***}&0.000\sym{**}&0.000\sym{***}&0.000\sym{**}&0.000\sym{***}&0.000\sym{**}&0.000\sym{***}&0.000\sym{**}\\
&(0.000)&(0.000)&(0.000)&(0.000)&(0.000)&(0.000)&(0.000)&(0.000)&(0.000)\\
\textit{education}&0.003\sym{***}&0.002\sym{***}&0.002\sym{***}&0.002\sym{***}&0.002\sym{***}&0.002\sym{***}&0.002\sym{***}&0.002\sym{***}&0.002\sym{***}\\
&(0.000)&(0.000)&(0.000)&(0.000)&(0.000)&(0.000)&(0.000)&(0.000)&(0.000)\\
\textit{paidwork}&&-0.002&-0.000&-0.002&-0.000&-0.002&-0.000&-0.002&0.000\\
&&(0.003)&(0.003)&(0.003)&(0.003)&(0.003)&(0.003)&(0.003)&(0.003)\\
\textit{pwbefore}&&-0.124&-0.181\sym{*}&-0.126&-0.181\sym{*}&-0.132&-0.189\sym{*}&-0.130&-0.189\sym{*}\\
&&(0.094)&(0.106)&(0.094)&(0.108)&(0.095)&(0.106)&(0.095)&(0.105)\\
\textit{partner}&&-0.004&-0.004&-0.005&-0.004&-0.005&-0.004&-0.004&-0.003\\
&&(0.003)&(0.003)&(0.003)&(0.003)&(0.003)&(0.003)&(0.003)&(0.003)\\
\textit{health}&&0.022\sym{***}&0.023\sym{***}&0.023\sym{***}&0.023\sym{***}&0.023\sym{***}&0.023\sym{***}&0.023\sym{***}&0.023\sym{***}\\
&&(0.002)&(0.002)&(0.002)&(0.002)&(0.002)&(0.002)&(0.002)&(0.002)\\
\textit{child}&&-0.020\sym{***}&-0.022\sym{***}&-0.020\sym{***}&-0.022\sym{***}&-0.019\sym{***}&-0.021\sym{***}&-0.020\sym{***}&-0.022\sym{***}\\
&&(0.003)&(0.003)&(0.003)&(0.003)&(0.003)&(0.003)&(0.003)&(0.003)\\
\textit{bornc}&&&-0.016\sym{***}&&-0.016\sym{***}&&-0.018\sym{***}&&-0.017\sym{***}\\
&&&(0.004)&&(0.004)&&(0.004)&&(0.004)\\
\textit{fathere}&&&0.005\sym{***}&&0.005\sym{***}&&0.005\sym{***}&&0.005\sym{***}\\
&&&(0.001)&&(0.001)&&(0.001)&&(0.001)\\
\textit{mothere}&&&0.002\sym{*}&&0.003\sym{**}&&0.003\sym{**}&&0.002\sym{*}\\
&&&(0.001)&&(0.001)&&(0.001)&&(0.001)\\
\textit{cons}&0.763\sym{***}&0.773\sym{***}&0.822\sym{***}&0.768\sym{***}&0.814\sym{***}&0.784\sym{***}&0.835\sym{***}&0.781\sym{***}&0.833\sym{***}\\
&(0.014)&(0.095)&(0.107)&(0.095)&(0.109)&(0.096)&(0.107)&(0.095)&(0.106)\\
\midrule
\(N\)&233,367&227,211&201,405&230,121&203,840&230,543&204,225&228,456&202,464\\
\textit{Adj. \(R^{2}\)}&0.074&0.079&0.081&0.080&0.082&0.080&0.081&0.080&0.081\\
\bottomrule
\end{tabular}
\begin{tablenotes}
\item Notes: Robust standard errors clustered at the level of instruments are in parentheses. All regressions include the full set of fixed effects. * $p<0.10$; **$p<0.05$; ***$p<0.01$.
\end{tablenotes}
\end{threeparttable}
}
\end{table}

\pagebreak
\begin{table}[H]\centering
\def\sym#1{\ifmmode^{#1}\else\(^{#1}\)\fi}
\caption*{\emph{Table C3. OLS Estimates: Religiosity and Being Free}} \label{free}
\resizebox{\columnwidth}{0.32\linewidth}{%
\begin{threeparttable}
\begin{tabular}{l*{9}{c}}
\toprule
\addlinespace
\textit{free}&\multicolumn{1}{c}{(1)}&\multicolumn{1}{c}{(2)}&\multicolumn{1}{c}{(3)}&\multicolumn{1}{c}{(4)}&\multicolumn{1}{c}{(5)}&\multicolumn{1}{c}{(6)}&\multicolumn{1}{c}{(7)}&\multicolumn{1}{c}{(8)}&\multicolumn{1}{c}{(9)}\\
\addlinespace
\midrule
\textit{religiosity}&-0.044\sym{***}&-0.038\sym{***}&-0.041\sym{***}&&&&&&\\
&(0.005)&(0.005)&(0.005)&&&&&&\\
\textit{degree}&&&&-0.024\sym{***}&-0.024\sym{***}&&&&\\
&&&&(0.004)&(0.004)&&&&\\
\textit{attendance}&&&&&&-0.048\sym{***}&-0.050\sym{***}&&\\
&&&&&&(0.004)&(0.005)&&\\
\textit{pray}&&&&&&&&-0.008\sym{***}&-0.010\sym{***}\\
&&&&&&&&(0.003)&(0.003)\\
\textit{gender}&-0.009\sym{***}&-0.011\sym{***}&-0.011\sym{***}&-0.012\sym{***}&-0.013\sym{***}&-0.012\sym{***}&-0.013\sym{***}&-0.012\sym{***}&-0.013\sym{***}\\
&(0.003)&(0.003)&(0.003)&(0.003)&(0.003)&(0.003)&(0.003)&(0.003)&(0.003)\\
\textit{age}&-0.003\sym{***}&0.000&0.001\sym{*}&0.001&0.001\sym{*}&0.000&0.001\sym{*}&0.001&0.001\sym{**}\\
&(0.001)&(0.001)&(0.001)&(0.001)&(0.001)&(0.001)&(0.001)&(0.001)&(0.001)\\
\textit{age2}&0.000\sym{***}&-0.000&-0.000&-0.000&-0.000&-0.000&-0.000&-0.000&-0.000\\
&(0.000)&(0.000)&(0.000)&(0.000)&(0.000)&(0.000)&(0.000)&(0.000)&(0.000)\\
\textit{education}&0.005\sym{***}&0.004\sym{***}&0.004\sym{***}&0.004\sym{***}&0.003\sym{***}&0.004\sym{***}&0.004\sym{***}&0.004\sym{***}&0.004\sym{***}\\
&(0.000)&(0.000)&(0.000)&(0.000)&(0.000)&(0.000)&(0.000)&(0.000)&(0.000)\\
\textit{paidwork}&&-0.003&0.000&-0.002&0.001&-0.002&0.000&-0.002&0.000\\
&&(0.003)&(0.003)&(0.003)&(0.003)&(0.003)&(0.003)&(0.003)&(0.003)\\
\textit{pwbefore}&&0.044&0.143&0.046&0.145&0.037&0.134&0.053&0.155\\
&&(0.144)&(0.130)&(0.145)&(0.129)&(0.143)&(0.129)&(0.146)&(0.128)\\
\textit{partner}&&-0.041\sym{***}&-0.040\sym{***}&-0.042\sym{***}&-0.040\sym{***}&-0.041\sym{***}&-0.040\sym{***}&-0.042\sym{***}&-0.040\sym{***}\\
&&(0.003)&(0.003)&(0.003)&(0.003)&(0.003)&(0.003)&(0.003)&(0.003)\\
\textit{health}&&0.010\sym{***}&0.010\sym{***}&0.011\sym{***}&0.011\sym{***}&0.011\sym{***}&0.011\sym{***}&0.011\sym{***}&0.011\sym{***}\\
&&(0.001)&(0.002)&(0.001)&(0.002)&(0.001)&(0.002)&(0.001)&(0.002)\\
\textit{child}&&-0.017\sym{***}&-0.017\sym{***}&-0.017\sym{***}&-0.017\sym{***}&-0.017\sym{***}&-0.017\sym{***}&-0.017\sym{***}&-0.018\sym{***}\\
&&(0.003)&(0.003)&(0.003)&(0.003)&(0.003)&(0.003)&(0.003)&(0.003)\\
\textit{bornc}&&&-0.005&&-0.004&&-0.005&&-0.003\\
&&&(0.004)&&(0.004)&&(0.004)&&(0.004)\\
\textit{fathere}&&&0.003\sym{***}&&0.003\sym{***}&&0.003\sym{***}&&0.003\sym{***}\\
&&&(0.001)&&(0.001)&&(0.001)&&(0.001)\\
\textit{mothere}&&&0.005\sym{***}&&0.005\sym{***}&&0.005\sym{***}&&0.005\sym{***}\\
&&&(0.001)&&(0.001)&&(0.001)&&(0.001)\\
\textit{cons}&0.791\sym{***}&0.667\sym{***}&0.543\sym{***}&0.662\sym{***}&0.535\sym{***}&0.669\sym{***}&0.546\sym{***}&0.648\sym{***}&0.517\sym{***}\\
&(0.012)&(0.145)&(0.131)&(0.146)&(0.130)&(0.144)&(0.130)&(0.147)&(0.129)\\
\midrule
\(N\)&233,529&227,361&201,521&230,303&203,980&230,708&204,350&228,610&202,580\\
\textit{Adj. \(R^{2}\)}&0.063&0.073&0.075&0.072&0.075&0.073&0.076&0.072&0.074\\
\bottomrule
\end{tabular}
\begin{tablenotes}
\item Notes: Robust standard errors clustered at the level of instruments are in parentheses. All regressions include the full set of fixed effects. * $p<0.10$; **$p<0.05$; ***$p<0.01$.
\end{tablenotes}
\end{threeparttable}
}
\end{table}

\begin{table}[H]\centering
\def\sym#1{\ifmmode^{#1}\else\(^{#1}\)\fi}
\caption*{\emph{Table C4. OLS Estimates: Religiosity and Being Adventurous}} \label{adventurous}
\resizebox{\columnwidth}{0.32\linewidth}{%
\begin{threeparttable}
\begin{tabular}{l*{9}{c}}
\toprule
\addlinespace
\textit{adventurous}&\multicolumn{1}{c}{(1)}&\multicolumn{1}{c}{(2)}&\multicolumn{1}{c}{(3)}&\multicolumn{1}{c}{(4)}&\multicolumn{1}{c}{(5)}&\multicolumn{1}{c}{(6)}&\multicolumn{1}{c}{(7)}&\multicolumn{1}{c}{(8)}&\multicolumn{1}{c}{(9)}\\
\addlinespace
\midrule
\textit{religiosity}&-0.017\sym{***}&-0.011\sym{*}&-0.010&&&&&&\\
&(0.006)&(0.006)&(0.006)&&&&&&\\
\textit{degree}&&&&-0.002&-0.003&&&&\\
&&&&(0.005)&(0.005)&&&&\\
\textit{attendance}&&&&&&-0.013\sym{**}&-0.011\sym{*}&&\\
&&&&&&(0.006)&(0.006)&&\\
\textit{pray}&&&&&&&&-0.006\sym{*}&-0.006\\
&&&&&&&&(0.003)&(0.004)\\
\textit{gender}&-0.075\sym{***}&-0.072\sym{***}&-0.071\sym{***}&-0.072\sym{***}&-0.071\sym{***}&-0.072\sym{***}&-0.072\sym{***}&-0.072\sym{***}&-0.071\sym{***}\\
&(0.003)&(0.003)&(0.003)&(0.003)&(0.003)&(0.003)&(0.003)&(0.003)&(0.003)\\
\textit{age}&-0.012\sym{***}&-0.008\sym{***}&-0.007\sym{***}&-0.008\sym{***}&-0.007\sym{***}&-0.008\sym{***}&-0.007\sym{***}&-0.008\sym{***}&-0.008\sym{***}\\
&(0.000)&(0.000)&(0.000)&(0.000)&(0.000)&(0.000)&(0.000)&(0.000)&(0.000)\\
\textit{age2}&0.000\sym{***}&0.000\sym{***}&0.000\sym{***}&0.000\sym{***}&0.000\sym{***}&0.000\sym{***}&0.000\sym{***}&0.000\sym{***}&0.000\sym{***}\\
&(0.000)&(0.000)&(0.000)&(0.000)&(0.000)&(0.000)&(0.000)&(0.000)&(0.000)\\
\textit{education}&0.002\sym{***}&0.002\sym{***}&0.001\sym{*}&0.002\sym{***}&0.001\sym{**}&0.002\sym{***}&0.001\sym{**}&0.002\sym{***}&0.001\sym{*}\\
&(0.000)&(0.000)&(0.000)&(0.000)&(0.000)&(0.000)&(0.000)&(0.000)&(0.000)\\
\textit{paidwork}&&0.002&0.004&0.003&0.004&0.002&0.004&0.002&0.004\\
&&(0.003)&(0.003)&(0.003)&(0.003)&(0.003)&(0.003)&(0.003)&(0.003)\\
\textit{pwbefore}&&-0.124&-0.045&-0.124&-0.045&-0.127&-0.048&-0.122&-0.042\\
&&(0.117)&(0.117)&(0.118)&(0.116)&(0.118)&(0.116)&(0.118)&(0.115)\\
\textit{partner}&&-0.037\sym{***}&-0.035\sym{***}&-0.037\sym{***}&-0.035\sym{***}&-0.037\sym{***}&-0.035\sym{***}&-0.037\sym{***}&-0.035\sym{***}\\
&&(0.003)&(0.003)&(0.003)&(0.003)&(0.003)&(0.003)&(0.003)&(0.003)\\
\textit{health}&&0.020\sym{***}&0.019\sym{***}&0.021\sym{***}&0.019\sym{***}&0.021\sym{***}&0.019\sym{***}&0.021\sym{***}&0.019\sym{***}\\
&&(0.001)&(0.001)&(0.001)&(0.001)&(0.001)&(0.001)&(0.001)&(0.001)\\
\textit{child}&&-0.028\sym{***}&-0.029\sym{***}&-0.029\sym{***}&-0.028\sym{***}&-0.028\sym{***}&-0.028\sym{***}&-0.028\sym{***}&-0.028\sym{***}\\
&&(0.003)&(0.003)&(0.003)&(0.003)&(0.003)&(0.003)&(0.003)&(0.003)\\
\textit{bornc}&&&-0.007\sym{*}&&-0.007\sym{*}&&-0.007\sym{*}&&-0.007\sym{*}\\
&&&(0.004)&&(0.004)&&(0.004)&&(0.004)\\
\textit{fathere}&&&0.006\sym{***}&&0.006\sym{***}&&0.006\sym{***}&&0.006\sym{***}\\
&&&(0.001)&&(0.001)&&(0.001)&&(0.001)\\
\textit{mothere}&&&0.006\sym{***}&&0.006\sym{***}&&0.006\sym{***}&&0.006\sym{***}\\
&&&(0.001)&&(0.001)&&(0.001)&&(0.001)\\
\textit{cons}&0.797\sym{***}&0.791\sym{***}&0.678\sym{***}&0.785\sym{***}&0.673\sym{***}&0.789\sym{***}&0.676\sym{***}&0.786\sym{***}&0.672\sym{***}\\
&(0.013)&(0.119)&(0.119)&(0.119)&(0.118)&(0.119)&(0.118)&(0.119)&(0.117)\\
\midrule
\(N\)&233,274&227,128&201,318&230,048&203,764&230,469&204,149&228,378&202,380\\
\textit{Adj. \(R^{2}\)}&0.163&0.173&0.175&0.172&0.174&0.172&0.174&0.172&0.174\\
\bottomrule
\end{tabular}
\begin{tablenotes}
\item Notes: Robust standard errors clustered at the level of instruments are in parentheses. All regressions include the full set of fixed effects. * $p<0.10$; **$p<0.05$; ***$p<0.01$..
\end{tablenotes}
\end{threeparttable}
}
\end{table}

\pagebreak[5]

\begin{table}[H]\centering
\def\sym#1{\ifmmode^{#1}\else\(^{#1}\)\fi}
\caption*{\emph{Table C5. OLS Estimates: Religiosity and Following Traditions}} \label{traditions}
\resizebox{\columnwidth}{0.32\linewidth}{%
\begin{threeparttable}
\begin{tabular}{l*{9}{c}}
\toprule
\addlinespace
\textit{traditions}&\multicolumn{1}{c}{(1)}&\multicolumn{1}{c}{(2)}&\multicolumn{1}{c}{(3)}&\multicolumn{1}{c}{(4)}&\multicolumn{1}{c}{(5)}&\multicolumn{1}{c}{(6)}&\multicolumn{1}{c}{(7)}&\multicolumn{1}{c}{(8)}&\multicolumn{1}{c}{(9)}\\
\addlinespace
\midrule
\textit{religiosity}&0.278\sym{***}&0.274\sym{***}&0.274\sym{***}&&&&&&\\
&(0.006)&(0.006)&(0.006)&&&&&&\\
\textit{degree}&&&&0.214\sym{***}&0.212\sym{***}&&&&\\
&&&&(0.006)&(0.006)&&&&\\
\textit{attendance}&&&&&&0.191\sym{***}&0.191\sym{***}&&\\
&&&&&&(0.006)&(0.006)&&\\
\textit{pray}&&&&&&&&0.120\sym{***}&0.120\sym{***}\\
&&&&&&&&(0.003)&(0.003)\\
\textit{gender}&0.001&0.002&0.002&0.006\sym{**}&0.006\sym{*}&0.014\sym{***}&0.013\sym{***}&0.005&0.004\\
&(0.003)&(0.003)&(0.003)&(0.003)&(0.003)&(0.003)&(0.003)&(0.003)&(0.003)\\
\textit{age}&0.002\sym{***}&0.001\sym{*}&0.001&0.001\sym{*}&0.001&0.001&0.000&0.001&0.000\\
&(0.000)&(0.000)&(0.001)&(0.000)&(0.000)&(0.000)&(0.000)&(0.001)&(0.001)\\
\textit{age2}&-0.000&0.000\sym{*}&0.000\sym{*}&0.000\sym{**}&0.000\sym{**}&0.000\sym{***}&0.000\sym{***}&0.000\sym{**}&0.000\sym{**}\\
&(0.000)&(0.000)&(0.000)&(0.000)&(0.000)&(0.000)&(0.000)&(0.000)&(0.000)\\
\textit{education}&-0.003\sym{***}&-0.002\sym{***}&-0.003\sym{***}&-0.002\sym{***}&-0.002\sym{***}&-0.003\sym{***}&-0.003\sym{***}&-0.003\sym{***}&-0.003\sym{***}\\
&(0.000)&(0.000)&(0.000)&(0.000)&(0.000)&(0.000)&(0.000)&(0.000)&(0.000)\\
\textit{paidwork}&&-0.001&0.002&-0.001&0.001&-0.002&0.001&-0.001&0.001\\
&&(0.003)&(0.003)&(0.003)&(0.003)&(0.003)&(0.003)&(0.003)&(0.003)\\
\textit{pwbefore}&&-0.184\sym{***}&-0.142\sym{***}&-0.200\sym{***}&-0.154\sym{***}&-0.201\sym{***}&-0.166\sym{***}&-0.233\sym{***}&-0.218\sym{***}\\
&&(0.038)&(0.022)&(0.042)&(0.020)&(0.030)&(0.022)&(0.028)&(0.034)\\
\textit{partner}&&0.027\sym{***}&0.024\sym{***}&0.029\sym{***}&0.026\sym{***}&0.027\sym{***}&0.024\sym{***}&0.031\sym{***}&0.027\sym{***}\\
&&(0.002)&(0.002)&(0.002)&(0.002)&(0.002)&(0.002)&(0.002)&(0.002)\\
\textit{health}&&0.004\sym{***}&0.004\sym{***}&0.004\sym{***}&0.004\sym{***}&0.003\sym{***}&0.003\sym{**}&0.006\sym{***}&0.005\sym{***}\\
&&(0.001)&(0.001)&(0.001)&(0.001)&(0.001)&(0.001)&(0.001)&(0.001)\\
\textit{child}&&0.002&0.001&0.004&0.003&0.005\sym{*}&0.004&0.004\sym{*}&0.003\\
&&(0.002)&(0.003)&(0.002)&(0.003)&(0.002)&(0.003)&(0.002)&(0.002)\\
\textit{bornc}&&&-0.019\sym{***}&&-0.026\sym{***}&&-0.029\sym{***}&&-0.025\sym{***}\\
&&&(0.004)&&(0.004)&&(0.004)&&(0.004)\\
\textit{fathere}&&&-0.000&&0.000&&-0.000&&-0.000\\
&&&(0.001)&&(0.001)&&(0.001)&&(0.001)\\
\textit{mothere}&&&-0.004\sym{***}&&-0.004\sym{***}&&-0.004\sym{***}&&-0.004\sym{***}\\
&&&(0.001)&&(0.001)&&(0.001)&&(0.001)\\
\textit{cons}&0.490\sym{***}&0.676\sym{***}&0.678\sym{***}&0.682\sym{***}&0.687\sym{***}&0.738\sym{***}&0.759\sym{***}&0.771\sym{***}&0.807\sym{***}\\
&(0.012)&(0.040)&(0.024)&(0.044)&(0.023)&(0.032)&(0.023)&(0.028)&(0.033)\\
\midrule
\(N\)&233,556&227,390&201,539&230,321&203,994&230,747&204,378&228,640&202,599\\
\textit{Adj. \(R^{2}\)}&0.229&0.230&0.232&0.225&0.226&0.212&0.214&0.213&0.214\\
\bottomrule
\end{tabular}
\begin{tablenotes}
\item Notes: Robust standard errors clustered at the level of instruments are in parentheses. All regressions include the full set of fixed effects. * $p<0.10$; **$p<0.05$; ***$p<0.01$.
\end{tablenotes}
\end{threeparttable}
}
\end{table}

\begin{table}[H]\centering
\def\sym#1{\ifmmode^{#1}\else\(^{#1}\)\fi}
\caption*{\emph{Table C6. OLS Estimates: Religiosity and Following Rules}} \label{rules}
\resizebox{\columnwidth}{0.32\linewidth}{%
\begin{threeparttable}
\begin{tabular}{l*{9}{c}}
\toprule
\addlinespace
\textit{rules}&\multicolumn{1}{c}{(1)}&\multicolumn{1}{c}{(2)}&\multicolumn{1}{c}{(3)}&\multicolumn{1}{c}{(4)}&\multicolumn{1}{c}{(5)}&\multicolumn{1}{c}{(6)}&\multicolumn{1}{c}{(7)}&\multicolumn{1}{c}{(8)}&\multicolumn{1}{c}{(9)}\\
\addlinespace
\midrule
\textit{religiosity}&0.081\sym{***}&0.079\sym{***}&0.075\sym{***}&&&&&&\\
&(0.006)&(0.006)&(0.006)&&&&&&\\
\textit{degree}&&&&0.051\sym{***}&0.048\sym{***}&&&&\\
&&&&(0.006)&(0.006)&&&&\\
\textit{attendance}&&&&&&0.068\sym{***}&0.066\sym{***}&&\\
&&&&&&(0.006)&(0.006)&&\\
\textit{pray}&&&&&&&&0.033\sym{***}&0.031\sym{***}\\
&&&&&&&&(0.004)&(0.004)\\
\textit{gender}&-0.019\sym{***}&-0.018\sym{***}&-0.017\sym{***}&-0.015\sym{***}&-0.015\sym{***}&-0.014\sym{***}&-0.014\sym{***}&-0.016\sym{***}&-0.016\sym{***}\\
&(0.003)&(0.003)&(0.003)&(0.003)&(0.003)&(0.003)&(0.003)&(0.003)&(0.003)\\
\textit{age}&0.000&-0.001\sym{**}&-0.002\sym{***}&-0.001\sym{**}&-0.002\sym{***}&-0.001\sym{**}&-0.002\sym{***}&-0.001\sym{**}&-0.002\sym{***}\\
&(0.000)&(0.001)&(0.001)&(0.001)&(0.001)&(0.001)&(0.001)&(0.001)&(0.001)\\
\textit{age2}&0.000\sym{***}&0.000\sym{***}&0.000\sym{***}&0.000\sym{***}&0.000\sym{***}&0.000\sym{***}&0.000\sym{***}&0.000\sym{***}&0.000\sym{***}\\
&(0.000)&(0.000)&(0.000)&(0.000)&(0.000)&(0.000)&(0.000)&(0.000)&(0.000)\\
\textit{education}&-0.004\sym{***}&-0.004\sym{***}&-0.004\sym{***}&-0.004\sym{***}&-0.003\sym{***}&-0.004\sym{***}&-0.004\sym{***}&-0.004\sym{***}&-0.004\sym{***}\\
&(0.000)&(0.000)&(0.000)&(0.000)&(0.000)&(0.000)&(0.000)&(0.000)&(0.000)\\
\textit{paidwork}&&0.006\sym{**}&0.007\sym{**}&0.006\sym{**}&0.007\sym{**}&0.006\sym{**}&0.007\sym{**}&0.006\sym{**}&0.007\sym{**}\\
&&(0.003)&(0.003)&(0.003)&(0.003)&(0.003)&(0.003)&(0.003)&(0.003)\\
\textit{pwbefore}&&-0.302\sym{***}&-0.273\sym{***}&-0.310\sym{***}&-0.280\sym{***}&-0.304\sym{***}&-0.274\sym{***}&-0.317\sym{***}&-0.295\sym{***}\\
&&(0.046)&(0.057)&(0.047)&(0.058)&(0.047)&(0.058)&(0.045)&(0.057)\\
\textit{partner}&&0.023\sym{***}&0.022\sym{***}&0.023\sym{***}&0.022\sym{***}&0.022\sym{***}&0.021\sym{***}&0.024\sym{***}&0.023\sym{***}\\
&&(0.002)&(0.002)&(0.002)&(0.002)&(0.002)&(0.002)&(0.002)&(0.003)\\
\textit{health}&&0.003\sym{**}&0.003\sym{**}&0.003\sym{**}&0.004\sym{**}&0.003\sym{*}&0.003\sym{**}&0.004\sym{**}&0.004\sym{***}\\
&&(0.001)&(0.001)&(0.001)&(0.001)&(0.001)&(0.001)&(0.001)&(0.001)\\
\textit{child}&&0.003&-0.000&0.004&0.000&0.004&0.001&0.004&0.000\\
&&(0.002)&(0.003)&(0.003)&(0.003)&(0.002)&(0.003)&(0.003)&(0.003)\\
\textit{bornc}&&&-0.043\sym{***}&&-0.045\sym{***}&&-0.045\sym{***}&&-0.045\sym{***}\\
&&&(0.005)&&(0.005)&&(0.005)&&(0.005)\\
\textit{fathere}&&&-0.001&&-0.001&&-0.001&&-0.002\\
&&&(0.001)&&(0.001)&&(0.001)&&(0.001)\\
\textit{mothere}&&&-0.004\sym{***}&&-0.004\sym{***}&&-0.004\sym{***}&&-0.004\sym{***}\\
&&&(0.001)&&(0.001)&&(0.001)&&(0.001)\\
\textit{cons}&0.529\sym{***}&0.833\sym{***}&0.870\sym{***}&0.844\sym{***}&0.884\sym{***}&0.844\sym{***}&0.884\sym{***}&0.862\sym{***}&0.907\sym{***}\\
&(0.012)&(0.047)&(0.058)&(0.048)&(0.060)&(0.048)&(0.059)&(0.046)&(0.059)\\
\midrule
\(N\)&232,561&226,405&200,723&229,292&203,139&229,694&203,509&227,635&201,768\\
\textit{Adj. \(R^{2}\)}&0.130&0.132&0.138&0.130&0.135&0.131&0.137&0.130&0.136\\
\bottomrule
\end{tabular}
\begin{tablenotes}
\item Notes: Robust standard errors clustered at the level of instruments are in parentheses. All regressions include the full set of fixed effects. * $p<0.10$; **$p<0.05$; ***$p<0.01$.
\end{tablenotes}
\end{threeparttable}
}
\end{table}

\pagebreak[5]
\begin{table}[H]\centering
	\def\sym#1{\ifmmode^{#1}\else\(^{#1}\)\fi}
	\caption*{\emph{Table C7. Summary Statistics for Religious Belonging}} \label{tabsum2}
	\resizebox{0.7\columnwidth}{0.25\linewidth}{%
	\begin{tabular}{l*{6}{ccccc}}
		\toprule
		 & obs & mean & std dev & min & max \\
		\midrule
		&& Believers \\
		\addlinespace
		\textit{religiosity}&  221,530&	0.51  & 0.24	& 	0	&	1\\
		\textit{degree}&	220,275&	0.61&    0.23&		0	&	1\\
		\textit{\textit{attendance}}	&220,549&	0.36&   0.25&		0	&	1\\
		\textit{pray}&	216,972&	0.57&  0 .39&		0	&	1\\
		\addlinespace
		\addlinespace
		&& Never-believers \\
		\addlinespace							
		\textit{religiosity}&	107,004&	0.14&	0.16&		0&		1\\
		\textit{degree}&	105,964	&	0.22&	0.24&		0	&	1\\
		\textit{\textit{attendance}}&	106,424	&	0.08&	0.14&	0	&	1\\
		\textit{pray}	&105,700&	0.10&	0.23&		0	&	1\\
		
		\addlinespace
		\addlinespace
		
		&&Once-believers \\
		\textit{religiosity}&	32,922	&	0.18&   0.19&		0	&	1\\
		\textit{degree}&	32,744&	0.29&   0.26&		0	&	1\\
		\textit{\textit{attendance}}&	32,872&		0.10& 0.15&		0	&	1\\
		\textit{pray}&	32,641	&	0.16&    0.29&		0	&	1\\
		
		\addlinespace
		\bottomrule
	\end{tabular}
	}
\end{table}

\begin{table}[H]\centering
	\def\sym#1{\ifmmode^{#1}\else\(^{#1}\)\fi}
	\caption*{\emph{Table C8.  Disaggregate OLS Results for Religious Belonging \label{tabage}}}
\resizebox{0.9\columnwidth}{0.25\linewidth}{%
\begin{threeparttable}
		\begin{tabular}{l*{8}{c}}
			\toprule
			\addlinespace
			&\multicolumn{2}{c}{Believers} && \multicolumn{2}{c}{ Never-believers}& & \multicolumn{2}{c}{ Once-believers} \\
			
			\addlinespace
			&\multicolumn{1}{c}{(\textit{inposav})}&\multicolumn{1}{c}{(\textit{innegav})}&&\multicolumn{1}{c}{(\textit{inposav})}&\multicolumn{1}{c}{(\textit{innegav})}&&\multicolumn{1}{c}{(\textit{inposav})}&\multicolumn{1}{c}{(\textit{innegav})} \\
			\midrule
			\textit{religiosity}& -0.009** & 0.177***  &  & 0.049***  &0.194***   &   & 0.046***  &0.176***  \\
			& (0.004) & (0.005)  &   & (0.009)  & (0.014)  &   & (0.010)  & (0.014) \\
			& 121,761  & 121,710&   & 59,187 & 59,137  &   & 20,968 &20,961 \\
			& 0.155 &0.185  &   & 0.137  &0.164  &   &0.130   & 0.152 \\
			\addlinespace
			\textit{degree}&0.000  & 0.154***  &   & 0.020***  &  0.102*** &   &  0.021*** &0.103***  \\
			&(0.004)  & (0.005)  &   &  (0.006) & (0.009)  &   & (0.007)  & (0.010) \\
			& 123,543& 123,486  &   & 59,758 & 59,703  &   &21,075&21,068  \\
			&0.156  & 0.178 &   &   0.137& 0.160  &   &0.129  &0.149  \\
			\addlinespace
			\textit{\textit{attendance}}& -0.019*** & 0.123***  &   & 0.033***  &0.178***   &   &0.024**   & 0.177*** \\
			&(0.004)  & (0.006)  &   &  (0.011) &  (0.014) &   & (0.012)  &(0.018)  \\
			& 123,722 &  123,662&   & 59,906 &59,854   &   & 21,129  & 21,122 \\
			& 0.156 &  0.170 &   &  0.138 &0.159  &   & 0.128 & 0.147\\
			\addlinespace
			
			\textit{pray}&0.000  & 0.081***  &   & 0.038***  & 0.075***  &   &0.030***   &0.055***  \\
			&(0.003)  & (0.003)  &   &  (0.007) & (0.009)  &   &  (0.006) & (0.009) \\
			& 122,329 & 122,273  &   & 59,611  & 59,559  &   &21,043   &21,036  \\
			& 0.155 &0.171   &   & 0.138  & 0.153 &   &  0.130 & 0.141 \\
			
			\bottomrule
			
		\end{tabular}
	
\begin{tablenotes}
\item Notes: OLS estimates for alternative measures of religiosity are reported. Robust standard errors clustered at the level of instruments are in parentheses. All regressions include the full set of fixed effects. * $p<0.10$; **$p<0.05$; ***$p<0.01$.
\end{tablenotes}
\end{threeparttable}
}
\end{table}

\pagebreak

\subsection*{D. Detailed IV Results}\label{section: appd}
\begin{table}[H]\centering
\def\sym#1{\ifmmode^{#1}\else\(^{#1}\)\fi}
\caption*{\emph{Table D1. IV Estimates: Religiosity and Creativity \label{tabd1}}}
\resizebox{\columnwidth}{0.305\linewidth}{%
\begin{threeparttable}
\begin{tabular}{l*{9}{c}}
\toprule
\addlinespace
\textit{creative}&\multicolumn{1}{c}{(1)}&\multicolumn{1}{c}{(2)}&\multicolumn{1}{c}{(3)}&\multicolumn{1}{c}{(4)}&\multicolumn{1}{c}{(5)}&\multicolumn{1}{c}{(6)}&\multicolumn{1}{c}{(7)}&\multicolumn{1}{c}{(8)}&\multicolumn{1}{c}{(9)}\\
\addlinespace
\midrule
\textit{religiosity}&-0.142&-0.105&-0.173\sym{*}&&&&&&\\
&(0.094)&(0.088)&(0.090)&&&&&&\\
\textit{degree}&&&&-0.070&-0.171&&&&\\
&&&&(0.132)&(0.133)&&&&\\
\textit{attendance}&&&&&&-0.126\sym{*}&-0.174\sym{**}&&\\
&&&&&&(0.074)&(0.074)&&\\
\textit{pray}&&&&&&&&-0.023&-0.071\\
&&&&&&&&(0.056)&(0.057)\\
\textit{gender}&-0.005&-0.005&-0.003&-0.007&-0.003&-0.008\sym{**}&-0.009\sym{***}&-0.009&-0.005\\
&(0.006)&(0.005)&(0.005)&(0.007)&(0.007)&(0.003)&(0.003)&(0.006)&(0.006)\\
\textit{age}&-0.001&-0.000&0.000&-0.000&0.000&-0.000&0.000&-0.000&0.001\\
&(0.001)&(0.001)&(0.001)&(0.001)&(0.001)&(0.001)&(0.001)&(0.001)&(0.001)\\
\textit{age2}&-0.000&-0.000&-0.000&-0.000&-0.000&-0.000&-0.000&-0.000&-0.000\sym{**}\\
&(0.000)&(0.000)&(0.000)&(0.000)&(0.000)&(0.000)&(0.000)&(0.000)&(0.000)\\
\textit{education}&0.005\sym{***}&0.005\sym{***}&0.004\sym{***}&0.005\sym{***}&0.004\sym{***}&0.005\sym{***}&0.004\sym{***}&0.005\sym{***}&0.004\sym{***}\\
&(0.000)&(0.000)&(0.000)&(0.000)&(0.000)&(0.000)&(0.000)&(0.000)&(0.000)\\
\textit{paidwork}&&-0.001&-0.001&-0.001&-0.000&-0.001&-0.001&-0.001&-0.001\\
&&(0.003)&(0.003)&(0.003)&(0.003)&(0.003)&(0.003)&(0.003)&(0.003)\\
\textit{pwbefore}&&-0.291\sym{***}&-0.273\sym{***}&-0.281\sym{***}&-0.283\sym{***}&-0.301\sym{***}&-0.281\sym{***}&-0.267\sym{***}&-0.224\sym{**}\\
&&(0.094)&(0.103)&(0.098)&(0.109)&(0.091)&(0.101)&(0.101)&(0.106)\\
\textit{partner}&&0.007\sym{**}&0.008\sym{***}&0.006\sym{*}&0.008\sym{**}&0.007\sym{**}&0.009\sym{***}&0.005\sym{**}&0.006\sym{**}\\
&&(0.003)&(0.003)&(0.003)&(0.003)&(0.003)&(0.003)&(0.003)&(0.003)\\
\textit{health}&&0.014\sym{***}&0.014\sym{***}&0.014\sym{***}&0.014\sym{***}&0.015\sym{***}&0.015\sym{***}&0.014\sym{***}&0.014\sym{***}\\
&&(0.002)&(0.002)&(0.002)&(0.002)&(0.002)&(0.002)&(0.002)&(0.001)\\
\textit{child}&&-0.008\sym{***}&-0.009\sym{***}&-0.009\sym{***}&-0.010\sym{***}&-0.008\sym{***}&-0.010\sym{***}&-0.010\sym{***}&-0.011\sym{***}\\
&&(0.003)&(0.003)&(0.003)&(0.003)&(0.002)&(0.003)&(0.003)&(0.003)\\
\textit{bornc}&&&-0.025\sym{***}&&-0.023\sym{**}&&-0.022\sym{***}&&-0.021\sym{***}\\
&&&(0.008)&&(0.009)&&(0.006)&&(0.007)\\
\textit{fathere}&&&0.007\sym{***}&&0.007\sym{***}&&0.008\sym{***}&&0.008\sym{***}\\
&&&(0.001)&&(0.001)&&(0.001)&&(0.001)\\
\textit{mothere}&&&0.004\sym{***}&&0.004\sym{***}&&0.004\sym{***}&&0.004\sym{***}\\
&&&(0.001)&&(0.001)&&(0.001)&&(0.001)\\
\midrule
\textit{N}&233,137&226,965&201,219&229,884&203,663&230,311&204,054&228,212&202,280\\
\textit{idp}&0.000&0.000&0.000&0.000&0.000&0.000&0.000&0.000&0.000\\
\textit{cdf}&2,325&2,363&2,141&825&749&2,707&2,499&2,017&1,731\\
\textit{widstat}&186&192&159&79&74&141&130&195&162\\
\bottomrule
\end{tabular}
\begin{tablenotes}
\item Notes: Robust standard errors clustered at the instruments level are in parentheses. All regressions include the full set of fixed effects.   *$p<0.10$; **$p<0.05$; ***$p<0.01$.
\end{tablenotes}
\end{threeparttable}
}
\end{table}

\begin{table}[H]\centering
\def\sym#1{\ifmmode^{#1}\else\(^{#1}\)\fi}
\caption*{\emph{Table D2. IV Estimates: Religiosity and Being Different \label{tabd2}}}
\resizebox{\columnwidth}{0.305\linewidth}{%
\begin{threeparttable}
\begin{tabular}{l*{9}{c}}
\toprule
\addlinespace
\textit{different}&\multicolumn{1}{c}{(1)}&\multicolumn{1}{c}{(2)}&\multicolumn{1}{c}{(3)}&\multicolumn{1}{c}{(4)}&\multicolumn{1}{c}{(5)}&\multicolumn{1}{c}{(6)}&\multicolumn{1}{c}{(7)}&\multicolumn{1}{c}{(8)}&\multicolumn{1}{c}{(9)}\\
\addlinespace
\midrule
\textit{religiosity}&-0.123&-0.097&-0.131&&&&&&\\
&(0.090)&(0.083)&(0.081)&&&&&&\\
\textit{degree}&&&&0.000&-0.069&&&&\\
&&&&(0.108)&(0.113)&&&&\\
\textit{attendance}&&&&&&-0.106&-0.124\sym{*}&&\\
&&&&&&(0.068)&(0.066)&&\\
\textit{pray}&&&&&&&&-0.070&-0.102\sym{**}\\
&&&&&&&&(0.052)&(0.051)\\
\textit{gender}&0.012\sym{*}&0.014\sym{**}&0.016\sym{***}&0.008&0.013\sym{*}&0.011\sym{***}&0.012\sym{***}&0.016\sym{**}&0.021\sym{***}\\
&(0.006)&(0.006)&(0.006)&(0.007)&(0.007)&(0.004)&(0.004)&(0.007)&(0.006)\\
\textit{age}&-0.006\sym{***}&-0.005\sym{***}&-0.004\sym{***}&-0.004\sym{***}&-0.004\sym{***}&-0.005\sym{***}&-0.004\sym{***}&-0.005\sym{***}&-0.004\sym{***}\\
&(0.001)&(0.001)&(0.001)&(0.001)&(0.001)&(0.001)&(0.001)&(0.001)&(0.001)\\
\textit{age2}&0.000\sym{***}&0.000\sym{***}&0.000\sym{***}&0.000\sym{***}&0.000\sym{***}&0.000\sym{***}&0.000\sym{***}&0.000\sym{***}&0.000\sym{***}\\
&(0.000)&(0.000)&(0.000)&(0.000)&(0.000)&(0.000)&(0.000)&(0.000)&(0.000)\\
\textit{education}&0.003\sym{***}&0.002\sym{***}&0.001\sym{***}&0.002\sym{***}&0.002\sym{***}&0.002\sym{***}&0.002\sym{***}&0.002\sym{***}&0.002\sym{***}\\
&(0.000)&(0.000)&(0.000)&(0.000)&(0.000)&(0.000)&(0.000)&(0.000)&(0.000)\\
\textit{paidwork}&&-0.003&-0.001&-0.002&-0.000&-0.003&-0.001&-0.003&-0.001\\
&&(0.003)&(0.003)&(0.003)&(0.003)&(0.003)&(0.003)&(0.003)&(0.003)\\
\textit{pwbefore}&&-0.162&-0.245\sym{**}&-0.135&-0.225\sym{**}&-0.169\sym{*}&-0.248\sym{**}&-0.150&-0.220\sym{**}\\
&&(0.102)&(0.106)&(0.100)&(0.112)&(0.103)&(0.105)&(0.099)&(0.109)\\
\textit{partner}&&-0.002&-0.001&-0.004&-0.002&-0.002&-0.001&-0.003&-0.002\\
&&(0.003)&(0.003)&(0.004)&(0.004)&(0.003)&(0.003)&(0.003)&(0.003)\\
\textit{health}&&0.023\sym{***}&0.023\sym{***}&0.023\sym{***}&0.023\sym{***}&0.024\sym{***}&0.024\sym{***}&0.022\sym{***}&0.022\sym{***}\\
&&(0.002)&(0.002)&(0.002)&(0.002)&(0.002)&(0.002)&(0.002)&(0.002)\\
\textit{child}&&-0.018\sym{***}&-0.020\sym{***}&-0.019\sym{***}&-0.020\sym{***}&-0.018\sym{***}&-0.020\sym{***}&-0.018\sym{***}&-0.019\sym{***}\\
&&(0.003)&(0.003)&(0.003)&(0.003)&(0.003)&(0.003)&(0.003)&(0.003)\\
\textit{bornc}&&&-0.027\sym{***}&&-0.022\sym{**}&&-0.024\sym{***}&&-0.029\sym{***}\\
&&&(0.008)&&(0.009)&&(0.006)&&(0.008)\\
\textit{fathere}&&&0.005\sym{***}&&0.005\sym{***}&&0.005\sym{***}&&0.005\sym{***}\\
&&&(0.001)&&(0.001)&&(0.001)&&(0.001)\\
\textit{mothere}&&&0.002\sym{*}&&0.003\sym{**}&&0.003\sym{**}&&0.002\\
&&&(0.001)&&(0.001)&&(0.001)&&(0.001)\\
\midrule
\textit{N}&233,367&227,211&201,405&230,121&203,840&230,543&204,225&228,456&202,464\\
\textit{idp}&0.000&0.000&0.000&0.000&0.000&0.000&0.000&0.000&0.000\\
\textit{cdf}&2,329&2,363&2,135&833&753&2,693&2,485&2,015&1,729\\
\textit{widstat}&185&191&157&81&74&142&130&193&161\\
\bottomrule
\end{tabular}
\begin{tablenotes}
\item Notes: Robust standard errors clustered at the instruments level are in parentheses. All regressions include the full set of fixed effects.   *$p<0.10$; **$p<0.05$; ***$p<0.01$.
\end{tablenotes}
\end{threeparttable}
}
\end{table}

\pagebreak[5]
\begin{table}[H]\centering
\def\sym#1{\ifmmode^{#1}\else\(^{#1}\)\fi}
\caption*{\emph{Table D3. IV Estimates: Religiosity and Being Free \label{tabd3}}}
\resizebox{\columnwidth}{0.32\linewidth}{%
\begin{threeparttable}
\begin{tabular}{l*{9}{c}}
\toprule
\addlinespace
\textit{free}&\multicolumn{1}{c}{(1)}&\multicolumn{1}{c}{(2)}&\multicolumn{1}{c}{(3)}&\multicolumn{1}{c}{(4)}&\multicolumn{1}{c}{(5)}&\multicolumn{1}{c}{(6)}&\multicolumn{1}{c}{(7)}&\multicolumn{1}{c}{(8)}&\multicolumn{1}{c}{(9)}\\
\addlinespace
\midrule
\textit{religiosity}&-0.304\sym{***}&-0.332\sym{***}&-0.366\sym{***}&&&&&&\\
&(0.071)&(0.071)&(0.068)&&&&&&\\
\textit{degree}&&&&-0.289\sym{**}&-0.342\sym{***}&&&&\\
&&&&(0.122)&(0.120)&&&&\\
\textit{attendance}&&&&&&-0.325\sym{***}&-0.338\sym{***}&&\\
&&&&&&(0.060)&(0.060)&&\\
\textit{pray}&&&&&&&&-0.177\sym{***}&-0.208\sym{***}\\
&&&&&&&&(0.046)&(0.045)\\
\textit{gender}&0.006&0.007&0.008\sym{*}&0.004&0.006&-0.005&-0.006\sym{*}&0.007&0.009\sym{*}\\
&(0.005)&(0.005)&(0.005)&(0.007)&(0.007)&(0.003)&(0.004)&(0.006)&(0.005)\\
\textit{age}&-0.003\sym{***}&-0.000&0.000&0.000&0.000&0.000&0.001&0.000&0.001\\
&(0.001)&(0.001)&(0.001)&(0.001)&(0.001)&(0.001)&(0.001)&(0.001)&(0.001)\\
\textit{age2}&0.000\sym{***}&0.000&0.000&0.000&0.000&0.000&-0.000&0.000&-0.000\\
&(0.000)&(0.000)&(0.000)&(0.000)&(0.000)&(0.000)&(0.000)&(0.000)&(0.000)\\
\textit{education}&0.004\sym{***}&0.004\sym{***}&0.003\sym{***}&0.004\sym{***}&0.003\sym{***}&0.004\sym{***}&0.004\sym{***}&0.004\sym{***}&0.004\sym{***}\\
&(0.000)&(0.000)&(0.000)&(0.000)&(0.000)&(0.000)&(0.000)&(0.000)&(0.000)\\
\textit{paidwork}&&-0.004&-0.001&-0.003&0.000&-0.003&-0.001&-0.004&-0.001\\
&&(0.003)&(0.003)&(0.003)&(0.003)&(0.003)&(0.003)&(0.003)&(0.003)\\
\textit{pwbefore}&&-0.038&0.015&-0.029&0.002&-0.049&0.011&0.014&0.104\\
&&(0.131)&(0.148)&(0.135)&(0.167)&(0.131)&(0.146)&(0.140)&(0.126)\\
\textit{partner}&&-0.037\sym{***}&-0.035\sym{***}&-0.037\sym{***}&-0.036\sym{***}&-0.035\sym{***}&-0.034\sym{***}&-0.040\sym{***}&-0.039\sym{***}\\
&&(0.003)&(0.003)&(0.004)&(0.004)&(0.003)&(0.003)&(0.003)&(0.003)\\
\textit{health}&&0.011\sym{***}&0.011\sym{***}&0.012\sym{***}&0.011\sym{***}&0.014\sym{***}&0.014\sym{***}&0.009\sym{***}&0.009\sym{***}\\
&&(0.001)&(0.002)&(0.002)&(0.002)&(0.002)&(0.002)&(0.002)&(0.002)\\
\textit{child}&&-0.012\sym{***}&-0.012\sym{***}&-0.013\sym{***}&-0.013\sym{***}&-0.013\sym{***}&-0.013\sym{***}&-0.013\sym{***}&-0.013\sym{***}\\
&&(0.003)&(0.003)&(0.003)&(0.003)&(0.003)&(0.003)&(0.003)&(0.003)\\
\textit{bornc}&&&-0.027\sym{***}&&-0.023\sym{***}&&-0.019\sym{***}&&-0.024\sym{***}\\
&&&(0.006)&&(0.009)&&(0.005)&&(0.006)\\
\textit{fathere}&&&0.003\sym{***}&&0.003\sym{***}&&0.003\sym{***}&&0.004\sym{***}\\
&&&(0.001)&&(0.001)&&(0.001)&&(0.001)\\
\textit{mothere}&&&0.004\sym{***}&&0.005\sym{***}&&0.005\sym{***}&&0.004\sym{***}\\
&&&(0.001)&&(0.001)&&(0.001)&&(0.001)\\
\midrule
\textit{N}&233,529&227,361&201,521&230,303&203,980&230,708&204,350&228,610&202,580\\
\textit{idp}&0.000&0.000&0.000&0.000&0.000&0.000&0.000&0.000&0.000\\
\textit{cdf}&2,333&2,374&2,148&836&761&2,729&2,522&2,018&1,731\\
\textit{widstat}&187&193&159&84&78&143&130&192&159\\
\bottomrule
\end{tabular}
\begin{tablenotes}
\item Notes: Robust standard errors clustered at the instruments level are in parentheses. All regressions include the full set of fixed effects.   *$p<0.10$; **$p<0.05$; ***$p<0.01$.
\end{tablenotes}
\end{threeparttable}
}
\end{table}

\begin{table}[H]\centering
\def\sym#1{\ifmmode^{#1}\else\(^{#1}\)\fi}
\caption*{\emph{Table D4. IV Estimates: Religiosity and Being Adventurous \label{tabd4}}}
\resizebox{\columnwidth}{0.32\linewidth}{%
\begin{threeparttable}
\begin{tabular}{l*{9}{c}}
\toprule
\addlinespace
\textit{adventurous}&\multicolumn{1}{c}{(1)}&\multicolumn{1}{c}{(2)}&\multicolumn{1}{c}{(3)}&\multicolumn{1}{c}{(4)}&\multicolumn{1}{c}{(5)}&\multicolumn{1}{c}{(6)}&\multicolumn{1}{c}{(7)}&\multicolumn{1}{c}{(8)}&\multicolumn{1}{c}{(9)}\\
\addlinespace
\midrule
\textit{religiosity}&0.004&-0.013&-0.074&&&&&&\\
&(0.071)&(0.062)&(0.061)&&&&&&\\
\textit{degree}&&&&0.051&-0.048&&&&\\
&&&&(0.085)&(0.085)&&&&\\
\textit{attendance}&&&&&&-0.021&-0.057&&\\
&&&&&&(0.057)&(0.056)&&\\
\textit{pray}&&&&&&&&-0.019&-0.066\\
&&&&&&&&(0.041)&(0.041)\\
\textit{gender}&-0.076\sym{***}&-0.072\sym{***}&-0.068\sym{***}&-0.075\sym{***}&-0.069\sym{***}&-0.072\sym{***}&-0.071\sym{***}&-0.071\sym{***}&-0.065\sym{***}\\
&(0.005)&(0.005)&(0.005)&(0.006)&(0.006)&(0.003)&(0.003)&(0.006)&(0.006)\\
\textit{age}&-0.012\sym{***}&-0.008\sym{***}&-0.008\sym{***}&-0.008\sym{***}&-0.008\sym{***}&-0.008\sym{***}&-0.008\sym{***}&-0.008\sym{***}&-0.008\sym{***}\\
&(0.000)&(0.000)&(0.000)&(0.000)&(0.000)&(0.000)&(0.000)&(0.000)&(0.000)\\
\textit{age2}&0.000\sym{***}&0.000\sym{***}&0.000\sym{***}&0.000\sym{***}&0.000\sym{***}&0.000\sym{***}&0.000\sym{***}&0.000\sym{***}&0.000\sym{***}\\
&(0.000)&(0.000)&(0.000)&(0.000)&(0.000)&(0.000)&(0.000)&(0.000)&(0.000)\\
\textit{education}&0.002\sym{***}&0.002\sym{***}&0.001&0.002\sym{***}&0.001&0.002\sym{***}&0.001\sym{**}&0.002\sym{***}&0.001\sym{*}\\
&(0.000)&(0.000)&(0.000)&(0.000)&(0.000)&(0.000)&(0.000)&(0.000)&(0.000)\\
\textit{paidwork}&&0.002&0.003&0.003&0.004&0.002&0.004&0.002&0.003\\
&&(0.003)&(0.003)&(0.003)&(0.003)&(0.003)&(0.003)&(0.003)&(0.003)\\
\textit{pwbefore}&&-0.124&-0.071&-0.109&-0.065&-0.130&-0.068&-0.125&-0.058\\
&&(0.118)&(0.122)&(0.124)&(0.127)&(0.118)&(0.120)&(0.117)&(0.115)\\
\textit{partner}&&-0.037\sym{***}&-0.034\sym{***}&-0.038\sym{***}&-0.035\sym{***}&-0.037\sym{***}&-0.034\sym{***}&-0.037\sym{***}&-0.035\sym{***}\\
&&(0.003)&(0.003)&(0.003)&(0.003)&(0.003)&(0.003)&(0.003)&(0.003)\\
\textit{health}&&0.020\sym{***}&0.019\sym{***}&0.020\sym{***}&0.019\sym{***}&0.021\sym{***}&0.020\sym{***}&0.021\sym{***}&0.019\sym{***}\\
&&(0.001)&(0.001)&(0.001)&(0.001)&(0.001)&(0.001)&(0.001)&(0.001)\\
\textit{child}&&-0.028\sym{***}&-0.028\sym{***}&-0.029\sym{***}&-0.028\sym{***}&-0.028\sym{***}&-0.028\sym{***}&-0.028\sym{***}&-0.027\sym{***}\\
&&(0.003)&(0.003)&(0.003)&(0.003)&(0.003)&(0.003)&(0.003)&(0.003)\\
\textit{bornc}&&&-0.011\sym{*}&&-0.009&&-0.009\sym{*}&&-0.013\sym{**}\\
&&&(0.006)&&(0.006)&&(0.005)&&(0.006)\\
\textit{fathere}&&&0.006\sym{***}&&0.006\sym{***}&&0.006\sym{***}&&0.006\sym{***}\\
&&&(0.001)&&(0.001)&&(0.001)&&(0.001)\\
\textit{mothere}&&&0.006\sym{***}&&0.006\sym{***}&&0.006\sym{***}&&0.006\sym{***}\\
&&&(0.001)&&(0.001)&&(0.001)&&(0.001)\\
\midrule
\textit{N}&233,274&227,128&201,318&230,048&203,764&230,469&204,149&228,378&202,380\\
\textit{idp}&0.000&0.000&0.000&0.000&0.000&0.000&0.000&0.000&0.000\\
\textit{cdf}&2,323&2,367&2,141&846&765&2,706&2,500&2,011&1,726\\
\textit{widstat}&187&194&160&82&77&143&131&192&159\\
\bottomrule
\end{tabular}
\begin{tablenotes}
\item Notes: Robust standard errors clustered at the instruments level are in parentheses. All regressions include the full set of fixed effects.   *$p<0.10$; **$p<0.05$; ***$p<0.01$.
\end{tablenotes}
\end{threeparttable}
}
\end{table}

\pagebreak[5]
\begin{table}[H]\centering
\def\sym#1{\ifmmode^{#1}\else\(^{#1}\)\fi}
\caption*{\emph{Table D5. IV Estimates: Religiosity and Following Traditions \label{tabd5}}}
\resizebox{\columnwidth}{0.32\linewidth}{%
\begin{threeparttable}
\begin{tabular}{l*{9}{c}}
\toprule
\addlinespace
\textit{traditions}&\multicolumn{1}{c}{(1)}&\multicolumn{1}{c}{(2)}&\multicolumn{1}{c}{(3)}&\multicolumn{1}{c}{(4)}&\multicolumn{1}{c}{(5)}&\multicolumn{1}{c}{(6)}&\multicolumn{1}{c}{(7)}&\multicolumn{1}{c}{(8)}&\multicolumn{1}{c}{(9)}\\
\addlinespace
\midrule
\textit{religiosity}&0.120\sym{**}&0.162\sym{***}&0.192\sym{***}&&&&&&\\
&(0.059)&(0.055)&(0.060)&&&&&&\\
\textit{degree}&&&&0.215\sym{***}&0.260\sym{***}&&&&\\
&&&&(0.074)&(0.084)&&&&\\
\textit{attendance}&&&&&&0.175\sym{***}&0.187\sym{***}&&\\
&&&&&&(0.047)&(0.050)&&\\
\textit{pray}&&&&&&&&0.089\sym{**}&0.111\sym{**}\\
&&&&&&&&(0.041)&(0.045)\\
\textit{gender}&0.010\sym{**}&0.009\sym{*}&0.007&0.006&0.003&0.014\sym{***}&0.013\sym{***}&0.008&0.005\\
&(0.005)&(0.005)&(0.005)&(0.005)&(0.006)&(0.003)&(0.003)&(0.006)&(0.006)\\
\textit{age}&0.002\sym{***}&0.001&0.001&0.001\sym{*}&0.001&0.001&0.000&0.001&0.000\\
&(0.000)&(0.001)&(0.001)&(0.000)&(0.001)&(0.000)&(0.001)&(0.001)&(0.001)\\
\textit{age2}&-0.000&0.000\sym{**}&0.000\sym{**}&0.000\sym{**}&0.000\sym{*}&0.000\sym{**}&0.000\sym{***}&0.000\sym{**}&0.000\sym{**}\\
&(0.000)&(0.000)&(0.000)&(0.000)&(0.000)&(0.000)&(0.000)&(0.000)&(0.000)\\
\textit{education}&-0.003\sym{***}&-0.003\sym{***}&-0.003\sym{***}&-0.002\sym{***}&-0.002\sym{***}&-0.003\sym{***}&-0.003\sym{***}&-0.003\sym{***}&-0.003\sym{***}\\
&(0.000)&(0.000)&(0.000)&(0.000)&(0.000)&(0.000)&(0.000)&(0.000)&(0.000)\\
\textit{paidwork}&&-0.001&0.001&-0.001&0.001&-0.002&0.001&-0.001&0.001\\
&&(0.003)&(0.003)&(0.003)&(0.003)&(0.003)&(0.003)&(0.003)&(0.003)\\
\textit{pwbefore}&&-0.215\sym{***}&-0.174\sym{***}&-0.200\sym{***}&-0.132\sym{***}&-0.206\sym{***}&-0.168\sym{***}&-0.240\sym{***}&-0.220\sym{***}\\
&&(0.032)&(0.033)&(0.047)&(0.043)&(0.032)&(0.031)&(0.030)&(0.036)\\
\textit{partner}&&0.029\sym{***}&0.025\sym{***}&0.029\sym{***}&0.025\sym{***}&0.028\sym{***}&0.024\sym{***}&0.031\sym{***}&0.027\sym{***}\\
&&(0.002)&(0.003)&(0.003)&(0.003)&(0.003)&(0.003)&(0.002)&(0.002)\\
\textit{health}&&0.004\sym{***}&0.004\sym{***}&0.004\sym{***}&0.004\sym{***}&0.003\sym{***}&0.003\sym{**}&0.006\sym{***}&0.005\sym{***}\\
&&(0.001)&(0.001)&(0.001)&(0.001)&(0.001)&(0.001)&(0.001)&(0.001)\\
\textit{child}&&0.004&0.003&0.004&0.002&0.005\sym{**}&0.004&0.005\sym{*}&0.003\\
&&(0.003)&(0.003)&(0.003)&(0.003)&(0.002)&(0.003)&(0.003)&(0.003)\\
\textit{bornc}&&&-0.025\sym{***}&&-0.023\sym{***}&&-0.029\sym{***}&&-0.026\sym{***}\\
&&&(0.006)&&(0.007)&&(0.005)&&(0.006)\\
\textit{fathere}&&&-0.000&&0.000&&-0.000&&-0.000\\
&&&(0.001)&&(0.001)&&(0.001)&&(0.001)\\
\textit{mothere}&&&-0.004\sym{***}&&-0.004\sym{***}&&-0.004\sym{***}&&-0.004\sym{***}\\
&&&(0.001)&&(0.001)&&(0.001)&&(0.001)\\
\midrule
\textit{N}&233,556&227,390&201,539&230,321&203,994&230,747&204,378&228,640&202,599\\
\textit{idp}&0.000&0.000&0.000&0.000&0.000&0.000&0.000&0.000&0.000\\
\textit{cdf}&2,322&2,364&2,140&847&768&2,705&2,502&2,000&1,718\\
\textit{widstat}&188&196&162&84&79&143&131&192&160\\
\bottomrule
\end{tabular}
\begin{tablenotes}
\item Notes: Robust standard errors clustered at the instruments level are in parentheses. All regressions include the full set of fixed effects.   *$p<0.10$; **$p<0.05$; ***$p<0.01$.
\end{tablenotes}
\end{threeparttable}
}
\end{table}


\begin{table}[H]\centering
\def\sym#1{\ifmmode^{#1}\else\(^{#1}\)\fi}
\caption*{\emph{Table D6. IV Estimates: Religiosity and Following Rules \label{tabd6}}}
\resizebox{\columnwidth}{0.32\linewidth}{%
\begin{threeparttable}
\begin{tabular}{l*{9}{c}}
\toprule
\addlinespace
\textit{rules} &\multicolumn{1}{c}{(1)}&\multicolumn{1}{c}{(2)}&\multicolumn{1}{c}{(3)}&\multicolumn{1}{c}{(4)}&\multicolumn{1}{c}{(5)}&\multicolumn{1}{c}{(6)}&\multicolumn{1}{c}{(7)}&\multicolumn{1}{c}{(8)}&\multicolumn{1}{c}{(9)}\\
\addlinespace
\midrule
\textit{religiosity}&0.164\sym{**}&0.199\sym{***}&0.225\sym{***}&&&&&&\\
&(0.066)&(0.064)&(0.070)&&&&&&\\
\textit{degree}&&&&0.158\sym{*}&0.176\sym{*}&&&&\\
&&&&(0.090)&(0.098)&&&&\\
\textit{attendance}&&&&&&0.240\sym{***}&0.244\sym{***}&&\\
&&&&&&(0.059)&(0.063)&&\\
\textit{pray}&&&&&&&&0.090\sym{**}&0.123\sym{**}\\
&&&&&&&&(0.045)&(0.049)\\
\textit{gender}&-0.024\sym{***}&-0.025\sym{***}&-0.025\sym{***}&-0.022\sym{***}&-0.023\sym{***}&-0.018\sym{***}&-0.018\sym{***}&-0.023\sym{***}&-0.026\sym{***}\\
&(0.005)&(0.005)&(0.005)&(0.006)&(0.007)&(0.003)&(0.004)&(0.006)&(0.006)\\
\textit{age}&0.000&-0.001\sym{*}&-0.001\sym{**}&-0.001\sym{*}&-0.001\sym{**}&-0.001&-0.001\sym{**}&-0.001\sym{**}&-0.001\sym{**}\\
&(0.000)&(0.001)&(0.001)&(0.001)&(0.001)&(0.001)&(0.001)&(0.001)&(0.001)\\
\textit{age2}&0.000\sym{***}&0.000\sym{***}&0.000\sym{***}&0.000\sym{***}&0.000\sym{***}&0.000\sym{***}&0.000\sym{***}&0.000\sym{***}&0.000\sym{***}\\
&(0.000)&(0.000)&(0.000)&(0.000)&(0.000)&(0.000)&(0.000)&(0.000)&(0.000)\\
\textit{education}&-0.004\sym{***}&-0.004\sym{***}&-0.003\sym{***}&-0.003\sym{***}&-0.003\sym{***}&-0.004\sym{***}&-0.004\sym{***}&-0.004\sym{***}&-0.004\sym{***}\\
&(0.000)&(0.000)&(0.000)&(0.000)&(0.000)&(0.000)&(0.000)&(0.000)&(0.000)\\
\textit{paidwork}&&0.007\sym{**}&0.008\sym{***}&0.006\sym{**}&0.007\sym{**}&0.007\sym{**}&0.008\sym{**}&0.006\sym{**}&0.008\sym{**}\\
&&(0.003)&(0.003)&(0.003)&(0.003)&(0.003)&(0.003)&(0.003)&(0.003)\\
\textit{pwbefore}&&-0.268\sym{***}&-0.215\sym{***}&-0.280\sym{***}&-0.223\sym{***}&-0.250\sym{***}&-0.198\sym{***}&-0.304\sym{***}&-0.271\sym{***}\\
&&(0.055)&(0.061)&(0.064)&(0.075)&(0.060)&(0.063)&(0.044)&(0.054)\\
\textit{partner}&&0.021\sym{***}&0.020\sym{***}&0.021\sym{***}&0.020\sym{***}&0.019\sym{***}&0.017\sym{***}&0.023\sym{***}&0.022\sym{***}\\
&&(0.003)&(0.003)&(0.003)&(0.003)&(0.003)&(0.003)&(0.002)&(0.003)\\
\textit{health}&&0.003\sym{*}&0.003\sym{**}&0.003\sym{*}&0.003\sym{**}&0.001&0.002&0.004\sym{***}&0.005\sym{***}\\
&&(0.001)&(0.001)&(0.001)&(0.001)&(0.002)&(0.002)&(0.001)&(0.002)\\
\textit{child}&&0.001&-0.003&0.002&-0.001&0.002&-0.002&0.002&-0.002\\
&&(0.003)&(0.003)&(0.003)&(0.003)&(0.003)&(0.003)&(0.003)&(0.003)\\
\textit{bornc}&&&-0.033\sym{***}&&-0.037\sym{***}&&-0.036\sym{***}&&-0.035\sym{***}\\
&&&(0.006)&&(0.007)&&(0.005)&&(0.007)\\
\textit{fathere}&&&-0.001&&-0.001&&-0.002&&-0.002\\
&&&(0.001)&&(0.001)&&(0.001)&&(0.001)\\
\textit{mothere}&&&-0.004\sym{***}&&-0.004\sym{***}&&-0.004\sym{***}&&-0.004\sym{***}\\
&&&(0.001)&&(0.001)&&(0.001)&&(0.001)\\
\midrule
\textit{N}&232,561&226,405&200,723&229,292&203,139&229,694&203,509&227,635&201,768\\
\textit{idp}&0.000&0.000&0.000&0.000&0.000&0.000&0.000&0.000&0.000\\
\textit{cdf}&2,313&2,356&2,140&811&747&2,723&2,515&2,007&1,727\\
\textit{widstat}&188&195&160&80&75&143&130&194&164\\
\bottomrule
\end{tabular}
\begin{tablenotes}
\item Notes: Robust standard errors clustered at the instruments level are in parentheses. All regressions include the full set of fixed effects.   *$p<0.10$; **$p<0.05$; ***$p<0.01$.
\end{tablenotes}
\end{threeparttable}
}
\end{table}

\pagebreak[5]

\printbibliography
\end{document}